\documentclass[amsmath,preprintnumbers,aps,nofootinbib,tightenlines,floatfix,showpacs,superscriptaddress]{revtex4}
\usepackage{amssymb,latexsym}
\usepackage{amsmath,amsbsy,bbm}
\usepackage{epsfig,bm}
\usepackage{graphicx,comment}
\usepackage{bm,comment}
\usepackage{epstopdf}
\usepackage{color}
\usepackage{subfigure}
\usepackage{slashed,array,mathtools,amsfonts,multirow,hyperref}

\newcommand{\avg}[1]{\left< #1 \right>} 

\newcommand{\mbraket}[3]{\left< #1 \vphantom{#2#3} \right|
 #2 \left| #3 \vphantom{#1#2} \right>} 
\newcommand{\tr}{\text{Tr}} 
\let\bar=\smallbar 
\newcommand{\bar}[1]{\overline{#1}} 
\newcommand{\fs}[1]{\slashed{#1}} 
\let\tilde=\widetilde

\renewcommand{\arraystretch}{1.2}
\begin{document}

\preprint{INT-PUB-15-014}

\title{Perturbative Renormalization of Neutron-Antineutron Operators}

\author{Michael I. Buchoff}
\affiliation{%
Institute for Nuclear Theory, Box 351550, Seattle, WA 98195-1550, USA}

\author{Michael Wagman}
\affiliation{%
Institute for Nuclear Theory, Box 351550, Seattle, WA 98195-1550, USA}
\affiliation{%
Department of Physics, University of Washington, Box 351560, Seattle, WA 98195, USA}



\begin{abstract}
  Two-loop anomalous dimensions and one-loop renormalization scheme matching factors are calculated for six-quark operators responsible for neutron-antineutron transitions. When combined with lattice QCD determinations of the matrix elements of these operators, our results can be used to reliably predict the neutron-antineutron vacuum transition time, $\tau_{n\bar{n}}$, in terms of basic parameters of baryon-number violating beyond-the-Standard-Model theories. The operators are classified by their chiral transformation properties, and a basis in which there is no operator mixing due to strong interactions is identified. Operator projectors that are required for non-perturbative renormalization of the corresponding lattice QCD six-quark operator matrix elements are constructed. A complete calculation of $\delta m = 1/\tau_{n\bar{n}}$ in a particular beyond-the-Standard-Model theory is presented as an example to demonstrate how operator renormalization and results from lattice QCD are combined with experimental  bounds on $\delta m$ to constrain the scale of new baryon-number violating physics. At the present computationally accessible lattice QCD matching scale of $\sim$ 2 GeV, the next-to-next-to-leading-order effects calculated in this work correct the leading-order plus next-to-leading-order $\delta m$ predictions of beyond-the-Standard-Model theories by $< 26\%$. Next-to-next-to-next-to-leading-order effects provide additional unknown corrections to predictions of $\delta m$ that are estimated to be $< 7\%$.
\end{abstract}
\maketitle

\section{Introduction}  \label{sec:intro}

The universe contains many more baryons than antibaryons~\cite{Ade:2013zuv}. Unless this baryon asymmetry is attributed to fine-tuning of the initial conditions of the universe, the baryon asymmetry must have been generated dynamically during the early universe. Any mechanism describing this process of baryogenesis must include violation of baryon-number $(B)$ conservation, violation of $C$ and $CP$, and departure from thermal equilibrium~\cite{Sakharov:1967dj}.  The Standard Model includes $B$ violation through non-perturbative electroweak processes that violate $B+L$ but preserve $B-L$~\cite{tHooft:1976up, tHooft:1976fv}. It also includes classical $C$ and $CP$ violation and departure from thermal equilibrium during the electroweak phase transition. However, the $B$ and $CP$ violating effects present in the Standard Model \textit{cannot} reproduce the observed magnitude of the baryon asymmetry~\cite{Kuzmin:1985mm, Cohen:1993nk, Moore:1998ge}.  As a result, Beyond-the-Standard-Model (BSM) physics is needed to explain baryogenesis.  BSM baryon number violation could occur in many different ways. Theories that allow $\Delta B = 1$ transitions can allow $B-L$ conserving proton decay,\footnote{See Ref.~\cite{Senjanovic:2009kr} for a recent review on proton decay} which has been experimentally constrained to a high degree~\cite{Nishino:2009aa,Regis:2012sn,Abe:2014mwa}. Other classes of BSM theories do not allow proton decay, but do allow other baryon number violating processes. These models often instead include the $\Delta B = 2$, $B-L$ violating, neutron-antineutron transition~\cite{glashow79,Mohapatra:1980qe,Chacko:1998td,Dvali:1999gf,Babu:2001qr,Nussinov:2001rb,Barbier:2004ez,Davoudiasl:2004gf,Bambi:2006mi,Dolgov:2006ay,Babu:2008rq,Winslow:2010wf,Csaki:2011ge,Arnold:2012sd, Babu:2012vc,Babu:2013yca,Herrmann:2014fha,Deppisch:2014zta,Gripaios:2014tna,Perez:2015rza,Addazi:2015ata,Brennan:2015psa,Dev:2015uca,Addazi:2015oba,Addazi:2014ila,Addazi:2015hka,Addazi:2015rwa,Addazi:2015eca}. 

In vacuum, neutron-antineutron ($n\bar{n}$) transitions would manifest themselves as oscillations between neutrons and antineutrons. The probability that a free neutron has transformed into an antineutron after time $t$ is given by $P_{n\bar{n}} = \sin^2(t/\tau_{n\bar{n}})$, where $\tau_{n\bar{n}}$ is the neutron-antineutron vacuum transition time. Experimental measurements of magnetically shielded cold neutron beams at the Institut Laue-Langevin (ILL) have established a limit of $\tau_{n\bar{n}} > 2.7\text{ years}$~\cite{baldoCeolin:1994jz}. There are also experimental bounds on the decay rate of neutrons bound in nuclei from large volume underground detectors. Super-K has bounded the transition time $\tau_{O^{16}}$ for $n\bar{n}$ transitions in oxygen, $\tau_{O^{16}} > 1.89 \times 10^{32}\text{ years}$~\cite{Abe:2011ky}. Nuclear structure calculations can be used to relate this nuclear transition time to the vacuum transition time $\tau_{n\bar{n}}$. This bound on the vacuum transition time is estimated to be a factor of four or five larger than the ILL bound, but the nuclear structure calculations introduce non-trivial systematic uncertainties.\footnote{In particular Ref.~\cite{Abe:2011ky} cites a derived bound of $\tau_{n\bar{n}} > 7.7\text{ years}$. More recent structure calculations in Ref.~\cite{Friedman:2008es} modify this bound to be $\tau_{n\bar{n}} > 10.9\text{ years}$, as noted in Ref.~\cite{Phillips:2014fgb}.} It is believed that improvements in neutron transport/optics and neutron moderation technologies since the 1994 ILL experiment would allow for new neutron beam experiments to improve the ILL bounds by an order of magnitude or more~\cite{Phillips:2014fgb}. There has been a recent push from both theoretical and experimental communities in support of new, state-of-the-art $n\bar{n}$ experiments.~\cite{Kronfeld:2013uoa,Babu:2013yww,Phillips:2014fgb}.

In order to constrain BSM theories predicting $n\bar{n}$ transitions, experimental results must be compared to theoretical predictions for $\tau_{n\bar{n}}$. Making reliable predictions for $\tau_{n\bar{n}}$ within a particular BSM theory is challenging. In particular, theoretical descriptions of the  $n\bar{n}$ transition process must include strong interaction physics as well as BSM physics.  These effects are important at very different scales. High-scale BSM physics gives rise to effectively local $\Delta B = 2$ interactions turning three quarks into three antiquarks. Comparatively low-scale strong interactions bind these quarks (antiquarks) into a neutron (antineutron). Theoretical descriptions of this high- and low-scale physics can be factorized by using a Standard Model effective field theory description of $n\bar{n}$ transitions. In this approach, the Hamiltonian governing $n\bar{n}$ transitions is described as a linear combination of operators built from Standard Model fields.

The most relevant Standard Model effective field theory operators contributing to $n\bar{n}$ transitions are dimension-nine six-quark operators. A complete basis of these six-quark operators can be constructed without specializing to a particular BSM theory. This construction was begun in Refs.~\cite{Chang:1980ey,Kuo:1980ew}, generalized and detailed in Refs.~\cite{Rao:1982gt,Rao:1983sd}, and completed in Ref.~\cite{Caswell:1982qs} where spin-color Fierz identities were used to remove redundant operators from the basis. Higher-order operators of potential interest have also been discussed~\cite{Gardner:2014cma,Babu:2015axa}. The effects of low-scale strong interaction physics on $n\bar{n}$ transitions are encoded in quantum chromodynamic (QCD) matrix elements of six-quark operators between initial neutron and final antineutron states. All high-scale physics and BSM model dependence is encoded in the particular numerical coefficients used to express the effective Hamiltonian for a given theory in a six-quark operator basis. These numerical coefficients can be calculated perturbatively in BSM matching calculations for particular theories of interest.

Testable predictions for $\tau_{n\bar{n}}$ cannot be made without reliable calculations of six-quark operator QCD matrix elements. Equivalently, experimental bounds on $\tau_{n\bar{n}}$ cannot be used to constrain BSM theory without reliable QCD matrix element calculations. These six-quark matrix elements have been estimated in the MIT bag model~\cite{Rao:1982gt,Rao:1983sd}, but model estimates introduce uncontrolled uncertainties into the relation between BSM parameters and experimental observables \cite{Fajfer:1983ja}. The only available method to determine hadronic matrix elements with controlled uncertainties is lattice QCD. Preliminary lattice QCD calculations of $n\bar{n}$ matrix elements are underway~\cite{Buchoff:2012bm}. Once completed, lattice QCD $n\bar{n}$ matrix elements can be non-perturbatively renormalized and then combined with BSM matching calculations performed with renormalized perturbation theory.

The need for perturbative $n\bar{n}$ operator renormalization arises because lattice QCD matrix elements can only be renormalized at scales smaller than the UV cutoff of the lattice, (typical calculations today use lattice matching scales of $p_0 \simeq 2\ \text{GeV}$~\cite{Aoki:2013ldr}) but renormalization scales that are currently accessible in lattice QCD simulations cannot be (directly) used for perturbative BSM matching calculations. These BSM matching calculations receive logarithmic corrections that become large enough to invalidate perturbation theory unless the renormalization scale chosen is comparable to high scales where BSM physics becomes important. For typical BSM theories, these scales are in the range $\Lambda_{BSM} = 10^2 - 10^{16}\ \text{GeV}$. To address this issue, renormalization group (RG) techniques can be used to sum these large logs and reliably relate matrix elements calculated with different renormalization scales. This RG evolution (``running'') and typical BSM matching calculations are both simplest in mass-independent, renormalization schemes such as modified minimal subtraction (NDR-$\bar{\text{MS}}$).\footnote{Naive dimensional regularization (NDR) prescribes that $\gamma_5$ anticommutes with $\gamma_\mu$ in $D$ dimensions. Since closed fermion loops do not appear in $n\bar{n}$ calculations, no complications arise from using the NDR prescription. In the remainder of this paper we abbreviate NDR-$\bar{\text{MS}}$ as $\bar{\text{MS}}$ for brevity.} The $\bar{\text{MS}}$ renormalization scheme can only be applied directly to dimensionally regularized matrix elements, and in particular cannot be applied directly to lattice regularized matrix elements. Instead, the Regularization-Independent-Momentum (RI-MOM) scheme can be introduced as an intermediate renormalization scheme~\cite{Martinelli:1994ty}. As long as the lattice matching scale $p_0$ used for non-perturbative renormalization is larger than hadronic scales where QCD becomes non-perturbative, it is possible to relate RI-MOM and $\bar{\text{MS}}$ renormalized matrix elements perturbatively (``matching''). The perturbative calculation of RG running and matching factors therefore allows non-perturbatively renormalized lattice QCD matrix elements to be combined with perturbative BSM matching calculations to provide testable predictions for $\tau_{n\bar{n}}$ in BSM theories of interest.

The largest corrections to $\tau_{n\bar{n}}$ arising from RG evolution are encoded in perturbative one-loop-running factors. These have been correctly calculated for $n\bar{n}$ operators in Ref.~\cite{Caswell:1982qs}. One-loop running provides an overall multiplicative correction to non-perturbatively renormalized matrix elements, see Eq.~\eqref{master}. Further RG corrections to this result can be organized as a power series in $\alpha_s(p_0)$. In order to verify that this perturbative expansion is well-controlled at a given $p_0$, it is necessary to determine the first term in this $\alpha_s(p_0)$ power series. This term is parametrically $O(\alpha_s(p_0))$, and includes one-loop-matching effects. When running to high scales $\mu$ where $\alpha_s(\mu)\ll \alpha_s(p_0)$, two-loop-running effects also contribute at $O(\alpha_s(p_0))$ and must be included as well, see Eq.~\eqref{master}. This work provides the first calculation of the one-loop-matching and two-loop-running factors needed to reliably estimate the convergence of RG relations between $n\bar{n}$ matrix elements at low scales $p_0$ accessible to lattice QCD simulations and high scales $\mu$ accessible to perturbative BSM matching calculations.

The remainder of this paper begins with a summary of our final results in Sec.~\ref{sec:renorm}. Results are presented for the fixed-flavor basis commonly used in the literature and for a new chiral basis that is diagonal under RG evolution. The construction of this chiral basis is presented in Sec.~\ref{sec:ops}. The RI-MOM renormalization scheme and associated operator projectors needed for perturbative and non-perturbative $n\bar{n}$ operator renormalization are defined in Sec.~\ref{sec:renormschemes}. Calculation of one-loop-matching factors relating RI-MOM and $\bar{\text{MS}}$ renormalized operators is discussed in Sec.~\ref{sec:matching}. Calculation of two-loop-running factors is discussed in Sec.~\ref{sec:running}. Both Sec.~\ref{sec:matching}-\ref{sec:running} discuss the careful treatment of evanescent operators vanishing in $D=4$ that is necessary for a correct calculation of RG effects. To demonstrate the phenomenological application of our results, a complete calculation of $\tau_{n\bar{n}}^{-1}$  and resulting experimental constraints are discussed for a simplified BSM model in Sec.~\ref{sec:pheno}. Physical results and implications are summarized in Sec.~\ref{sec:concl}. Analogous one-loop-matching and two-loop-running calculations have been performed for four-quark weak matrix elements~\cite{Altarelli:1974exa,Gaillard:1974nj,Buras:1989xd,Buras:1992tc,Buras:2000if,Aoki:2007xm,Allton:2008pn,Aoki:2010pe,Buras:2012fs} and proton decay~\cite{Ellis:1981tv,Pivovarov:1991nk,Nihei:1994tx,Aoki:2006ib,Aoki:2008ku,Gockeler:2008we, Kraenkl:2011qb, Aoki:2013yxa}, the latter of which has also recently been analyzed at the level of two-loop-matching and three-loop-running~\cite{Gracey:2012gx}. These calculations provide useful techniques as well as cross-checks for intermediate results. We avoid discussion of established techniques for multi-loop diagram evaluation in the main text, but for readers unfamiliar with two-loop diagram evaluation we present a pedagogical discussion in Appendices~\ref{app:algebra}~-~\ref{app:integrals}. Our explicit evanescent operator basis (technically required for a full definition of $\bar{\text{MS}}$ operator renormalization) is presented in Appendix~\ref{app:evops}, and some intermediate results are shown in Appendix~\ref{app:tables}.

\section{Summary of Results} \label{sec:renorm}

The neutron-antineutron vacuum transition time $\tau_{n\bar{n}}$ predicted by a particular BSM theory can be calculated from matrix elements of the Hamiltonian density
\begin{equation}
	\mathcal{H}^{n\bar{n}}_{eff} = \sum_I C_I(\mu)Q_I(\mu)
\end{equation}
where the $Q_I(\mu)$ form a complete basis of dimension-nine local six-quark operators with non-vanishing matrix elements $\mbraket{\bar{n}}{Q_I(\mu)}{n}$ between initial neutron and final antineutron states, the $C_I(\mu)$ are Wilson coefficients, and $\mu$ is a renormalization scale. The Wilson coefficients are renormalization scheme and scale dependent and will differ between BSM theories. They can be calculated by matching tree- or one-loop- level $n\bar{n}$ amplitudes between the full BSM theory and an effective theory containing only Standard Model degrees of freedom. Hadronic matrix elements of $Q_I(\mu)$ are independent of the BSM theory used to calculate the $C_I(\mu)$ but renormalization scheme and scale dependent.

Lattice QCD first determines matrix elements of bare, lattice regularized operators. By subsequent lattice QCD calculations, these bare matrix elements can be non-perturbatively renormalized in the RI-MOM scheme described in Sec.~\ref{sec:renormschemes} at a lattice matching scale $p_0$. Provided $\alpha_s(p_0)\ll 1$, dimensionally regularized perturbation theory can be used to relate RI-MOM renormalized matrix elements to $\bar{\text{MS}}$ renormalized matrix elements. Introduction of RI-MOM as an intermediate renormalization scheme is necessary because the $\bar{\text{MS}}$ scheme can only be directly applied to dimensionally regularized (and not, for instance, lattice regularized) matrix elements. Setting the $\bar{\text{MS}}$ renormalization scale $\mu=p_0$ removes large logarithms from the RI-MOM matching calculation. Perturbative calculations of $C_I^{\bar{\text{MS}}}(\mu)$ in a particular BSM theory typically introduce additional logarithmic corrections $\ln(\mu/\Lambda_{BSM})$. Since lattice QCD computational limits demand $p_0 \ll \Lambda_{BSM}$, Wilson coefficients calculated at $\mu = \Lambda_{BSM}$ must be RG evolved to $\mu = p_0$ and then combined with $\bar{\text{MS}}$ renormalized matrix elements to include all large logs in BSM theory predictions of $\tau_{n\bar{n}}$.

\renewcommand{\arraystretch}{1.6}
\begin{table}
\begin{tabular}{|c|c|c|c|c|}\hline
	Chiral Basis & Flavor Basis & $\gamma_{I}^{(0)}$ & $\gamma_{I}^{(1)}$ & $r_{I}^{(0)}$  \\\hline
	$Q_1$ & $\mathcal{O}^3_{RRR},\;\mathcal{O}^3_{LLL}$ & \hspace{10pt}$4$\hspace{10pt} & \hspace{10pt}$335/3-34N_f/9$\hspace{10pt} & \hspace{10pt}$101/30 + 8/15\ln 2$\hspace{10pt} \\\hline
	$Q_2$ & $\mathcal{O}^3_{LRR},\;\mathcal{O}^3_{RLR},\;\mathcal{O}^3_{RLL},\;\mathcal{O}^3_{LRL}$ & $-4$ & $91/3 - 26N_f/9$ & $-31/6 + 88/15\ln 2$  \\\hline
	$Q_3$ & $\mathcal{O}^3_{LLR}$,\;$\mathcal{O}^3_{RRL}$ & $0$ & $64-10N_f/3$ & $-9/10+16/5\ln 2$  \\\hline
	\multirow{2}{*}{$Q_4$} & $\left( 4/5\;\mathcal{O}^2_{RRR} + 1/5\;\mathcal{O}^1_{RRR} \right),$ & \multirow{2}{*}{$24$} & \multirow{2}{*}{$229-46N_f/3$} & \multirow{2}{*}{$177/10-64/5\ln 2$} \\
	& $\left(4/5\;\mathcal{O}^2_{LLL} + 1/5\;\mathcal{O}^1_{LLL}  \right)$ &&& \\\hline
	\multirow{6}{*}{$Q_5$} & $\mathcal{O}^1_{RLL},\;\mathcal{O}^1_{LRR},\; \mathcal{O}^2_{RLL},$ & \multirow{6}{*}{$12$} & \multirow{6}{*}{$238-14N_f$} & \multirow{6}{*}{$49/10 - 24/5\ln 2$} \\
	& $\mathcal{O}^2_{LRL},\;\mathcal{O}^2_{LRR},\;\mathcal{O}^2_{RLR},$ &&& \\
	& $(2/3\;\mathcal{O}^2_{LLR} + 1/3\;O^1_{LLR}),$ &&& \\
	& $(2/3\;\mathcal{O}^2_{LLR} + 1/3\;O^1_{LRL}),$ &&& \\
	& $(2/3\;\mathcal{O}^2_{RRL} + 1/3\;O^1_{RRL}),$ &&& \\
	& $(2/3\;\mathcal{O}^2_{RRL} + 1/3\;O^1_{RLR})$ &&& \\\hline
	\multirow{2}{*}{$\tilde{Q}_1$} & $(1/3\;\mathcal{O}^2_{RRR}- 1/3\;\mathcal{O}^1_{RRR}),$ & \multirow{2}{*}{$4$} & \multirow{2}{*}{$797/3-118N_f/9$} & \multirow{2}{*}{$-109/30+8/15\ln 2$} \\
	& $(1/3\;\mathcal{O}^2_{LLL}- 1/3\;\mathcal{O}^1_{LLL})$ &&&\\\hline
	\multirow{2}{*}{$\tilde{Q}_3$} & $(1/3\;\mathcal{O}^2_{LLR}- 1/3\;\mathcal{O}^1_{LLR}),$ & \multirow{2}{*}{$0$} & \multirow{2}{*}{$218-38N_f/3$} & \multirow{2}{*}{$-79/10+16/5\ln 2$} \\
	& $(1/3\;\mathcal{O}^2_{RRL}- 1/3\;\mathcal{O}^1_{RRL})$ &&&\\\hline
\end{tabular}
\caption{Summary of results. The left-most column lists the chiral basis operators $Q_I$ with independent NNLO operator renormalization factors. The second column lists the corresponding fixed-flavor basis operators used in Ref.~\cite{Chang:1980ey,Kuo:1980ew,Rao:1982gt,Rao:1983sd,Caswell:1982qs} that renormalize identically to $Q_I$, see Sec.~\ref{sec:ops}. Each $Q_I$ is proportional to the first fixed-flavor basis operator listed in the corresponding row of the second column with proportionality factor $1$ for $Q_5$ and $(-4)$ for all other $Q_I$. The other fixed-flavor basis operators listed may not be directly proportional to $Q_I$ but share the same one-loop $\bar{\text{MS}}$ anomalous dimension $\gamma_I^{(0)}$ (third column), two-loop $\bar{\text{MS}}$ anomalous dimension $\gamma_I^{(1)}$ (fourth column), and one-loop Landau gauge RI-MOM matching factor $r_I^{(0)}$ (fifth column) appearing in Eq.~\eqref{master}. $\gamma_I^{(1)}$ and $r_I^{(0)}$ depend on the evanescent operator basis used to extend $D=4$ Fierz relations to $D$-dimensional operator relations in dimensional regularization. Our evanescent operator basis is presented in Appendix~\ref{app:evops}. One-loop BSM matching calculations must use the same evanescent operator basis for consistency. Tree-level BSM matching calculations are unaffected, see Sec.~\ref{sec:pheno} for an example matching calculation.\label{tab:summary}}
\end{table}

The renormalization scale dependence of the Wilson coefficients is encoded in the $\bar{\text{MS}}$ anomalous dimension matrix $\gamma_{IJ}$, defined in Sec.~\ref{sec:running}. In Sec.~\ref{sec:ops}, we use chiral flavor symmetry to construct an operator basis where the anomalous dimension matrix is diagonal. The RG equations relating Wilson coefficients at different renormalization scales can be solved perturbatively in this diagonal chiral basis. Including one-loop-matching and two-loop-running effects, the relation between the desired Hamiltonian $\mathcal{H}_{eff}^{n\bar{n}}$, the BSM matching coefficients $C_I^{\bar{\text{MS}}}(\mu)$ at arbitrary scale $\mu$, and the non-perturbatively renormalized operators $Q_I^{\text{RI}}(p_0)$ used in lattice QCD simulations is
\begin{equation}
  \begin{split}
    \mathcal{H}_{eff}^{n\bar{n}} &= \sum_I C_I^{\bar{\text{MS}}}(\mu) U_I(\mu,p_0) Q_I^{\text{RI}}(p_0),\\\\
   U_I(\mu,p_0) &= \begin{cases}U_I^{N_f=6}(\mu,m_t)U_I^{N_f=5}(m_t,m_b)U_I^{N_f=4}(m_b,p_0) & \text{for} \quad m_c < p_0 < m_b \\  U_I^{N_f=6}(\mu,m_t)U_I^{N_f=5}(m_t,p_0) &  \text{for} \quad m_b < p_0 < m_t  \end{cases} \\\\
  U_I^{N_f}(\mu_1,\mu_2) &= \left( \frac{\alpha_s(\mu_2)}{\alpha_s(\mu_1)} \right)^{-\gamma_{I}^{(0)}/2\beta_0}\left[ 1 - \delta_{\mu_2, p_0} r_{I}^{(0)}\frac{\alpha_s(p_0)}{4\pi} + \left( \frac{\beta_1 \gamma_{I}^{(0)}}{2\beta_0^2} - \frac{\gamma_{I}^{(1)}}{2\beta_0} \right)\frac{\alpha_s(\mu_2)-\alpha_s(\mu_1)}{4\pi} + O(\alpha_s^2)\right],
\end{split}
	\label{master}
\end{equation}
where $r_{I}^{(0)}$ is a one-loop-matching factor defined in Sec.~\ref{sec:matching}, $\gamma_{I}^{(0)}$ and $\gamma^{(1)}_{I}$ are one-loop- and two-loop-running factors defined in Sec.~\ref{sec:running}, and $\beta_0$ and $\beta_1$ are well-known perturbative coefficients of the QCD $\beta$-function presented for reference in Eq.~\eqref{betacoff}.  Only $\beta_0$, $\beta_1$, and $\gamma^{(1)}_{I}$ depend on the number of active quark flavors, $N_f$. Matching between theories with different $N_f$ at quark thresholds is included in the same manner as in RG evolution of weak matrix elements without penguin contributions \cite{Buras:1998raa} since no penguin diagrams exist for $n\bar{n}$ operators.

Ignoring QCD effects on RG evolution gives the leading-order (LO) result $U_I(\mu,p_0) = 1$. Next-to-leading-order (NLO) QCD effects give a multiplicative correction to $U_I(\mu,p_0)$ whose size is determined by the one-loop-running factor $\gamma_I^{(0)}$ correctly calculated in Ref.~\cite{Caswell:1982qs}. Higher-order corrections due to matching and running provide additive corrections that can be perturbatively expanded in powers of $\alpha_s(p_0)$ and $\alpha_s(\mu)$. For high scales $\mu$ where $\alpha_s(\mu) \ll \alpha_s(p_0)$, Eq.~\eqref{master} shows that one-loop-matching and two-loop-running effects receive similar $O(\alpha_s(p_0))$ suppression. Both one-loop-matching and two-loop-running effects must therefore be included in a next-to-next-to-leading-order (NNLO) calculation of $U_I(\mu,p_0)$. Next-to-next-to-next-to-leading-order (N$^3$LO) corrections not included in Eq.~\eqref{master} arise from two-loop-matching and three-loop-running effects that are both $O(\alpha_s(p_0)^2)$ suppressed.

The NNLO operator renormalization factors $r_{I}^{(0)}$ and $\gamma_{I}^{(1)}$ are calculated for the first time here and summarized in Table \ref{tab:summary}. The relative size of NNLO to NLO corrections to $U_I(\mu,p_0)$ depends on $\mu$ and differs between operators. Taking $p_0 = 2\text{ GeV}$ and using the four-loop parametrization of $\alpha_s(\mu)$ in Ref.~\cite{Bethke:2009jm}, NNLO corrections to NLO+LO results for $\delta m \equiv 1/\tau_{n\bar{n}}$ are $< 26\%$ for all $\mu \geq p_0$ and may be significantly smaller in some BSM theories. Sec.~\ref{sec:pheno} presents an example calculation of the $n\bar{n}$ vacuum transition rate for one of the simplified models of Ref.~\cite{Arnold:2012sd}. In this model the relative size of NNLO to NLO corrections to $\delta m$ is $14\%$. Estimating that unknown N$^3$LO $O(\alpha_s(p_0)^2)$ corrections are comparable to the square of NNLO $O(\alpha_s(p_0))$ corrections allows systematic uncertainty due to unknown N$^3$LO corrections to be quantified as $< 7\%$ generically and $2\%$ in the model discussed in Sec.~\ref{sec:pheno}.


\section{Chiral Operator Basis}\label{sec:ops}

The operators relevant for $n\bar{n}$ transitions are Lorentz, color, and electromagnetic singlet six-quark operators of dimension nine. Since hadronic matrix elements must be calculated in lattice QCD simulations that only maintain approximate chiral symmetry at best, operators that are not singlets of the full electroweak gauge group should be considered. Even so, operator renormalization is most simply performed in the limit of massless up and down quarks. Classifying operators according to the $SU(2)_L\times SU(2)_R$ chiral symmetry of QCD in this limit proves quite useful.\footnote{We thank Brian Tiburzi for very helpful insights on these chiral transformation properties.} In this section we construct a basis of irreducible chiral tensor operators that do not mix under perturbative QCD interactions. Fierz relations and symmetries of the color, spin, and flavor tensors used throughout this section are detailed in Appendix \ref{app:algebra}. Our notational conventions are as follows: we use $i,j,k,\ldots$ as fundamental color indices, $\mu,\nu,\rho,\ldots$ as 4-vector Lorentz indices, $\alpha,\beta,\gamma,\ldots$ as Lorentz spinor indices, $a,b,c,\dots$ as flavor spinor indices, $I,J,K,\dots$ as operator basis labels, and $\chi$'s as chirality labels $L,R$. We will use $A,B,C,\dots$ to denote adjoint indices in both color and flavor. $\mathfrak{su}(3)_c$ color generators will be denoted by $t^A$ and normalized to $\tr(t^At^B) = \frac{1}{2}\delta^{AB}$ while $\mathfrak{su}(2)_L$ and $\mathfrak{su}(2)_R$ flavor generators will be denoted by $\tau^A$ and normalized as Pauli matrices $\tr(\tau^A\tau^B) = 2\delta^{AB}$ with $\tau^\pm \equiv \frac{1}{2}(\tau^1 \pm i \tau^2)$. We use Euclidean $(++++)$ metric signature and will not distinguish between raised and lowered indices. Final results are valid in Minkowski signature; intermediate steps are not. Summation convention applies to all indices but not to operator basis $I,J,K,\dots$ and chirality $\chi$ labels.

Two quarks can be combined into a spin-singlet diquark by contraction with the antisymmetric charge conjugation matrix $C$ and projected onto definite chirality by including $P_{L,R} = \frac{1}{2}(1\mp \gamma_5)$. In $D=4$, spin Fierz identities can be used to express any product of vector diquarks containing $\gamma_\mu$ or tensor diquarks containing $\sigma_{\mu\nu} = \frac{i}{2}[\gamma_\mu,\gamma_\nu]$ as a product of scalar diquarks. Denoting flavor doublet quark fields by $\psi^\alpha_{i a} = (u_i^\alpha,d_i^\alpha)$, only operators containing three products of scalar diquarks $\psi^\alpha_{i a}[CP_\chi]^{\alpha\beta}\psi^\beta_{j b}$ need to be considered.

Flavor Fierz identities allow us to only consider operators where each diquark is either a flavor singlet contracted with the antisymmetric tensor $i\tau^2_{ab}$ or a flavor vector contracted with the symmetric tensor $[i\tau^2\tau^A]_{ab}$,
\begin{equation}
	\mathcal{D}_\chi \equiv (\psi CP_\chi i\tau^2 \psi),\hspace{20pt} \mathcal{D}_\chi^A \equiv (\psi CP_\chi i\tau^2 \tau^A \psi),
\end{equation}
where we have suppressed free color indices. Irreducible $\mathfrak{su}(2)_\chi$-spin-two and $\mathfrak{su}(2)_\chi$-spin-three chiral tensor operators can then be defined as
\begin{equation}
  \begin{split}
	\mathcal{D}_\chi^{AB} &\equiv \mathcal{D}_\chi^{\{A}\mathcal{D}_\chi^{B\}} - \frac{1}{3}\delta^{AB}\mathcal{D}_\chi^C \mathcal{D}_\chi^C,\\
	\mathcal{D}_\chi^{ABC} &\equiv \mathcal{D}_\chi^{\{A}\mathcal{D}_\chi^B\mathcal{D}_\chi^{C\}} - \frac{1}{5}\left[ \delta^{AB}\mathcal{D}_\chi^{\{C}\mathcal{D}_\chi^D\mathcal{D}_\chi^{D\}} + \delta^{AC}\mathcal{D}_\chi^{\{B}\mathcal{D}_\chi^D\mathcal{D}_\chi^{D\}} + \delta^{BC}\mathcal{D}_\chi^{\{A}\mathcal{D}_\chi^D\mathcal{D}_\chi^{D\}} \right].
  \end{split}
\end{equation}
Since operators contributing to $n\bar{n}$ transitions must lower the third $SU(2)_V$ isospin component $I_3$ isospin by one unit,\footnote{In particular, $n\bar{n}$ transitions only involve operators with negative parity, $\Delta I = 1$, and $\Delta I_3 = -1$. We only explicitly enforce the latter constraint $\Delta I_3 = -1$  in order to simplify the perturbative calculations presented here.} at least one diquark must be contracted with $[i\tau^2\tau^+]_{ab}$ to form a $d_i^\alpha d_j^\beta$ diquark. The other two diquarks must combine to have no net effect on $I_3$. Taking this $d_i^\alpha d_j^\beta$ combination to be our third diquark for convenience and enforcing antisymmetry under quark exchange, the only available tensors for constructing color singlet six-quark operators are
\begin{equation}
  \begin{split}
	T^{SSS}_{\{ij\}\{kl\}\{mn\}} &= \varepsilon_{ikm}\varepsilon_{jln} + \varepsilon_{jkm}\varepsilon_{iln} + \varepsilon_{ilm}\varepsilon_{jkn} + \varepsilon_{ikn}\varepsilon_{jlm},\\
	T^{AAS}_{[ij][kl]\{mn\}} &= \varepsilon_{ijm}\varepsilon_{kln} + \varepsilon_{ijn}\varepsilon_{klm},
  \end{split}
\end{equation}
where $\{\;\}$ denotes index symmetrization and $[\;\;]$ denotes index antisymmetrization. From here onward we suppress explicit quark indices and use the diquark notation
\begin{equation}
  (\psi_i CP_R i\tau^2 \psi_j) \equiv \psi^\alpha_{i a}[CP_R]^{\alpha\beta}[i\tau^2]_{ab}\psi^\beta_{j b}.
\end{equation}
We further suppress color indices in diquark products, e.g. $(\psi\psi)(\psi\psi)(\psi\psi)T^{AAS} \equiv (\psi_i\psi_j)(\psi_k\psi_l)(\psi_m\psi_n)T^{AAS}_{[ij][kl]\{mn\}}$.

\renewcommand{\arraystretch}{1.5}
\begin{table}
  \begin{tabular}{|c|c|c|c|}
	\hline
	Chiral Basis & Fixed-Flavor Basis & Chiral Tensor Structure & Chiral Irrep  \\\hline 
	$Q_1$ & $\mathcal{O}^3_{RRR}$ & $\mathcal{D}_R\mathcal{D}_R\mathcal{D}_R^+ T^{AAS}$ &  $(\mathbf{1}_L,\mathbf{3}_R)$\\
	$Q_2$ & $\mathcal{O}^3_{LRR}$ & $\mathcal{D}_L\mathcal{D}_R\mathcal{D}_R^+T^{AAS}$ & $(\mathbf{1}_L,\mathbf{3}_R)$ \\
	$Q_3$ & $\mathcal{O}^3_{LLR}$ & $\mathcal{D}_L\mathcal{D}_L\mathcal{D}_R^+T^{AAS}$ &  $(\mathbf{1}_L,\mathbf{3}_R)$ \\
	$Q_4$ & $4/5\; \mathcal{O}^2_{RRR} + 1/5\; \mathcal{O}^1_{RRR}$ & $\mathcal{D}_R^{33+}T^{SSS}$ & $(\mathbf{1}_L,\mathbf{7}_R)$ \\
	$Q_5$ & $\mathcal{O}^1_{RLL}$ & $\mathcal{D}_R^-\mathcal{D}_L^{++}T^{SSS}$ &  $(\mathbf{5}_L,\mathbf{3}_R)$ \\\hline
	$Q_6$ & $\mathcal{O}^2_{RLL}$ & $\mathcal{D}_R^3\mathcal{D}_L^{3+}T^{SSS}$ & $(\mathbf{5}_L,\mathbf{3}_R)$ \\
	$Q_7$ & $2/3\; \mathcal{O}^2_{LLR} + 1/3\; \mathcal{O}^1_{LLR}$ & $\mathcal{D}_R^+\mathcal{D}_L^{33}T^{SSS}$ & $(\mathbf{5}_L,\mathbf{3}_R)$ \\\hline
	$\tilde{Q}_1$ & $1/3\;\mathcal{O}^2_{RRR} - 1/3\;\mathcal{O}^1_{RRR}$ & $\mathcal{D}_R\mathcal{D}_R\mathcal{D}_R^+T^{SSS}$ & $(\mathbf{1}_L,\mathbf{3}_R)$  \\
	$\tilde{Q}_3$ & $1/3\;\mathcal{O}^2_{LLR} - 1/3\;\mathcal{O}^1_{LLR}$ & $\mathcal{D}_L\mathcal{D}_L\mathcal{D}_R^+T^{SSS}$ &  $(\mathbf{1}_L,\mathbf{3}_R)$ \\\hline
  \end{tabular}
  \caption{The chiral basis operators $Q_I$ shown in the first column are proportional to the corresponding fixed-flavor basis operator combinations shown in the second column with proportionality factor $1$ for $Q_5$ and $(-4)$ for all other $Q_I$. Each chiral basis operator is equal to a color contraction of the tensor operators $\mathcal{D}_\chi^{A\dots}$ shown in the third column. The corresponding chiral irrep of each operator is shown in the last column. $Q_1,\ldots,Q_7$ and their parity conjugates ($L\leftrightarrow R$) form a complete basis for $n\bar{n}$ transition operators in $D=4$. Since they are components of the same chiral tensor operator $\mathcal{D}_R^A\mathcal{D}_L^{BC}$, $Q_6$ and $Q_7$ renormalize identically to $Q_5$ and are redundant for our purposes. $\tilde{Q}_1$ and $\tilde{Q}_3$ are equal to $Q_1$ and $Q_3$ in $D=4$, but they renormalize independently in $\bar{\text{MS}}$ at NNLO. \label{chiraltab}}
\end{table}

Using these building blocks and neglecting operators that have $\Delta I_3 \neq -1$ or vanish by quark anticommutivity, we find that at NLO there are five chiral tensor operators with independent renormalization properties,
\begin{subequations}
  \begin{align}
	Q_1 &= (\psi CP_R i\tau^2 \psi)(\psi CP_R i\tau^2 \psi)(\psi CP_R i\tau^2 \tau^+ \psi)T^{AAS},\\\nonumber\\
	Q_2 &= (\psi CP_L i\tau^2 \psi)(\psi CP_R i\tau^2 \psi)(\psi CP_R i\tau^2 \tau^+ \psi)T^{AAS},\\\nonumber\\
	Q_3 &= (\psi CP_L i\tau^2 \psi)(\psi CP_L i\tau^2 \psi)(\psi CP_R i\tau^2 \tau^+ \psi)T^{AAS},\\\nonumber\\
  Q_4 &= (\psi CP_R i\tau^2\tau^3 \psi)(\psi CP_R i\tau^2\tau^3 \psi)(\psi CP_R i\tau^2 \tau^+ \psi)T^{SSS}\\\nonumber
  &\hspace{20pt} - \frac{1}{5}(\psi CP_R i\tau^2\tau^A \psi)(\psi CP_R i\tau^2\tau^A \psi)(\psi CP_R i\tau^2 \tau^+ \psi)T^{SSS}, \\\nonumber\\
  Q_5 &= (\psi CP_R i\tau^2\tau^- \psi)(\psi CP_L i\tau^2\tau^+ \psi)(\psi CP_L i\tau^2 \tau^+ \psi)T^{SSS}.
\end{align}
  \label{chiralbasis}
\end{subequations}
Symmetries of $T^{SSS}$ and $T^{AAS}$ under diquark exchange ensure that all products of flavor vector diquarks are totally symmetric. $Q_4$ includes a flavor trace subtraction. This ensures that all operators are irreducible chiral tensor operators.

There are two additional operators that cannot be expressed as linear combinations of $Q_1,\dots,Q_5$,
\begin{subequations}
  \begin{align}
	Q_6 &= (\psi CP_R i\tau^2\tau^3\psi)(\psi CP_L i\tau^2\tau^3\psi)(\psi CP_L i\tau^2\tau^+ \psi)T^{SSS},\\\nonumber\\
	Q_7 &= (\psi CP_L i\tau^2\tau^3\psi)(\psi CP_L i\tau^2\tau^3\psi)(\psi CP_R i\tau^2\tau^+ \psi)T^{SSS}\\\nonumber
	&\hspace{20pt} - \frac{1}{3}(\psi CP_L i\tau^2\tau^A \psi)(\psi CP_L i\tau^2\tau^A \psi)(\psi CP_R i\tau^2\tau^+ \psi)T^{SSS}.
  \end{align}
  \label{extrabasis}
\end{subequations}
These two operators and $Q_5$ are different components of the same chiral tensor operator $\mathcal{D}_R^A\mathcal{D}_L^{BC}$. This implies that $Q_5$, $Q_6$, and $Q_7$ have identical anomalous dimensions and matching factors in renormalization schemes respecting chiral symmetry. In $D=4$, $Q_1,\ldots,Q_7$ and their seven parity conjugates found by taking $L\leftrightarrow R$ everywhere and including a relative minus sign form a complete basis of dimension-nine operators contributing to $n\bar{n}$ transitions.

We also consider two more operators $\tilde{Q}_1$ and $\tilde{Q}_3$ that in $D=4$ are equal to $Q_1$ and $Q_3$ respectively by Fierz identities,
\begin{subequations}
  \begin{align}
	\tilde{Q}_1 &= \frac{1}{3}(\psi CP_R i\tau^2\tau^A  \psi)(\psi CP_R i\tau^2\tau^A \psi)(\psi CP_R i\tau^2 \tau^+ \psi)T^{SSS},\\
	\tilde{Q}_3 &= \frac{1}{3}(\psi CP_L i\tau^2\tau^A \psi)(\psi CP_L i\tau^2\tau^A \psi)(\psi CP_R i\tau^2 \tau^+ \psi)T^{SSS}.
  \end{align}
  \label{fierzbasis}
\end{subequations}
The Fierz relations $Q_1=\tilde{Q}_1$ and $Q_2=\tilde{Q}_3$ are broken in dimensional regularization, and $\tilde{Q}_1$ and $\tilde{Q}_3$ are independent operators in $D$ dimensions. In principle, we could choose our physical operator basis to be $Q_1,\dots,Q_5$ and include $Q_1-\tilde{Q}_1$ and $Q_3-\tilde{Q}_3$ as additional evanescent operators vanishing in $D=4$ but present in $D$ dimensions. In practice, it is much easier to directly determine matrix elements of $\tilde{Q}_1$ and $\tilde{Q}_3$ and explicitly include them in the physical operator basis. For the purposes of NNLO operator renormalization we take our chiral basis operators $Q_I$ to include $Q_1,\dots,Q_5,\;\tilde{Q}_1,\;\tilde{Q}_3$.

The basis commonly used in the literature involves fixed-flavor quark fields \cite{Chang:1980ey,Kuo:1980ew,Rao:1982gt,Rao:1983sd,Caswell:1982qs},
\begin{equation}
  \begin{split}
  \mathcal{O}^1_{\chi_1\chi_2\chi_3} &= (u C P_{\chi_1} u)(d CP_{\chi_2} d)(d C P_{\chi_3} d)T^{SSS},\\
  \mathcal{O}^2_{\chi_1\chi_2\chi_3} &= (u C P_{\chi_1} d)(u CP_{\chi_2} d)(d C P_{\chi_3} d)T^{SSS},\\
  \mathcal{O}^3_{\chi_1\chi_2\chi_3} &= (u C P_{\chi_1} d)(u CP_{\chi_2} d)(d C P_{\chi_3} d)T^{AAS}.
\end{split}
  \label{flavorbasis}
\end{equation}
These fixed-flavor basis operators satisfy the relations $\mathcal{O}^1_{\chi LR} = \mathcal{O}^1_{\chi RL}$ and $\mathcal{O}^{2,3}_{LR\chi} = \mathcal{O}^{2,3}_{RL\chi}$. In $D=4$, they also satisfy the Fierz identities $\mathcal{O}^2_{\chi\chi\chi^\prime}-\mathcal{O}^1_{\chi\chi\chi^\prime} = 3\mathcal{O}^3_{\chi\chi\chi^\prime}$. These relations reduce the number of linearly independent operators to 14. It is straightforward to evaluate the flavor contractions of $\psi = (u,d)$ in the $Q_I$ and verify that $Q_1,\dots,Q_7$ and their parity conjugates form 14 linearly independent combinations of fixed-flavor basis operators. One can similarly verify that the Fierz relations $\tilde{Q}_1 = Q_1$ and $\tilde{Q}_3 = Q_3$ are equivalent to the fixed-flavor basis Fierz relation above. The precise relations between the chiral basis and fixed-flavor basis operators and their explicit chiral tensor structures are shown in Table~\ref{chiraltab}.

\section{Renormalization Schemes}\label{sec:renormschemes}

The commonly used $\bar{\text{MS}}$ renormalization scheme simplifies RG evolution, preserves important symmetries of chiral gauge theories such as the Standard Model, and is technically simple to implement in perturbative calculations performed with dimensional regularization. $\bar{\text{MS}}$ is limited, however, in that its defining renormalization condition can only be applied to regularized matrix elements calculated with dimensional regularization and not other with regularization schemes such as lattice. The RI-MOM renormalization scheme~\cite{Martinelli:1994ty}, while not as technically simple to apply to dimensionally regularized matrix elements, has the advantage of a regularization independent renormalization condition. In this section we construct a RI-MOM operator renormalization condition for the $Q_I$ that can be applied both non-perturbatively to lattice QCD matrix elements and perturbatively to dimensionally regularized matrix elements. This is an essential intermediate step in connecting lattice regularized and $\bar{\text{MS}}$ renormalized $n\bar{n}$ matrix elements.

In this section we explicitly display the spacetime and renormalization scale dependence of the quark fields $\psi^\alpha_{ia}(x,\mu)$ and six-quark operators $Q_I(x,\mu)$. These renormalized quark fields and six-quark operators should be distinguished from their bare (regularized) counterparts, defined for quark fields of flavor $q=u,d$ by
\begin{equation}
  q_i^\alpha(x,\mu) = Z_q^{-1/2}(\mu)[q^0]_i^\alpha(x),\hspace{20pt} Q_I(x,\mu) = \sum_J Z_{IJ}(\mu)Q_J^0(x),
  \label{Zdef}
\end{equation}
where the wavefunction renormalization factor $Z_q(\mu)$ and operator renormalization matrix $Z_{IJ}(\mu)$ are formally defined by the renormalization conditions of a particular renormalization scheme. We denote perturbative expansion coefficients for either renormalization factor by
\begin{equation}
  Z(\mu) = 1 + \left( \frac{\alpha_s(\mu)}{4\pi} \right)Z^{(1)} + \left( \frac{\alpha_s(\mu)}{4\pi} \right)^2Z^{(2)} + O(\alpha_s^3).
  \label{Zpertdef}
\end{equation}

The RI-MOM scheme wavefunction renormalization factor $Z_q^{\text{RI}}(\mu)$ is defined by~\cite{Martinelli:1994ty} 
\begin{equation}
  \begin{split}
	1 =& \frac{-i}{48}\tr \left( \gamma^\mu \frac{\partial S_q(p_0,\mu=p_0)^{-1}}{\partial p^\mu} \right)\\
	=& \frac{-i}{48}Z_q^{\text{RI}}(p_0)\tr \left( \gamma^\mu \frac{\partial S_q^0(p_0)^{-1}}{\partial p^\mu} \right),
  \label{ZqRIdef}
\end{split}
\end{equation}
where $\tr$ denotes a trace over color and spin (quark flavor $q$ is held fixed), $p_0$ is the lattice matching scale, and $S_q^0(p_0)$ and $S_q(p_0,\mu)$  are bare and renormalized quark propagators respectively. The normalization is chosen such that $Z_q^{\text{RI}}(\mu) = 1 + O(\alpha_s)$. Eq.~\eqref{ZqRIdef} assumes that a gauge fixing condition has been imposed so that the quark propagator is non-vanishing. In this work we consider a general $R_\xi$ gauge where the tree-level gluon propagator is $\delta^{AB}\left(g_{\mu\nu}/p^2 - (1-\xi)p_\mu p_\nu/p^4\right)$. One-loop matching is performed in Landau gauge, $\xi=0$. Two-loop-running results are gauge invariant, and for simplicity the two-loop calculation is performed in Feynman gauge, $\xi = 1$.

A one-loop calculation of the quark self-energy in $D=4-2\varepsilon$ dimensions shows that the counterterm needed to renormalize the bare propagator according to the renormalization condition Eq.~\eqref{ZqRIdef} is~\cite{Martinelli:1994ty}
\begin{equation}
  Z_q^{\text{RI},(1)} = -\frac{4}{3}\left( \frac{\xi}{\bar{\varepsilon}} + \frac{\xi}{2}\right),
	\label{ZqRI1}
\end{equation}
where $1/\bar{\varepsilon} = 1/\varepsilon - \gamma_E + \ln 4\pi$. The $\bar{\text{MS}}$ wavefunction renormalization factor is defined by the condition that $Z_q$ remove precisely the poles in $1/\bar{\varepsilon}$ from the quark propagator. At one-loop
\begin{equation}
  Z_q^{\bar{\text{MS}},(1)} = -\frac{4}{3}\left(\frac{\xi}{\bar{\varepsilon}}\right).
	\label{Zq1}
\end{equation}
At two-loop order the quark propagator includes diagrams with divergent one-loop subdiagrams. These diagrams include non-local divergences proportional to $\ln(\mu^2/p^2)/\bar{\varepsilon}$. Renormalizability guarantees that these non-local two-loop divergences cancel after including counterterm diagrams in which divergent subdiagrams are replaced by their one-loop counterterms \cite{Hooft:1972fi}.\footnote{For a comprehensive review of renormalization theory with further references to the original literature, see Ref.~\cite{Collins:105730}} The remaining local divergences are removed by a two-loop counterterm that in Feynman gauge is given by~\cite{Egorian:1978zx}
\begin{equation}
  Z_q^{\bar{\text{MS}},(2)} = \frac{44}{9\bar{\varepsilon}^2} + \frac{-47+2N_f}{3\bar{\varepsilon}},
	\label{Zq2}
\end{equation}
where $N_f$ is the number of active quark flavors.

A regularization independent definition of $Z_{IJ}^{\text{RI}}(\mu)$ can be given in terms of a renormalization condition applied to vertex functions for each $Q_I$. These vertex functions can be constructed, perturbatively or non-perturbatively, by Wick contracting $Q_I$ with interpolating operators for initial neutron and final antineutron states. A vertex function with $Q_I$ inserted at zero-momentum can be constructed by including three external antiquark fields carrying momentum $p$ and three external antiquark fields momentum $-p$. These antiquark fields act as interpolating operators capable of creating initial neutron and final antineutron states. In order to simplify the non-perturbative construction of this vertex function in lattice QCD calculations, it is convenient to work with interpolating operators built from fixed-flavor quark fields. The quark fields must be assigned momenta symmetrically in order for the RI-MOM scheme defined using the vertex function to preserve chiral symmetry. A suitable definition is given by~\cite{Syritsyn:2015}
\begin{equation}
	\begin{split}
		[\Lambda_I]^{\alpha\beta\gamma\delta\eta\zeta}_{ijklmn}(p) &=  \left. \frac{1}{5}\avg{ Q_I(0)\, \bar{u}^\alpha_i(p) \bar{u}^\beta_j(p) \bar{d}^\gamma_k(p) \bar{d}^\delta_l(-p) \bar{d}^\eta_m(-p) \bar{d}^\zeta_n(-p) }\right|_{amp}\\
		&\hspace{10pt}+  \left. \frac{3}{5} \avg{ Q_I(0)\, \bar{u}^\alpha_i(p) \bar{u}^\beta_j(-p) \bar{d}^\gamma_k(p) \bar{d}^\delta_l(p) \bar{d}^\eta_m(-p) \bar{d}^\zeta_n(-p) }\right|_{amp} \\
		&\hspace{10pt}+  \left. \frac{1}{5} \avg{ Q_I(0)\, \bar{u}^\alpha_i(-p) \bar{u}^\beta_j(-p) \bar{d}^\gamma_k(p) \bar{d}^\delta_l(p) \bar{d}^\eta_m(p) \bar{d}^\zeta_n(-p) }\right|_{amp},
	\label{vertexdef}
\end{split}
\end{equation}
where antisymmetrization of all antiquark fields of the same flavor is implied, renormalization scale dependence is suppressed, and the subscript amp refers to the prescription of amputating external legs with the replacement
\begin{equation}
  \bar{q}_i^\alpha(p,\mu) \rightarrow \bar{q}_{i^\prime}^{\alpha^\prime}(p,\mu)[S_q^{-1}(p,\mu)]_{i^\prime i}^{\alpha^\prime \alpha}.
\end{equation}
When the Wick contractions for $\Lambda_I$ are enumerated, the 20 distinct momentum routings available for Feynman diagrams with six indistinguishable external legs carrying momentum $\{p,p,p,-p,-p,-p\}$ each appear with equal weight. Perturbative contributions to $\Lambda_I$ are defined by
\begin{equation}
  \Lambda_I(p,\mu) = \Lambda_I^{(0)} + \left( \frac{\alpha_s(\mu)}{4\pi} \right)\Lambda_I^{(1)}(p,\mu) + \left( \frac{\alpha_s(\mu)}{4\pi} \right)^2 \Lambda_I^{(2)}(p,\mu).
\end{equation}
It should also be noted that $\Lambda_I$ is defined with an ``exceptional momentum configuration'' where the momenta of some subsets of external fields add to zero. However, vertex functions for purely baryonic operators like $Q_I$ are not expected to suffer from the non-perturbative chiral symmetry breaking artifacts that affect mesonic operators in exceptional momentum configurations. In particular, infrared divergences in the chiral limit arising from pseudo-Goldstone poles can lead to enhanced non-perturbative chiral symmetry violating mesonic operator mixing in lattice QCD simulations with chiral fermions, see for example~\cite{Aoki:2007xm}. Enhancements arise from diagrams in which an external quark and antiquark can be combined in a subdiagram with zero external momentum. For the purely baryonic operators considered here as well as for proton decay operators, there are no pseudo-Goldstone pole enhancements from diagrams in which two external quarks with positive baryon number are combined in a zero-momentum subdiagram.

The RI-MOM scheme is defined by a renormalization condition on $\Lambda_I(p,\mu)$,
\begin{equation}
  \begin{split}
  \delta_{IJ} =& \tr\left[ \mathcal{P}_I \Lambda_J(p_0,\mu=p_0) \right]  \\
  =& \sum_K Z_{IK}^{\text{RI}}(p_0)\left( Z_q^{\text{RI}}(p_0) \right)^{3} \tr\left[ \mathcal{P}_J \Lambda_K^0(p_0) \right],
  \label{RIdef}
\end{split}
\end{equation}
where $\Lambda_K^0(p_0)$ is a bare vertex function built from bare operators and amputated with bare propagators, $\tr\left[ \mathcal{P}_I \Lambda_J \right] \equiv \mathcal{P}^{\alpha\beta\gamma\delta\eta\zeta}_{ijklmn}\Lambda^{\alpha\beta\gamma\delta\eta\zeta}_{ijklmn}$, and the operator projectors $\mathcal{P}_I$ are defined by
\begin{equation}
	\tr\left[ \mathcal{P}_I \Lambda_J^{(0)} \right] = \delta_{IJ},
    \label{projectordef}
\end{equation}
Each external quark in Eq.~\eqref{RIdef} carries momentum $\pm p_0$ and the renormalization scale is identified with this lattice matching scale $\mu=p_0$. As discussed above, $p_0$ must be chosen to be much larger than hadronic scales to allow for perturbative matching but much smaller than the inverse lattice spacing used for non-perturbative renormalization to control discretization errors. For many quantities, these constraints are satisfied at $p_0\simeq 2\text{ GeV}$. Comparison of the size of $O(\alpha_s(p_0))$ NNLO corrections to the NLO result in Eq. \eqref{master} should provide an estimate of perturbative convergence with a chosen $p_0$.

A set of projectors satisfying Eq. \eqref{projectordef} for the chiral basis operators is given by
\begin{subequations}
    \label{projectors}
  \begin{align}
	(\mathcal{P}_1)^{\alpha \beta\gamma\delta\eta\zeta}_{ijklmn}&= -\frac{1}{92160}\left(-T^{SSS}_{\{ij\}\{kl\}\{mn\}}(CP_R)^{\alpha \beta}(CP_R)^{\gamma \delta}  (CP_R)^{\eta \zeta}+ 2T^{AAS}_{[ij][kl]\{mn\}}(CP_R)^{\alpha \delta}(CP_R)^{\gamma \beta}  (CP_R)^{\eta \zeta}\right),\\
	(\mathcal{P}_2)^{\alpha \beta\gamma\delta\eta\zeta}_{ijklmn}&= -\frac{1}{18432}\left( -T^{SSS}_{\{ij\}\{kl\}\{mn\}}(CP_L)^{\alpha \delta}(CP_R)^{\gamma \beta}  (CP_R)^{\eta \zeta}+ 2T^{AAS}_{[ij][kl]\{mn\}}(CP_L)^{\alpha \delta}(CP_R)^{\gamma \zeta}  (CP_R)^{\eta \beta}\right),\\
	(\mathcal{P}_3)^{\alpha \beta\gamma\delta\eta\zeta}_{ijklmn}&= -\frac{1}{36864}\left( -T^{SSS}_{\{ij\}\{kl\}\{mn\}}(CP_L)^{\alpha \beta}(CP_L)^{\gamma \delta}  (CP_R)^{\eta \zeta}+ 2T^{AAS}_{[ij][kl]\{mn\}}(CP_L)^{\alpha \delta}(CP_L)^{\gamma \beta}  (CP_R)^{\eta \zeta}\right),\\
	(\mathcal{P}_4)^{\alpha \beta\gamma\delta\eta\zeta}_{ijklmn}&= -\frac{1}{221184}\left( T^{SSS}_{\{ij\}\{kl\}\{mn\}}(CP_R)^{\alpha \beta}(CP_R)^{\gamma \delta}  (CP_R)^{\eta \zeta}+3T^{AAS}_{[ij][kl]\{mn\}}(CP_R)^{\alpha \delta}(CP_R)^{\gamma \beta}  (CP_R)^{\eta \zeta}\right), \\
    (\mathcal{P}_5)^{\alpha \beta\gamma\delta\eta\zeta}_{ijklmn}&= -\frac{1}{221184}\left( T^{SSS}_{\{ij\}\{kl\}\{mn\}}(CP_R)^{\alpha \beta}(CP_L)^{\gamma \delta}  (CP_L)^{\eta \zeta}\right), \\
	(\mathcal{P}_6)^{\alpha \beta\gamma\delta\eta\zeta}_{ijklmn}&= -\frac{1}{55296}\left( T^{SSS}_{\{ij\}\{kl\}\{mn\}}(CP_R)^{\alpha \delta}(CP_L)^{\gamma \beta}  (CP_L)^{\eta \zeta}+6T^{AAS}_{[ij][kl]\{mn\}}(CP_R)^{\alpha \delta}(CP_L)^{\gamma \zeta}  (CP_L)^{\eta \beta}\right), \\
	(\mathcal{P}_7)^{\alpha \beta\gamma\delta\eta\zeta}_{ijklmn}&= -\frac{1}{73728}\left( T^{SSS}_{\{ij\}\{kl\}\{mn\}}(CP_L)^{\alpha \beta}(CP_L)^{\gamma \delta}  (CP_R)^{\eta \zeta}+2T^{AAS}_{[ij][kl]\{mn\}}(CP_L)^{\alpha \delta}(CP_L)^{\gamma \beta}  (CP_R)^{\eta \zeta}\right).
  \end{align}
\end{subequations}  
We explicitly include projectors for $Q_6$ and $Q_7$ since they must be analyzed separately in lattice QCD calculations without exact chiral symmetry. The seven parity conjugates of $Q_1,\dots,Q_7$ should be analyzed separately in lattice QCD calculations; projectors for these operators are found by taking $L\leftrightarrow R$ everywhere and including a relative minus sign. $\mathcal{P}_1$ and $\mathcal{P}_3$ are suitable projectors for $\tilde{Q}_1$ and $\tilde{Q}_3$ since they are equal to $Q_1$ and $Q_3$ at tree level. Projectors and for fixed-flavor basis operators differ from those of Eq.~\eqref{projectors} in their overall normalizations and are described in more detail in Ref. \cite{Buchoff:2015wwa}. 

$Z_{IJ}^{\bar{\text{MS}}}$ can be defined through a renormalization condition for dimensionally regularized vertex functions: at each order of renormalized perturbation theory, add counterterms that remove precisely the $1/\bar{\varepsilon}$ poles proportional to $\Lambda_J$ from $\Lambda_I$. A more precise definition of both the RI-MOM and $\bar{\text{MS}}$ renormalization conditions for dimensionally regularized amplitudes requires a careful treatment of evanescent operators. This is postponed to Sec. \ref{sec:matchingev}.


\section{One-Loop Matching}\label{sec:matching}

RI-MOM and $\bar{\text{MS}}$ renormalized operators with renormalization scale $\mu = p_0$ are related by Eq. \eqref{Zdef},
\begin{equation}
  Q^{\text{RI}}_I(p_0) = \sum_{J,K} Z_{IJ}^{\text{RI}}(p_0) \left[\left(Z^{\bar{\text{MS}}}\right)^{-1}_{JK}(p_0)\right]Q_K^{\bar{\text{MS}}}(p_0) \equiv \sum_J r_{IJ}Q_J^{\bar{\text{MS}}}(p_0).
  \label{rdef}
\end{equation}
The matching factor $r_{IJ}$ relates renormalized operators and is therefore a finite quantity. $r_{IJ}$ can be consistently calculated perturbatively in terms of $Z^{\text{RI}}$ and $Z^{\bar{\text{MS}}}$ as long as both contain the same UV divergences and in particular are calculated with the same regularization. This allows us to perturbatively express $r_{IJ}$ as
\begin{equation}
	\begin{split}
  r_{IJ}(\alpha_s) &= 1 +  \frac{\alpha_s(p_0)}{4\pi}\left( Z_{IJ}^{\text{RI},(1)} - Z_{IJ}^{\bar{\text{MS}},(1)} \right) + O(\alpha_s^2)\\
  &\equiv 1 +  \frac{\alpha_s(p_0)}{4\pi}r_{IJ}^{(0)} + O(\alpha_s^2).
  \end{split}
  \label{rPT}
\end{equation}
Since the chiral basis operators do not mix under renormalization, $Z_{IJ}$ and $r_{IJ}$ are diagonal and we further define $r_I^{(0)}$ to be the diagonal elements $r_{IJ}^{(0)} = \delta_{IJ}r_I^{(0)}$ (no summation on $I$). Differences between definitions of the renormalized coupling constant $\alpha_s(p_0)$ in different schemes are formally $O(\alpha_s^2)$ and can therefore be neglected in Eq.~\eqref{rPT}. When calculating numerical results in Sec.~\ref{sec:pheno}, we use an $\bar{\text{MS}}$ coupling constant definition for both two-loop running and one-loop matching. This defines the one-loop-matching factor $r_I^{(0)}$ appearing in Eq. \eqref{master} in terms of $Z_{IJ}^{(1)}$. The remainder of this section describes the diagrammatic evaluation of $Z_{IJ}^{(1)}$ from one-loop corrections to $\Lambda_I^{(0)}$.

\subsection{Diagram Evaluation}\label{sec:matchingdiag}

\begin{figure}
	\includegraphics[scale=.52]{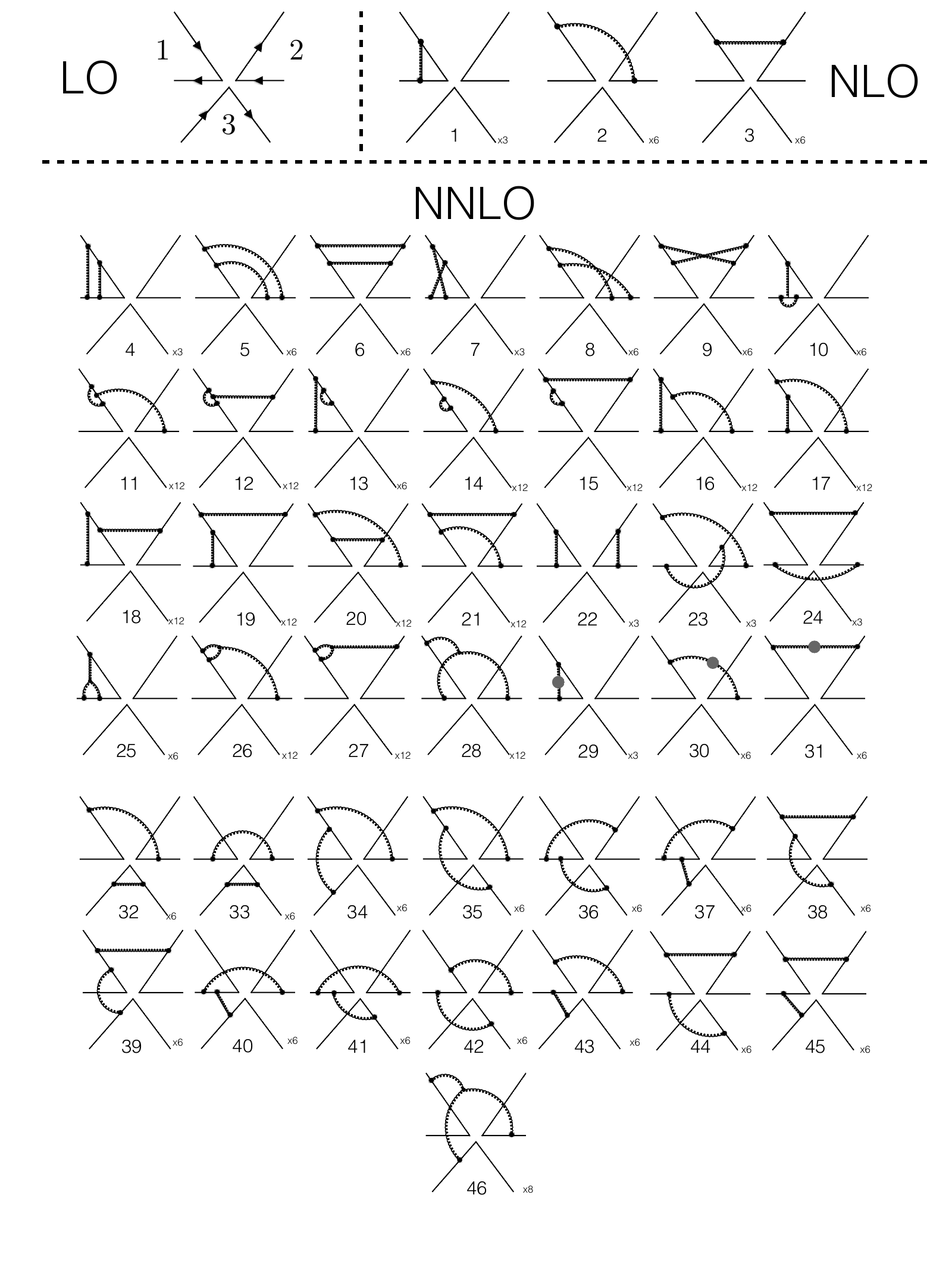}
        \caption{The tree-level operator diagram, 15 one-loop diagrams, and 320 two-loop diagrams evaluated in this work (350 in the counting of qgraf \cite{Nogueira:1991ex}, see Sec. \ref{sec:running}). Operator insertions are represented by a six-point vertex joining three quarks and three conjugate antiquarks. The operator insertions are local; the separate solid lines represent propagators for quarks contracted into separate spin-singlet diquarks as indicated by the fermion charge arrows. The external quarks are assigned momenta $\{p,p,p,-p,-p,-p\}$. Diagrams are organized into classes that share the same loop integrals and Dirac structures. The number of diagrams in each class is shown. All two-loop diagrams with divergent subdiagrams are accompanied by a one-loop counterterm diagram, not shown. The curly lines represent gluon propagators, and the gluon self-energy bubble shown in diagrams 29-31 includes quark, gluon, and ghost loops. Diagrams 1-31 contribute to four-quark operator renormalization and are numbered consistently with Refs.~\cite{Buras:1989xd,Buras:2000if}. Diagrams 32-46 are considered for the first time here. The $1/\bar{\varepsilon}$ pole structure of each diagram is summarized in Tables~\ref{diag_results}~-~\ref{results_S}. \label{diagrams}}
\end{figure}

Feynman diagrams representing corrections to $Q_I$ involve six quark lines carrying baryon number into a local vertex where the quark lines are joined to form three spin-singlet diquarks. To simplify the structure of these diagrams it is convenient to introduce charge conjugate quarks $(\psi^C)^\alpha_{i a} = (C\bar{\psi})^\alpha_{i a}$. Expressing all diquarks in $Q_I$ as $(\psi CP_\chi \psi) = (\bar{\psi}^C P_\chi \psi)$ removes the need to introduce spin-transposed propagators and explicit factors of $C$ at the six-quark vertex. With this approach, one quark line contained in each spin-singlet diquark is replaced with a conjugate-antiquark line with it's fermion charge arrow pointing out of the six-quark vertex, as shown in Fig.~\ref{diagrams}. These obey standard Feynman rules for quark lines carrying fermion charge away from the vertex, except that conjugate quark-gluon vertices receive an extra minus sign and transposition of $t^A$ because $(\bar{\psi}^C_i \gamma^\mu t^A_{ij} \psi^C_j) = -(\bar{\psi}_i \gamma^\mu t^A_{ji} \psi_j)$.

Matching between RI-MOM and $\bar{\text{MS}}$ is performed in the limit of massless quarks where $SU(2)_L\times SU(2)_R$ chiral symmetry guarantees that loop-level operator corrections will contain diquarks with the same chiral structure as the tree-level operator. When calculating diagrams with no closed fermion loops in NDR, explicit factors of $P_\chi$ can therefore always be moved from the operator vertex to one end of the quark line representing each spin-singlet diquark. In addition, the tree-level flavor structure of each operator is preserved diagram-by-diagram because of the flavor-blind nature of QCD. With these considerations, the only non-trivial tensor structure that needs to be inserted at each six-point operator vertex is $(1\otimes 1\otimes 1)T^{AAS}$ for $Q_1$, $Q_2$, and $Q_3$ and $(1\otimes 1\otimes 1)T^{SSS}$ for $Q_4$, $Q_5$, $\tilde{Q}_1$, and $\tilde{Q}_3$. Diagrams with six-quark vertex factors of $(1\otimes 1\otimes 1)T^{AAS}$ and $(1\otimes 1\otimes 1)T^{SSS}$ represent amplitudes denoted by $\mathcal{M}^A$ and $\mathcal{M}^S$ respectively. These amplitudes provide loop-level operator corrections to $Q_I$ once factors of $CP_\chi$, flavor tensors, and contractions with quark fields are included. As a concrete example, diagrammatic operator corrections to $Q_1$ are found from the amplitude $\mathcal{M}^A$ by making the replacement
\begin{equation}
	(\Gamma_1\otimes \Gamma_2\otimes \Gamma_3) T \rightarrow (\psi CP_R \Gamma_1 i\tau^2 \psi)(\psi CP_R \Gamma_2 i\tau^2 \psi)(\psi CP_R \Gamma_3 i\tau^2 \tau^+ \psi)T^{AAS}.
  \label{rosetta}
\end{equation}

To determine finite $O(\varepsilon^0)$ contributions to $\Lambda_I^{(0)}$, all contributing diagrams should be calculated with all distinct momentum routings assigning incoming momenta $\{p,p,p,-p,-p,-p\}$ to the external legs. $\Lambda_I^{(0)}$ is then found by adding external quark fields to build the correlation functions of Eq.~\eqref{vertexdef}, performing Wick contractions, and amputating external legs. The contribution of each Wick contraction is represented by the sum of all contributing amputated diagrams with a particular momentum routing. $1/\bar{\varepsilon}$ pole terms are momentum independent, and can be determined from any momentum routing free of infrared divergences.

The topologically distinct classes of one- and two-loop Feynman diagrams contributing to $\Lambda_I^{(1)}$ and $\Lambda_I^{(2)}$ are shown in Fig.~\ref{diagrams}. Calculating $r_I^{(1)}$ requires evaluating the 15 one-loop diagrams in classes $d$ = 1-3. For each of the one-gluon-exchange diagrams in $d$ = 1-3, the two distinct momentum routings correspond to gluon exchange between quarks with equal momenta and gluon exchange between quarks with opposite momenta. With $\Lambda_I$ defined by Eq.~\eqref{vertexdef}, a gluon is exchanged between quarks with equal momenta in 2/5 of the Wick contractions and between quarks with opposite momenta in 3/5 of the Wick contractions. Since external quark fields of the same flavor are antisymmetrized when constructing $\Lambda_I$, each Wick contraction contributes to $\Lambda_I$ with equal weight. 

The amplitudes $\mathcal{M}^A_d$ and $\mathcal{M}^S_d$ for diagrams of class $d$ can be evaluated using standard techniques summarized in Appendices~\ref{app:algebra}~-~\ref{app:tables}. After expressing the resulting spin-color tensors as linear combinations of the basis tensors introduced in Appendix~\ref{app:algebra}, it is straightforward to verify that most spin-color tensors contributing to $\mathcal{M}^A_d$ and $\mathcal{M}^S_d$ have index exchange symmetries different from the symmetries of the tree-level operator insertion. These contributions vanish after making the replacement of Eq.~\eqref{rosetta} and can be neglected. The remaining contributions to the one-loop amplitudes $\mathcal{M}^A_d$ for the relevant combination of 2/5 the equal momentum routing amplitude plus 3/5 the opposite momentum routing amplitude are given by
\begin{subequations}\label{matchingA}\begin{align}
		\mathcal{M}^A_1 &= \frac{\alpha_s(\mu)}{4\pi}\left( \frac{\mu^2}{p^2} \right)^\varepsilon\left( \frac{3+\xi}{\bar{\varepsilon}} + 4 + 2\xi - \frac{24}{5}\ln 2 - \frac{8\xi}{5}\ln 2\right)[1\otimes 1\otimes 1]T^{AAS},\\\nonumber\\
		\mathcal{M}^A_2 &= \frac{\alpha_s(\mu)}{4\pi}\left( \frac{\mu^2}{p^2} \right)^\varepsilon\left\lbrace \left( \frac{3\xi}{2\bar{\varepsilon}} + \frac{29}{20} + \frac{19\xi}{10} - \frac{8}{5}\ln 2 - \frac{8\xi}{5}\ln 2 \right)[1\otimes 1\otimes 1]T^{AAS}\right.\\\nonumber
	&\hspace{10pt} + \left( -\frac{1}{8\bar{\varepsilon}} - \frac{17}{60} + \frac{2}{15}\ln 2 \right)\left[ (\sigma_{\mu\nu}\otimes\sigma_{\mu\nu}\otimes 1) T^{SSS} + (1\otimes\sigma_{\mu\nu}\otimes\sigma_{\mu\nu}) T^{ASA} + (\sigma_{\mu\nu}\otimes 1\otimes \sigma_{\mu\nu}) T^{SAA}\right]\\\nonumber
	&\hspace{10pt} + \frac{1}{p^2}\left( \frac{1}{15} - \frac{11\xi}{60} + \frac{2}{15}\ln 2 + \frac{2\xi}{15}\ln 2 \right)\left[ (\gamma_\mu\fs{p}\otimes\fs{p}\gamma_\mu\otimes 1)\left( T^{SSS} - \frac{1}{3}T^{AAS} \right)\right.\\\nonumber
	&\hspace{30pt} \left.\left.	+ (1\otimes\gamma_\mu\fs{p}\otimes\fs{p}\gamma_\mu)\left( T^{ASA} + \frac{5}{3}T^{AAS} \right) + (\gamma_\mu\fs{p}\otimes 1\otimes\fs{p}\gamma_\mu)\left( T^{SAA} + \frac{5}{3}T^{AAS} \right) \right],\right\rbrace \\\nonumber\\
	\mathcal{M}^A_3 &= \frac{\alpha_s(\mu)}{4\pi}\left( \frac{\mu^2}{p^2} \right)^\varepsilon\left\lbrace \left( \frac{3\xi}{2\bar{\varepsilon}} + \frac{37}{20} + \frac{4\xi}{5} - \frac{4}{5}\ln 2 - \frac{4\xi}{5}\ln 2 \right)[1\otimes 1\otimes 1]T^{AAS} \right.\\\nonumber
	&\hspace{10pt} + \left( -\frac{1}{8\bar{\varepsilon}} - \frac{17}{60} + \frac{2}{15}\ln 2 \right)\left[ (\sigma_{\mu\nu}\otimes\sigma_{\mu\nu}\otimes 1) T^{SSS} + (1\otimes\sigma_{\mu\nu}\otimes\sigma_{\mu\nu}) T^{ASA} + (\sigma_{\mu\nu}\otimes 1\otimes \sigma_{\mu\nu}) T^{SAA}\right]\\\nonumber
	&\hspace{10pt} + \frac{1}{p^2}\left( \frac{1}{15} - \frac{11\xi}{60} + \frac{2}{15}\ln 2 + \frac{2\xi}{15}\ln 2 \right)\left[ (\gamma_\mu\fs{p}\otimes\fs{p}\gamma_\mu\otimes 1)\left( T^{SSS} + \frac{1}{3}T^{AAS} \right) + (1\otimes\gamma_\mu\fs{p}\otimes\fs{p}\gamma_\mu)\left( T^{ASA} - \frac{5}{3}T^{AAS} \right)\right.\\\nonumber
	&\hspace{30pt} \left.\left. + (\gamma_\mu\fs{p}\otimes 1\otimes\fs{p}\gamma_\mu)\left( T^{SAA} - \frac{5}{3}T^{AAS} \right) \right]\right\rbrace ,
\end{align}\end{subequations}
where the color tensors $T^{ASA}$ and $T^{SAA}$ are defined in Appendix~\ref{app:algebra}. Similarly, the one-loop contributions to $\mathcal{M}^S_d$ with the correct index symmetries are
\begin{subequations}\label{matchingS}\begin{align}
		\mathcal{M}^S_1 &= \frac{\alpha_s(\mu)}{4\pi}\left( \frac{\mu^2}{p^2} \right)^\varepsilon \left( -\frac{3+\xi}{\bar{\varepsilon}} - 4 - 2\xi + \frac{24}{5}\ln 2 + \frac{8\xi}{5}\ln 2 \right)[1\otimes 1\otimes 1]T^{SSS},\\\nonumber\\
	\mathcal{M}^S_2 &= \frac{\alpha_s(\mu)}{4\pi}\left( \frac{\mu^2}{p^2} \right)^\varepsilon\left\lbrace\left( \frac{5\xi}{2\bar{\varepsilon}} + \frac{29}{12} + \frac{19\xi}{6} - \frac{8}{3}\ln 2 - \frac{8\xi}{3}\ln 2 \right)[1\otimes 1\otimes 1]T^{SSS}\right.\\\nonumber
	&\hspace{10pt} + \left( -\frac{3}{8\bar{\varepsilon}} - \frac{17}{20} + \frac{2}{5}\ln 2 \right)\left[ (\sigma_{\mu\nu}\otimes\sigma_{\mu\nu}\otimes 1) T^{AAS} + (1\otimes\sigma_{\mu\nu}\otimes\sigma_{\mu\nu}) T^{SAA} + (\sigma_{\mu\nu}\otimes 1\otimes \sigma_{\mu\nu}) T^{ASA}\right]\\\nonumber
	&\hspace{10pt} + \frac{1}{p^2}\left( \frac{1}{5} - \frac{11\xi}{20} + \frac{2}{5}\ln 2 + \frac{2\xi}{5}\ln 2 \right)\left[ (\gamma_\mu\fs{p}\otimes\fs{p}\gamma_\mu\otimes 1)\left( T^{AAS} + \frac{5}{9}T^{SSS} \right) \right.\\\nonumber
	&\hspace{30pt} \left.\left. + (1\otimes\gamma_\mu\fs{p}\otimes\fs{p}\gamma_\mu)\left( T^{SAA} + \frac{5}{9}T^{SSS} \right) + (\gamma_\mu\fs{p}\otimes 1\otimes\fs{p}\gamma_\mu)\left( T^{ASA} + \frac{5}{9}T^{SSS} \right) \right]\right\rbrace,\\\nonumber \\
	\mathcal{M}^S_3 &=\frac{\alpha_s(\mu)}{4\pi}\left( \frac{\mu^2}{p^2} \right)^\varepsilon\left\lbrace \left( \frac{5\xi}{2\bar{\varepsilon}} + \frac{37}{12} + \frac{4\xi}{3} - \frac{4}{3}\ln 2 - \frac{4\xi}{3}\ln 2 \right)[1\otimes 1\otimes 1]T^{SSS}\right.\\\nonumber
	&\hspace{10pt} + \left( -\frac{3}{8\bar{\varepsilon}} - \frac{17}{20} + \frac{2}{5}\ln 2 \right)\left[ (\sigma_{\mu\nu}\otimes\sigma_{\mu\nu}\otimes 1) T^{AAS} + (1\otimes\sigma_{\mu\nu}\otimes\sigma_{\mu\nu}) T^{SAA} + (\sigma_{\mu\nu}\otimes 1\otimes \sigma_{\mu\nu}) T^{ASA}\right]\\\nonumber
	&\hspace{10pt} + \left( \frac{1}{5} - \frac{11\xi}{20} + \frac{2}{5}\ln 2 + \frac{2\xi}{5}\ln 2 \right)\left[ (\gamma_\mu\fs{p}\otimes\fs{p}\gamma_\mu\otimes 1)\left( T^{AAS} - \frac{5}{9}T^{SSS} \right) + (1\otimes\gamma_\mu\fs{p}\otimes\fs{p}\gamma_\mu)\left( T^{SAA} - \frac{5}{9}T^{SSS} \right)\right.\\\nonumber
	&\hspace{30pt} \left.\left. + (\gamma_\mu\fs{p}\otimes 1\otimes\fs{p}\gamma_\mu)\left( T^{ASA} - \frac{5}{9}T^{SSS} \right) \right]\right\rbrace.
\end{align}\end{subequations}
To complete our calculation of $r_I^{(0)}$, we need to precisely define operator counterterms that renormalize the vertex functions associated with these amplitudes. Subtleties arise at this step. These subtleties and their resolution are discussed in the following section.

\subsection{Evanescent Operators}\label{sec:matchingev}

In order to precisely define operator counterterms suitable for RI-MOM or $\bar{\text{MS}}$ renormalization, we must address the issue that our operator basis is complete in $D=4$ but incomplete in arbitrary $D$. This issue also arises for four-quark operators in weak matrix element calculations. For four-quark operators it has been consistently resolved through the introduction of evanescent operators vanishing in $D=4$~\cite{Buras:1989xd,Dugan:1990df,Herrlich:1994kh}. Following this approach, in this section we precisely define evanescent operator counterterms for the $Q_I$. It would be possible to present complete one-loop-matching results without these precise definitions, but the definitions and notation introduced in this section will prove essential for calculating the non-trivial evanescent contributions to two-loop running in Sec.~\ref{sec:running}.

Renormalized vertex functions include counterterms that remove the $1/\bar{\varepsilon}$ poles in Eq.~\eqref{matchingA}-\eqref{matchingS}. The physical operators $Q_I$ mix under RG evolution with the operators used to construct these counterterms, so a complete operator basis must include all operators used to construct counterterms. Dirac structures involving $\sigma\otimes \sigma$ appear in the pole terms above, but have been eliminated from our complete basis in $D=4$ by means of the Fierz transformation
\begin{equation}
  [CP_\chi \sigma_{\mu\nu}]^{\alpha\beta}[CP_{\chi^\prime} \sigma_{\mu\nu}]^{\gamma\delta} \stackrel{D=4}{=} \delta_{\chi\chi^\prime}\left( 8P_\chi^{\alpha\delta}P_\chi^{\gamma\beta} - 4P_\chi^{\alpha\beta}P_{\chi^\prime}^{\gamma\delta} \right).
  \label{2sigmafierz}
\end{equation}
This relation follows from completeness of a basis of 16 Dirac matrices in $D=4$ and cannot be uniquely continued to an analytic function of $D$. In particular, one could prescribe that in the dimensionally regularized theory
\begin{equation}
  [CP_\chi \sigma_{\mu\nu}]^{\alpha\beta}[CP_{\chi^\prime} \sigma_{\mu\nu}]^{\gamma\delta} \rightarrow \delta_{\chi\chi^\prime}\left( (8+a_1\varepsilon)P_\chi^{\alpha\delta}P_\chi^{\gamma\beta} - (4+a_2\varepsilon)P_\chi^{\alpha\beta}P_{\chi^\prime}^{\gamma\delta} \right).
  \label{2sigmafierzprescrip}
\end{equation}
with $a_1$ and $a_2$ arbitrary. The choice $a_1=a_2=0$ ensures that $\sigma_{\mu\nu}\otimes\sigma_{\mu\nu}$ is kept equal to its $D=4$ Fierz transform. Conversely, $\gamma_\mu \gamma_\nu \otimes \gamma_\mu \gamma_\nu$ is not kept equal to its $D=4$ Fierz transform with $a_1=a_2=0$. The necessity of breaking one or the other Fierz relation follows from the well-known property that contraction of a tensor operator with $g_{\mu\nu}$ does not commute with renormalization of the dimensionally regularized tensor operator \cite{Collins:105730}.

Working in a dimensionally regularized theory, it is important to distinguish between counterterms in the span of the $D=4$ basis operators $Q_I$ in $D$ dimensions, and counterterms that are in linearly independent in $D$ dimensions. A convenient basis is found by including $Q_I$ along with a set of evanescent operators $E_I$ that vanish in $D=4$ but are needed as counterterms to renormalize matrix elements of $Q_I$. For example, renormalization of $Q_1$ requires a counterterm insertion of
\begin{equation}\label{E1a}
	E_1^a = (\psi CP_R \sigma_{\mu\nu} i\tau_2\psi)(\psi CP_R \sigma_{\mu\nu} i\tau_2\psi)(\psi CP_R i\tau_2\tau_+ \psi)T^{SSS} - 12Q_1.
\end{equation}
The evanescent operator $E_1^a$ vanishes in $D=4$ by Eq.~\eqref{2sigmafierz}, color-flavor Fierz relations, and the $D=4$ Fierz relation $Q_1\stackrel{D=4}{=}\tilde{Q}_1$. Because Eq.~\eqref{2sigmafierz} is broken in dimensional regularization, it is possible that loop-level corrections will introduce $O(\alpha_s)$ contributions to matrix elements of $E_1^a$ that do not vanish in $D=4$. Explicit calculation demonstrates that this possibility is realized. The non-vanishing one-loop contributions are $O(\varepsilon^0)$ and arise from $1/\bar{\varepsilon}$ poles in one-loop integrals multiplied by $O(\varepsilon)$ suppressed differences in Dirac algebra for the two terms on the RHS of Eq.~\eqref{E1a}. In a naive definition of the $\bar{\text{MS}}$ renormalization scheme, matrix elements of RG evolved physical operators will include non-vanishing contributions from renormalized evanescent operators.

Following Ref.~\cite{Buras:1989xd}, we adopt a definition of the $\bar{\text{MS}}$ renormalization scheme in which the renormalized evanescent operators $E_I$ mixing with $Q_I$ under RG evolution are defined to include finite $O(\varepsilon^0)$ counterterms. These counterterms are chosen to make loop-level matrix elements of $E_I$ vanish in $D=4$ at a particular scale $\mu$. It is proven in Refs.~\cite{Dugan:1990df,Herrlich:1994kh} that this is sufficient to make renormalized matrix elements of generic four-quark evanescent operators vanish at all scales. Extension of this proof to six-quark operators is straightforward and discussed in Sec.~\ref{sec:running}. The basis used here for the $E_I$ needed as one-loop counterterms for $Q_I$ is explicitly presented in Appendix~\ref{app:evops}. Physical observables are independent of evanescent basis, but renormalized Wilson coefficients and matrix elements separately are not. It is therefore imperative that this same basis is used for loop-level BSM matching calculations. This subtlety is irrelevant for tree-level BSM matching calculations.

Inclusion of $\bar{\text{MS}}$ counterterms in this scheme is equivalent to replacing all terms involving $\sigma\otimes\sigma/\bar{\varepsilon}$ with their $D=4$ Fierz transforms plus the evanescent operators of Appendix~\ref{app:evops}, for instance
\begin{equation}
	\frac{1}{\bar{\varepsilon}}(\psi CP_R \sigma_{\mu\nu} i\tau_2\psi)(\psi CP_R \sigma_{\mu\nu} i\tau_2\psi)(\psi CP_R i\tau_2\tau_+)T^{SSS}  + \left(\bar{\text{MS}} \text{ counterterm}\right) = \frac{1}{\bar{\varepsilon}}12 Q_1 + \frac{1}{\bar{\varepsilon}}E_1^a.
\end{equation}
Inclusion of these counterterms leads to mixing between $Q_I$ and $E_I$ and we must enlarge our basis of renormalized operators to
\begin{equation}
  \begin{pmatrix} Q_I(\mu) \\ E_I(\mu) \\ \vdots \end{pmatrix} = \begin{pmatrix} Z_{II}(\mu) & Z_{IE_I}(\mu) & \hdots \\ Z_{E_I I}(\mu) & Z_{E_I E_I}(\mu) \\ \vdots & & \ddots \end{pmatrix} \begin{pmatrix} Q_I^0 \\ E_I^0 \\ \vdots \end{pmatrix} = \hat{Z}_{I}(\mu) \begin{pmatrix} Q_I^0 \\ E_I^0 \\ \vdots \end{pmatrix},
  \label{Zev}
\end{equation}
where the ellipses indicate that increasingly many evanescent operators are required to form a complete basis for RG evolution at increasingly high loop order. We are specializing to the case of no mixing between the $Q_I$ or between the $E_I$, extension to the general case is straightforward.

The one-loop vertex function $\Lambda_I^{(1)}$ can now be expressed in terms of $\Lambda_I^{(0)}$, tree-level vertex functions $\Lambda_{E_I}^{(0)}$ built from $E_I$, operator counterterms $\delta_{II}^{(1)}$, and vertex functions $\Phi_I(p,\mu)$ built from the non-local finite terms in Eq.~\eqref{matchingA}~-~\eqref{matchingS} including $ \gamma_\mu \fs{p}\otimes \fs{p}\gamma_\mu$ and $\ln(p^2/\mu^2)$,
\begin{equation}
  \Lambda_I^{(1)}(p,\mu) = \left( \frac{L_{II}^{(1),1}}{\bar{\varepsilon}} + L_{II}^{(1),0} + \delta_{II}^{(1)} \right)\Lambda_I^{(0)} + \left(\frac{L_{IE_I}^{(1),1}}{\bar{\varepsilon}}  + \delta_{IE_I}^{(1)}\right)\Lambda_{E_I}^{(0)} + L_{I\Phi_I}^{(1),0}\Phi_I(p,\mu) + O(\varepsilon),
  \label{oneloopQ}
\end{equation}
where all the loop diagram coefficients $L^{(1)}$ are pure numbers independent of $\mu$, $p$, and $\varepsilon$. $L^{(1)}$ and $\Phi_I(p,\mu)$ are simply obtained by expressing vertex functions constructed from Eqs.~\eqref{matchingA}~-~\eqref{matchingS} in the form of Eq.~\eqref{oneloopQ}. The one-loop vertex function $\Lambda_{E_I}^{(1)}$ for $E_I$ can similarly be expressed as
\begin{equation}
  \Lambda_{E_I}^{(1)}(p,\mu) = \left( \frac{L_{E_I E_I}^{(1),1}}{\bar{\varepsilon}} + \delta_{E_I E_I}^{(1)} \right)\Lambda_{E_I}^{(0)} + \left(L_{E_I I}^{(1),0} + \delta_{E_I I}^{(1)} \right)\Lambda_I^{(0)} + \left( \frac{L_{E_IF_I}^{(1),1}}{\bar{\varepsilon}} + \delta_{E_I F_I}^{(1)} \right)\Lambda_{F_I}^{(0)} + O(\varepsilon),
  \label{oneloopE}
\end{equation}
where $F_I$ is a new evanescent operator not included in the $E_I$ that should be included in the $\dots$'s in Eq. \eqref{Zev}. $\bar{\text{MS}}$ counterterms can be defined as
\begin{equation}
  \delta_{II}^{\bar{\text{MS}},(1)} = -\frac{L_{II}^{(1),1}}{\bar{\varepsilon}} ,\hspace{20pt} \delta_{IE_I}^{\bar{\text{MS}},(1)} = -\frac{L_{IE_I}^{(1),1}}{\bar{\varepsilon}},\hspace{20pt} \delta_{E_I I}^{\bar{\text{MS}},(1)} = -L_{E_I I}^{(1),0}.
\end{equation}
The bare operator $Q_I^0$ includes UV singularities due to the presence of six bare quark fields as well as the vertex function singularities above. In order to remove all UV singularities from $Q_I(\mu)$, define
\begin{equation}
  Z_{II}^{\bar{\text{MS}}}(\mu) = \left( Z_q^{\bar{\text{MS}}}(\mu) \right)^{-3}\left[ 1 + \delta_{II}^{\bar{\text{MS}},(1)}\left( \frac{\alpha_s(\mu)}{4\pi} \right) + \delta_{II}^{\bar{\text{MS}},(2)}\left( \frac{\alpha_s(\mu)}{4\pi} \right)^2 + O(\alpha_s^3) \right],
  \label{ZMSbar}
\end{equation}
where $\delta_{II}^{\bar{\text{MS}},(2)}$ represents two-loop counterterms that will be explicitly constructed in Sec.~\ref{sec:running}. For future use, define
\begin{equation}
  \begin{split}
	Z_{IE_I}^{\bar{\text{MS}}}(\mu) &= \left( Z_q^{\bar{\text{MS}}}(\mu) \right)^{-3}\left[ \delta_{IE_I}^{\bar{\text{MS}},(1)}\left( \frac{\alpha_s(\mu)}{4\pi} \right) + \delta_{IE_I}^{\bar{\text{MS}},(2)}\left( \frac{\alpha_s(\mu)}{4\pi} \right)^2 + O(\alpha_s^3) \right],\\
  Z_{E_II}^{\bar{\text{MS}}}(\mu) &= \left( Z_q^{\bar{\text{MS}}}(\mu) \right)^{-3}\left[ \delta_{E_II}^{\bar{\text{MS}},(1)}\left( \frac{\alpha_s(\mu)}{4\pi} \right) + \delta_{E_II}^{\bar{\text{MS}},(1)}\left( \frac{\alpha_s(\mu)}{4\pi} \right)^2 + O(\alpha_s^3) \right],\\
  Z_{E_IE_I}^{\bar{\text{MS}}}(\mu) &= \left( Z_q^{\bar{\text{MS}}}(\mu) \right)^{-3}\left[ 1 + \delta_{E_IE_I}^{\bar{\text{MS}},(1)}\left( \frac{\alpha_s(\mu)}{4\pi} \right) + \delta_{E_IE_I}^{\bar{\text{MS}},(2)}\left( \frac{\alpha_s(\mu)}{4\pi} \right)^2 + O(\alpha_s^3) \right].
\end{split}
  \label{ZevMSbar}
\end{equation}
This completes our definition of $\bar{\text{MS}}$ operator renormalization factors in terms of diagrammatic counterterms.

The RI-MOM operator renormalization condition should also be modified so that RI-MOM renormalized evanescent operators have vanishing matrix elements in $D=4$. This is accomplished by adding a supplemental RI-MOM renormalization condition
\begin{equation}
  \tr\left( \mathcal{P}_I \Lambda_{E_J} \right) = 0.
  \label{RIMOMevdef}
\end{equation}
Combining this with the RI-MOM condition Eq.~\eqref{RIdef} expanded to $O(\alpha_s)$ gives
\begin{equation}
  \delta_{II}^{\text{RI},(1)} = -\frac{L_{II}^{(1),1}}{\bar{\varepsilon}} - L_{II}^{(1),0} - L_{I\Phi_I}^{(1),0}\tr\left[ \mathcal{P}_I \Phi_I(p_0,\mu=p_0) \right],
\end{equation}
where in analogy to Eq.~\eqref{ZMSbar},
\begin{equation}
  Z_{II}^{\text{RI}}(\mu) = \left( Z_q^{\text{RI}}(\mu) \right)^{-3}\left[ 1 + \delta_{II}^{\text{RI},(1)} \left( \frac{\alpha_s(\mu)}{4\pi} \right) + O(\alpha_s^2) \right].
\end{equation}
The one-loop-matching factor $r_I^{(0)}$ defined in Eq.~\eqref{rdef} therefore has the diagrammatic expansion
\begin{equation}
  \begin{split}
	  r_{I}^{(0)} &= \delta_{II}^{\text{RI},(1)} - \delta_{II}^{\bar{\text{MS}},(1)} - 3Z_q^{\text{RI},(1)}(p_0) + 3Z_q^{\bar{\text{MS}},(1)}(p_0)\\
	&= -L_{II}^{(1),0} - L_{I\Phi_I}^{(1),0}\tr\left[ \mathcal{P}_I \Phi_I(p_0,\mu=p_0) \right] - 3Z_q^{\text{RI},(1)}(p_0) + 3Z_q^{\bar{\text{MS}},(1)}(p_0).
  \end{split}
\end{equation}
Applying the diagrammatic results of the last section gives
\begin{subequations}
  \begin{align}
	r^{(0)}_{1} &=  \frac{101}{30} - \frac{13\xi}{15} + \frac{8}{15}\ln 2 + \frac{8\xi}{3}\ln 2,\\
	r^{(0)}_{2} &= -\frac{31}{6} - \frac{7\xi}{3} + \frac{88}{15}\ln 2 + \frac{56\xi}{15}\ln 2,\\
	r^{(0)}_{3} &= -\frac{9}{10} - \frac{8\xi}{5} + \frac{16}{5}\ln 2 + \frac{16\xi}{5} \ln 2,\\
	r^{(0)}_{4} &= \frac{177}{10} + \frac{14\xi}{5} - \frac{64}{5}\ln 2,\\
	r^{(0)}_{5} &= \frac{49}{10} + \frac{3\xi}{5} - \frac{24}{5}\ln 2 + \frac{8\xi}{5} \ln 2,\\
	\tilde{r}^{(0)}_{1} &= -\frac{109}{30} - \frac{13\xi}{15} + \frac{8}{15}\ln 2 + \frac{8\xi}{3} \ln 2,\\
	\tilde{r}^{(0)}_{3} &= -\frac{79}{10} - \frac{8\xi}{5} + \frac{16}{5}\ln 2 + \frac{16\xi}{5} \ln 2.
  \end{align}
	\label{matchingres}
\end{subequations}
The final one-loop-matching results in Table~\ref{tab:summary} are obtained after choosing Landau gauge, $\xi = 0$.

\section{Two-Loop Running}\label{sec:running}

In order to simultaneously remove large logarithms from perturbative calculations of RI-MOM matching factors and BSM Wilson coefficients, RG evolution can be used to relate Wilson coefficients calculated at different renormalization scales.\footnote{See Ref.~\cite{Buras:1998raa} for a nice review of RG evolution for weak matrix elements including discussion of evanescent operators.} We perform this RG evolution in the $\bar{\text{MS}}$ scheme for simplicity, and all quantities in this section with suppressed renormalization scheme labels are in the $\bar{\text{MS}}$ scheme. The renormalization scale dependence of the Wilson coefficients can be determined from the $\bar{\text{MS}}$ anomalous dimension matrix
\begin{equation}
  \begin{split}
	\gamma_{IJ}(\alpha_s) &= \frac{1}{C_I(\mu)} \frac{d}{d\ln \mu}C_J(\mu) = \sum_K Z_{IK}(\mu) \frac{d}{d \ln \mu}Z^{-1}_{KJ}(\mu)\\
  &\equiv \gamma_{IJ}^{(0)} \left( \frac{\alpha_s(\mu)}{4\pi} \right) + \gamma_{IJ}^{(1)} \left( \frac{\alpha_s(\mu)}{4\pi} \right)^2 + O(\alpha_s^3),
\end{split}
  \label{gammadef}
\end{equation}
where the first equality follows from renormalization scale independence of $\mathcal{H}_{eff}^{n\bar{n}}$ and the second defines the perturbative expansion coefficients $\gamma^{(0)}$ and $\gamma^{(1)}$ appearing in Eq. \eqref{master}. The other factors appearing in Eq. \eqref{master} are related to the QCD $\beta$-function, defined by
\begin{equation}
  \begin{split}
	\frac{d}{d \ln\mu}\alpha_s(\mu) &= 2\beta(\alpha_s,\varepsilon)\alpha_s(\mu) = \left( -2\varepsilon + 2\beta(\alpha_s) \right)\alpha_s(\mu)\\
	&= \left( -2\varepsilon - 2\beta_0\frac{\alpha_s(\mu)}{4\pi} - 2\beta_1\left( \frac{\alpha_s(\mu)}{4\pi} \right)^2 + O(\alpha_s^3) \right)\alpha_s(\mu).
  \end{split}
\end{equation}
The $D$-independent piece of the $\beta$-function has a perturbative expansion that is conventionally written as
\begin{equation}
  \beta(\alpha_s) = -\beta_0\left( \frac{\alpha_s(\mu)}{4\pi} \right) - \beta_1\left( \frac{\alpha_s(\mu)}{4\pi} \right)^2 + O(\alpha_s^3),
\end{equation}
where for QCD with $N_f$ active quark flavors the well-known perturbative coefficients are~\cite{Gross:1973id,Politzer:1973fx,Jones:1974mm,Caswell:1974gg}
\begin{equation}
	\beta_0 = 11 - \frac{2}{3}N_f,\hspace{20pt} \beta_1 = 102 - \frac{38}{3}N_f.
	\label{betacoff}
\end{equation}
In $\bar{\text{MS}}$ and other minimal subtraction schemes, $\mu$ dependence of $Z_{IJ}$ and $\gamma_{IJ}$ only enters through dependence on $\alpha_s(\mu)$. The differential equation in Eq.~\eqref{gammadef} can be readily solved in a diagonal operator basis where $\gamma_{IJ}= \delta_{IJ}\gamma_I$,
\begin{equation}
  \begin{split}
	\frac{C_I(\mu_2)}{C_I(\mu_1)} &= \exp\left[ \int_{\mu_1}^{\mu_2} \gamma_{I}\left(\alpha_s(\mu^\prime)\right) \frac{d\mu^\prime}{\mu^\prime} \right] = \exp\left[ \int_{\alpha_s(\mu_1)}^{\alpha_s(\mu_2)} \frac{\gamma_{I}(\alpha_s^\prime)}{2 \beta(\alpha_s^\prime)} \frac{d\alpha_s^\prime}{\alpha_s^\prime} \right]\\
  &= \left( \frac{\alpha_s(\mu_2)}{\alpha_s(\mu_1)} \right)^{-\gamma_{I}^{(0)}/(2\beta_0)}\left[ 1 + \left( \frac{\beta_1\gamma_{I}^{(0)}}{2\beta_0^2} - \frac{\gamma_{I}^{(1)}}{2\beta_0} \right)\frac{\alpha_s(\mu_2)-\alpha_s(\mu_1)}{4\pi} + O(\alpha_s^2) \right].
\end{split}
  \label{Crunning}
\end{equation}
This equation can be used to RG evolve BSM-scale Wilson coefficients between quark mass thresholds at which the number of active flavors $N_f$ decreases.  In Sec.~\ref{sec:matching} we introduced $r_{IJ}(\mu)$ as the renormalization scheme matching factor relating $Q_I^{\bar{\text{MS}}}$ and $Q_I^{\text{RI}}$. The renormalization scheme invariance of $\mathcal{H}_{eff}^{n\bar{n}}$ allows $C_I^{\bar{\text{MS}}}$ and $C_I^{\text{RI}}$ to be related using $r_{IJ}(\mu)$. This allows us to express $U_I^{N_f}(\mu,p_0)$ appearing in Eq.~\eqref{master} as
\begin{equation}
  \begin{split}
	  U_I^{N_f}(\mu,p_0) &= \frac{C_I^{\text{RI}}(p_0)}{C_I^{\bar{\text{MS}}}(\mu)} = \frac{C_I^{\bar{\text{MS}}}(p_0)}{C_I^{\bar{\text{MS}}}(\mu)}\left( 1 - r_I^{(0)}\frac{\alpha_s(p_0)}{4\pi} + O(\alpha_s^2) \right)\\
  &= \left( \frac{\alpha_s(p_0)}{\alpha_s(\mu)} \right)^{-\gamma_{I}^{(0)}/2\beta_0}\left[ 1 - r_{I}^{(0)}\frac{\alpha_s(p_0)}{4\pi} + \left( \frac{\beta_1 \gamma_{I}^{(0)}}{2\beta_0^2} - \frac{\gamma_{I}^{(1)}}{2\beta_0} \right)\frac{\alpha_s(p_0)-\alpha_s(\mu)}{4\pi} + O(\alpha_s^2) \right].
\end{split}
\end{equation}
The remainder of this section discusses the diagrammatic evaluation of $\gamma_I^{(0)}$, $\gamma_I^{(1)}$.

In Sec.~\ref{sec:matchingev} we discussed the need to remove dimensional regularization artifacts by adding finite counterterms proportional to evanescent operators $E_I$ to physical operators $Q_I$ and vice versa. Without these counterterms the renormalized $E_I$ would contribute to physical observables and therefore to BSM matching calculations of Wilson coefficients. With these counterterms, $Q_I$ mixes under renormalization with $E_I$ and the assumption of a diagonal anomalous dimension matrix taken above is invalidated. We must instead consider the renormalization scale dependence of the infinite dimensional matrix of Eq. \eqref{Zev} and define
\begin{equation}
  \hat{\gamma} = \begin{pmatrix} \gamma_{IJ} & \gamma_{I E_I} & \hdots \\ \gamma_{E_I I} & \gamma_{E_I E_I} & \\ \vdots & & \ddots \end{pmatrix}.
\end{equation}
Eq. \eqref{Crunning} is preserved if and only if $\gamma_{E_I I}=0$ to two-loop order. A proof that $\gamma_{E_I I}$ vanishes to all-orders for generic four-quark operators is given in Refs.~\cite{Dugan:1990df,Herrlich:1994kh} and applies to our six-quark operators as well. This is discussed in detail at the end of this section.

Since $\mu$ dependence of $\hat{Z}$ only enters through explicit dependence on $\alpha_s(\mu)$, the anomalous dimension matrix $\hat{\gamma}$ is given by
\begin{equation}
  \hat{\gamma} = -\left(\mu\frac{d}{d\mu}\hat{Z}\right)\cdot \hat{Z}^{-1} = -\left( 2\beta(\alpha_s,\varepsilon)\alpha_s(\mu) \frac{\partial}{\partial \alpha_s}\hat{Z}\right)\cdot\hat{Z}^{-1},
  \label{fullgammadef}
\end{equation}
where $\cdot$ denotes matrix multiplication. Perturbative coefficients of $\hat{Z}$ defined in analogy to Eq.~\eqref{gammadef} are given by
\begin{equation}
  \begin{split}
  \hat{\gamma}^{(0)} &= 2\varepsilon \hat{Z}^{(1)},\\
  \hat{\gamma}^{(1)} &= 4\varepsilon \hat{Z}^{(2)} - 2\varepsilon \hat{Z}^{(1)} \cdot \hat{Z}^{(1)} +2\beta_0 \hat{Z}^{(1)}.
\end{split}
  \label{fullgamma}
\end{equation}
The anomalous dimensions of the physical operators $Q_I$ are therefore
\begin{equation}
  \begin{split}
  \gamma_{I}^{(0)} &= 2\varepsilon Z_{II}^{(1)},\\
  \gamma_{I}^{(1)} &= 4\varepsilon Z_{II}^{(2)} - 2\varepsilon\left( Z_{II}^{(1)^2} + Z_{IE_I}^{(1)}Z_{E_I I}^{(1)} \right) + 2\beta_0 Z_{II}^{(1)}.
\end{split}
  \label{gamma1II}
\end{equation}
The non-trivial effect of evanescent counterterm subtraction is the appearance of $Z_{IE_I}^{(1)}Z_{E_I I}^{(1)}$ in $\gamma_I^{(1)}$. 

The factors above are simply related to diagrammatic counterterms. The one-loop anomalous dimension is determined by the counterterms of Sec. \ref{sec:matchingev} as
\begin{equation}
  \gamma_{I}^{(0)} = 2\varepsilon \left(\delta_{II}^{(1)} - 3Z_q^{(1)}\right),
\end{equation}
which is finite at $D=4$. Calculation of $\gamma_{I}^{(1)}$ requires the $1/\bar{\varepsilon}$ pole contributions to the two-loop $Q_I$ vertex functions
\begin{equation}
	\Lambda_I^{(2)}(p,\mu) = \left( \frac{L_{II}^{(2),2}}{\bar{\varepsilon}^2} + \frac{L_{II}^{(2),1}}{\bar{\varepsilon}} + \delta^{(2)}_{II} \right)\Lambda_I^{(0)} + \left( \frac{L_{IE_I}^{(2),2}}{\bar{\varepsilon}^2} \right)\Lambda_{E_I}^{(0)} + \left( \frac{L_{IF_I}^{(2),2}}{\bar{\varepsilon}^2} \right)\Lambda_{F_I}^{(0)} + O(\varepsilon^0).
	\label{Lambda2}
\end{equation}
Including one-loop counterterm diagrams with insertions of $\delta_{II}^{(1)}$ as well as quark self-energy, gluon self-energy, and quark-gluon-vertex counterterms ensures that non-local divergences are cancelled and $L_{II}^{(2),1}$ is a pure number. The two-loop $\bar{\text{MS}}$ counterterm is then defined as
\begin{equation}
  \delta_{II}^{(2)} = -\frac{L_{II}^{(2),2}}{\bar{\varepsilon}^2} - \frac{L_{II}^{(2),1}}{\bar{\varepsilon}}.
\end{equation}
We can then use Eqs.~\eqref{ZMSbar}~-~\eqref{ZevMSbar} to express the $Z$ factors appearing in Eq.~\eqref{gamma1II} in terms of these counterterms, 
\begin{equation}
  \gamma_{I}^{(1)} = 4\varepsilon\left( \delta_{II}^{(2)} - 3Z_q^{(2)} \right) - 2\varepsilon\left( \delta_{II}^{(1)^2} + \delta_{IE_I}^{(1)}\delta_{E_I I}^{(1)} - 3Z_q^{(1)^2} \right) + 2\beta_0\left( \delta_{II}^{(1)} - 3Z_q^{(1)} \right).
  \label{gamma1}
\end{equation}
Two-loop counterterms include $1/\bar{\varepsilon}^2$ contributions, so the various terms in Eq. \eqref{gamma1} are divergent in $D=4$. Renormalizability of composite operators in the $\bar{\text{MS}}$ scheme guarantees that matrix elements of $Q_I(\mu)$ are free of UV divergences at all renormalization scales and therefore that $\gamma_I$ is finite order-by-order \cite{Collins:105730}. This means that divergences must cancel between the terms of Eq. \eqref{gamma1}.\footnote{A potential point of confusion: if one naively takes Eq. \eqref{fullgamma} with $\hat{Z}$ replaced by $Z_q^{-1}$ as a formula for the two-loop quark field anomalous dimension and inserts Eq.~\eqref{Zq2}, $1/\bar{\varepsilon}^2$ divergences do not cancel. The subtlety is that $Z_q$ depends on the gauge parameter $\xi$, which in turn depends on the renormalization scale \cite{Collins:105730}. When Eq. \eqref{fullgammadef} is modified to include this additional source of renormalization scale dependence, the resulting quark field anomalous dimension is indeed finite.} After this cancellation, the anomalous dimension is given by
\begin{equation}
	\gamma_{I}^{(1)} = -4L_{II}^{(2),1} - 12Z_q^{(2),1} + 2L_{I E_I}^{(1),1}L_{E_I I}^{(1),0} \equiv -4[L_{tot}]_{II}^{(2),1} - 12Z_q^{(2),1}
  \label{gamma1fin}
\end{equation}
where $Z_q^{(2),1}$ is the $1/\bar{\varepsilon}$ piece of Eq.~\eqref{Zq2}.

It was noticed in Ref.~\cite{Buras:1989xd} that the finite contributions to $\gamma_{I}^{(1)}$ from mixing with evanescent operators contribute exactly like an additional counterterm diagrams apart from the relative factor of $(-1/2)$ between $L_{II}^{(2),1}$ and $L_{IE_I}^{(1),1}L_{E_I I}^{(1),0}$ in Eq.~\eqref{gamma1fin}. As discussed after Eq.~\eqref{Lambda2}, $L_{II}^{(2)}$ includes contributions with one-loop counterterm diagrams containing insertions of $\delta_{II}^{(1)}$.  Suppose for each of these one-loop counterterm diagrams we include an additional counterterm diagram with an insertion of $(1/2) \delta_{E_I I}^{(1)} = (-1/2) L_{E_I I}^{(1),0}$. These diagrams make a $1/\bar{\varepsilon}$ pole contribution of $(-1/2) L_{I E_I}^{(1),1}L_{E_I I}^{(1),0}/\bar{\varepsilon}$. Including these additional one-loop counterterm diagrams with insertion of $(1/2) \delta_{E_I I}^{(1)}$ therefore shifts the $1/\bar{\varepsilon}$ single-pole part of $\Lambda_I^{(2)}$ to
\begin{equation}
	[L_{tot}]_{II}^{(2),1} = L_{II}^{(2),1} - \frac{1}{2}L_{I E_I}^{(1),1}L_{E_I I}^{(1),0},
  \label{Ltot}
\end{equation}
the factor appearing directly in Eq.~\eqref{gamma1fin}. 

To ensure a proper treatment of two-loop subdivergences and verify cancellation of non-local divergences diagram-by-diagram, each two-loop diagram should be combined with a one-loop counterterm diagram in which any divergent one-loop subdiagram present is replaced by a one-loop counterterm that cancels the subdivergence. To provide this cancellation diagram-by-diagram, the one-loop counterterm must have the same color structure as the one-loop subdiagram. For subdiagrams with the topology of a one-loop self-energy or vertex correction, the color structure of the subdiagram is a simple multiple of the corresponding tree-level color structure, and this procedure is straightforward. For subdiagrams with the topology of a one-loop operator correction $d$ = 1-3, the color structure of the subdiagram differs from the tree-level operator color structure, and care must be taken. In particular, the physical operator counterterm associated with each $d$ = 1-3 subdiagram must precisely reproduce that subdiagram's contribution to $\delta^{(1)}_{II}$ in $D$ dimensions. This means that the operator used for each physical operator counterterm must be proportional to $Q_I$ in $D$ dimensions. 

To produce physical operator counterterms proportional to $Q_I$ in $D$ dimensions, it is necessary but not sufficient that all appearances of $\sigma\otimes \sigma$ in $d$ = 1-3 subdiagrams are replaced in one-loop counterterm diagrams by trivial spin structures containing only the identity matrix. A convenient prescription that meets this necessary criterion is to use Eq.~\eqref{2sigmafierzprescrip} with $a_1=a_2=0$. This prescription amounts to replacing all appearances of $\sigma\otimes\sigma$ in $d$ = 1-3 subdiagrams with their $D=4$ Fierz transforms and clearly produces one-loop counterterm diagrams with the same color structures as each two-loop diagram. However, one-loop counterterms defined by this prescription are not proportional to $Q_I$ in $D$ dimensions and instead differ by terms proportional to $(Q_1-\tilde{Q}_1)$ and $(Q_3-\tilde{Q}_3)$. Applying the prescription of Eq.~\eqref{2sigmafierzprescrip} with $a_1=a_2=0$ to the one-loop amplitude defines a basis $E_I^\prime$ different from the one-loop counterterm basis $E_I$. For example, the operator $E_1^{a\prime}$ is given by
\begin{equation}
	\begin{split}
		&(\psi CP_R \sigma_{\mu\nu} i\tau_2\psi)(\psi CP_R \sigma_{\mu\nu} i\tau_2\psi)(\psi CP_R i\tau_2\tau_+ \psi)T^{SSS}\\
		&\equiv 8(\psi^\alpha [CP_R]^{\alpha\delta} i\tau_2 \psi^\beta)(\psi^\gamma [CP_R]^{\gamma\beta} i\tau_2 \psi^\delta)(\psi^\eta [CP_R]^{\eta\zeta} i\tau_2\tau_+ \psi^\zeta)T^{SSS} + E_1^{a\prime}\\
		&= 12 Q_1 - 6(Q_1 - \tilde{Q}_1) + E_1^{a\prime},
	\end{split}\label{E1aprime}
\end{equation}
where the last equality follows from color-flavor algebra. Operators $E_1^{a\prime}$, $E_3^{a\prime}$, $\tilde{E}_1^\prime$, and $\tilde{E}_3^\prime$ differ from their unprimed counterparts by factors of $(Q_1-\tilde{Q}_1)$ and $(Q_3-\tilde{Q}_3)$ as seen for $E_1^{a\prime}$ by comparing Eq.~\eqref{E1aprime} and Eq.~\eqref{E1a}. The remaining operators needed to define the $E_I^\prime$ basis and it relation to the $E_I$ basis are explicitly presented in Appendix~\ref{app:evops}.

In the $E_I^\prime$ basis, the total physical plus evanescent one-loop operator counterterm associated with two-loop diagrams $d$ = 4-6, 16-24, and 32-45 containing $d$ = 1-3 subdiagrams is simply given by minus the pole part of the subdiagram with appearances of $\sigma\otimes \sigma$ replaced by
\begin{equation}
	\begin{split}
		\sigma\otimes\sigma \rightarrow \frac{1}{2}\left( \sigma\otimes \sigma + F[\sigma\otimes\sigma] \right),
	\end{split}\label{evCTprescrip}
\end{equation}
where $F[\sigma\otimes\sigma]$ is given by the RHS of Eq.~\eqref{2sigmafierzprescrip} with $a_1=a_2=0$. The total $1/\bar{\varepsilon}$ single-pole contribution given by a diagrammatic two-loop calculation using this prescription we denote $[L_{tot}^\prime]_{II}^{(2),1}$ and is related to $[L_{tot}]_{II}^{(2),1}$ by
\begin{equation}
	\begin{split}
		[L_{tot}]_{II}^{(2),1)} &= [L_{tot}^\prime]_{II}^{(2),1} - \frac{1}{2}L_{I E_I}^{(1)}\left( L_{E_I I}^{(1),0} - L_{E_I^\prime I}^{(1),0} \right).
	\end{split}\label{Ltotprime}
\end{equation}
The change of evanescent basis factors appearing in Eq.~\eqref{Ltotprime} can be immediately obtained from one-loop results for $(r_1^{(0)} - \tilde{r}_1^{(0)})$ and $(r_3^{(0)} - \tilde{r}_3^{(0)})$ and are given in Appendix~\ref{app:evops}. We have explicitly verified that after including these change of evanescent basis factors the total contribution of one-loop physical operator counterterm diagrams is equal to the sum of one-loop counterterm diagrams with insertions of $\delta^{(1)}_{II}$.

With this evanescent-counterterm diagram prescriptions in hand, diagrammatic calculation of the $E_I^\prime$ basis contributions to $\gamma_I^{(1)}$ proceeds as described in Sec.~\ref{sec:matchingdiag} and Appendix~\ref{app:algebra}~-~\ref{app:tables}. There are 320 contributing two-loop diagrams, organized into independent classes $d$ = 4-46 in Fig.~\ref{diagrams}. The total number of two-loop diagrams can be determined through straightforward combinatoric arguments, for example there are ${6\choose 3} = 20$ diagrams involving a three-gluon vertex with gluon lines attached to three separate quark lines and ${6\choose 2}^2 - \frac{1}{2}{6\choose 2}{4\choose 2} = 180$ diagrams involving planar one-gluon-exchange between two quark pairs. The remaining diagram types can be grouped in multiples of ${6 \choose 2} = 15$, the number of one-loop diagrams. As a check on the completeness of the set of diagrams included in this work, we have verified that the number of diagrams in all classes shown in Fig. 1 agrees with the results of the automated Feynman diagram generation program qgraf \cite{Nogueira:1991ex}.\footnote{For reference, we note that a two-loop qgraf analysis of a process with six incoming quark fields interacting with standard QCD Feynman rules plus a six-quark vertex provides 350 one-particle-irreducible diagrams excluding ``tadpoles'' and ``snails.'' 45 of these are gluon-self-energy diagrams that are only counted as 15 diagrams, each containing an insertion of the complete one-loop gluon-self-energy bubble, in Fig. 1. Organizing a qgraf analysis with three incoming quarks and three incoming conjugate antiquarks requires more care; a qgraf analysis including three incoming quarks and three incoming antiquarks interacting with a six-point vertex produces an additional 90 spurious penguin diagrams with baryon-number-violating quark-conjugate-antiquark annihilation into gluons.} 

Diagrams 4-31 contribute to NNLO renormalization of scalar four-quark operators, and many also contribute to renormalization of three-quark operators. We have adopted the same numbering scheme for these diagrams used in Ref.~\cite{Buras:1989xd,Buras:2000if}, and have verified that our results for these diagrams agree with the scalar four-quark operator results of Ref.~\cite{Buras:2000if} after changing to the appropriate evanescent operator basis (the ``Greek projection'' basis, see Appendix \ref{app:algebra}). Diagrams 32-46 are new. For future calculations, it is interesting to note that there are no additional two-loop diagram classes appearing for operators with more than six quarks. In principle the two-loop anomalous dimension of any $\Delta B = N$ operator composed of a product of scalar diquarks could be computed from the results of Table~\ref{diag_results}, combinatorics, and group theory.

After including all appropriate one-loop counterterms, including the $E_I^\prime$ evanescent counterterms defined by Eq.~\eqref{evCTprescrip}, the $1/\bar{\varepsilon}$ pole part of each diagram can be decomposed into a color factor for each diagram times a linear combination of Dirac structures that is identical up to diquark permutations for all diagrams in the class. The pole parts of these Dirac structures are shown in Table~\ref{diag_results}, and the corresponding color factor for a representative diagram is shown in Table~\ref{color_factors}. The $1/\bar{\varepsilon}$ pole coefficients of the combined spin-color tensors with non-vanishing contributions to $T^{AAS}$ operators are shown in Table~\ref{results_A}. The corresponding pole coefficients contributing to $T^{SSS}$ operators are shown in Table~\ref{results_S}. After re-introducing quark fields and flavor tensors for a given operator by Eq.~\eqref{rosetta}, the resulting pole structures for each operator are related by Eq.~\eqref{evbasis} to a multiple of the original operator plus irrelevant contributions to $L_{IE_I}^{(2),2}$.

When the dust settles, these diagrammatic contributions sum to $L_{II}^{(2),2}/\bar{\varepsilon}^2 + [L_{tot}^\prime]_{II}^{(2),1}/\bar{\varepsilon}$. We have explicitly verified that the $1/\bar{\varepsilon}^2$ contributions to $\delta_{II}^{(2)}$ cancel with the other divergent terms in Eq. \eqref{gamma1}. This provides a highly non-trivial check on the calculation. The physical anomalous dimensions are then given by Eq. \eqref{gamma1fin} as
\begin{subequations}
  \begin{align}
	  \gamma^{(1)}_{1} &= \frac{335}{3} - \frac{34N_f}{9}, \\
	\gamma^{(1)}_{2} &= \frac{91}{3} - \frac{26N_f}{9}, \\
	\gamma^{(1)}_{3} &= 64 - \frac{10N_f}{3}, \\
	\gamma^{(1)}_{4} &= 229 - \frac{46N_f}{3}, \\
	\gamma^{(1)}_{5} &= 238 - 14N_f, \\
	\tilde{\gamma}^{(1)}_{1} &= \frac{797}{3} - \frac{118N_f}{9}, \\
	\tilde{\gamma}^{(1)}_{3} &= 218 - \frac{38N_f}{3}.
  \end{align}
	\label{runningres}
\end{subequations}

In the $\bar{\text{MS}}$ scheme used in this work, Fierz-conjugate operators $Q_1$, $\tilde{Q}_1$ and $Q_3$, $\tilde{Q}_3$ are equal in $D=4$ but do not have identical two-loop anomalous dimensions. Conversely, to be regularization independent the RI-MOM scheme cannot assign different anomalous dimensions to operators identical in $D=4$.\footnote{We thank Sergey Syritsyn for bringing this point to our attention.} Two-loop RI-MOM anomalous dimensions are gauge dependent and depend on the external state appearing in the RI-MOM renormalization condition, but $\gamma_I^{(1),\text{RI}}$ should be independent of evanescent basis and equal for Fierz-conjugate operators. The two-loop RI-MOM anomalous dimension
\begin{equation}
	\begin{split}
		\gamma_I^{(1),\text{RI}} = \gamma_I^{(1),\bar{\text{MS}}} + 2\beta_0 r_I^{(0)},
	\end{split}\label{gammaRIdef}
\end{equation}
can be shown to be independent of the renormalization scheme and evanescent basis used to define counterterms~\cite{Buras:1998raa,Chetyrkin:1997gb}. This provides a pair of consistency conditions
\begin{equation}
	\begin{split}
		\gamma_1^{\text{RI},(1)} = \tilde{\gamma}_1^{\text{RI},(1)},\hspace{20pt} \gamma_3^{\text{RI},(1)} = \tilde{\gamma}_3^{\text{RI},(1)},
	\end{split}\label{RIconsistency}
\end{equation}
that can be readily verified to hold for the results of Eq.~\eqref{matchingres} and Eq.~\eqref{runningres} in a general $R_\xi$ gauge. Eq.~\eqref{RIconsistency} provides a useful check on our calculation, and particularly on evanescent contributions to $\gamma_I^{(1)}$ that cannot spoil the $1/\bar{\varepsilon}^2$ cancellation consistency check. In particular, we observe that violations of Eq.~\eqref{RIconsistency} and $1/\bar{\varepsilon}^2$ poles cancel independently from the following sets of diagrams: $\{$4-6, 16-24, 32-45$\}$ containing $d$ = 1-3 subdiagrams; $\{$7-9$\}$ containing crossed gluon lines; $\{28, 46\}$ containing three-gluon vertices and no divergent subdiagrams; and $\{$10-15, 25-27, 29-31$\}$ combined with the one-loop contribution $2\beta_0 r_I^{(0)}$ and one- and two-loop wavefunction renormalization.

It remains to verify that $\gamma_{E_I I}=0$. At one-loop order
\begin{equation}
	\gamma_{E_I I}^{(0)} = 2\varepsilon \delta_{E_I I}^{(1)} .
\end{equation}
Since $\delta_{E_I I}^{(1)} = -L_{E_I I}^{(1),0}$ is a finite counterterm, $\gamma^{(0)}_{E_I I}$ trivially vanishes at $D=4$. At two-loop order
\begin{equation}
	\gamma_{E_I I}^{(1)} = 4\varepsilon \delta_{E_I I}^{(2)} - 2\varepsilon\left( \delta_{E_I I}^{(1)}\delta_{II}^{(1)} + \delta_{E_I E_I}^{(1)}\delta_{E_I I}^{(1)} \right) + 2\beta_0 \delta_{E_I I}^{(1)}.
	\label{gammaEII}
\end{equation}
Vanishing of $\gamma_{E_I I}^{(1)}$ is less trivial. Refs.~\cite{Dugan:1990df,Herrlich:1994kh} prove that for generic four-quark operators a strictly stronger statement is in fact true: the analog of $\hat{\gamma}$ is upper-triangular to all orders. The argument of Herrlich and Nierste in Ref.~\cite{Herrlich:1994kh} is quite general and only relies on cancellation of non-local divergences and $O(\varepsilon)$ suppression of diagrams with evanescent operator insertions. We briefly review their argument to demonstrate that it applies to our six-quark operators without modification.

The evanescent operators $E_I$ vanish in $D=4$ by spin-color-flavor Fierz transformations. The one-loop counterterm $\delta^{(1)}_{E_I I}$ therefore includes $O(\varepsilon)$-suppressed tensor algebra and can only include contributions non-vanishing in $D=4$ from terms that have received a $1/\bar{\varepsilon}$ enhancement from loop integrals. $\delta^{(1)}_{E_I I}$ is therefore $O(\varepsilon^0)$, and one-loop counterterm diagrams with insertions of $\delta^{(1)}_{E_I I}$ only make $1/\bar{\varepsilon}$ pole contributions to $\gamma_{E_I I}^{(1)}$ from terms that receive additional $1/\bar{\varepsilon}$ integral enhancements. This implies that the only $1/\bar{\varepsilon}$ pole contributions to $\gamma_{E_I I}^{(1)}$ from $\delta_{E_I I}^{(1)}\delta_{II}^{(1)}$ and $\delta_{E_I E_I}^{(1)}\delta_{E_I I}^{(1)}$ are single poles and arise from integral contributions with $1/\bar{\varepsilon}^2$ double-pole enhancements and $O(\varepsilon)$ tensor-algebra suppression. The sum of these one-loop counterterm diagram contributions is explicitly given by $(\mu^2/p^2)^\varepsilon(\delta_{E_I I}^{(1)}\delta_{II}^{(1)} + \delta_{E_I E_I}^{(1)}\delta_{E_I I}^{(1)} - \delta_{E_I I}^{(1)}\beta_0/\bar{\varepsilon})$. Without the $O(\varepsilon)$-suppressed tensor algebra, this expression would include non-local divergences arising from $1/\bar{\varepsilon}^2$ factors multiplied by $(\mu^2/p^2)^\varepsilon$. Cancellation of non-local divergences is independent of the tensor structure of operator insertions, and these would-be non-local divergences must cancel with $1/\bar{\varepsilon}^2$ two-loop integrals multiplied by $(\mu^2/p^2)^{2\varepsilon}$ and the same $O(\varepsilon)$-suppressed tensor algebra factors. This gives $\delta_{E_I I}^{(2)} = \frac{1}{2}(\delta_{E_I I}^{(1)}\delta_{II}^{(1)} + \delta_{E_I E_I}^{(1)}\delta_{E_I I}^{(1)} - \delta_{E_I I}^{(1)}\beta_0/\bar{\varepsilon})$, which when inserted in Eq.~\eqref{gammaEII} gives $\gamma_{E_I I}^{(1)} = 0$. The remaining inductive step needed to prove that $\hat{\gamma}$ is upper-triangular to all orders does not rely on a particular evanescent operator definition \cite{Dugan:1990df,Herrlich:1994kh} and applies here as well.


\section{Phenomenological Applications: An illustrative Example} \label{sec:pheno}

The phenomenological consequences of neutron-antineutron operator renormalization are encoded in the effective Hamiltonian $\mathcal{H}_{eff}^{n\bar{n}}$ of Eq.~\eqref{master} and the operator renormalization factors $\gamma_I^{(0)}$, $\gamma_I^{(1)}$, and $r_I^{(0)}$ collected in Table~\ref{tab:summary}. These operator renormalization factors govern the relations between matrix elements of $Q_I$ with different renormalization scheme and scale choices. Non-perturbative lattice QCD determinations of the renormalized QCD matrix elements\footnote{Lattice matching scales of $p_0 \simeq 2\ \text{GeV}$ are typically large enough for matching to be perturbative but small enough that unphysical UV cutoff effects are minimal ($\Lambda_{QCD} < p_0 < a^{-1}$, where $a$ is the lattice spacing).}  $\mbraket{\bar{n}}{Q_I^{RI}(p_0)}{n}$ can be combined with these operator renormalization factors to determine QCD matrix elements $\mbraket{\bar{n}}{Q_I^{\bar{MS}}(\mu)}{n}$ at high scales $\mu$, where BSM physics is usually assumed to be perturbative. Once high-scale QCD matrix elements have been calculated with fully quantified uncertainties, perturbative BSM matching for a particular BSM theory of interest can be used to predict $n\bar{n}$ transition matrix elements
\begin{equation}
 \frac{1}{\tau_{n\bar{n}}}= \delta m = \mbraket{\bar{n}}{\mathcal{H}_{eff}^{n\bar{n}}}{n},
\end{equation}
in terms of basic BSM parameters. Experimental constrains on the neutron-antineutron vacuum  transition probability $P_{n\bar{n}}(t) = \sin^2(|\delta m|t)$ then unambiguously constrain the parameter space of BSM theories predicting neutron-antineutron transitions.

The extraction of phenomenological predictions from Eq.~\eqref{master} is best explained via an example of one specific BSM model. Several broad classes of simplified models with classical baryon number violation but no proton decay were recently discussed by Arnold, Fornal, and Wise~\cite{Arnold:2012sd}. The BSM field content of these models consists of a pair of colored scalar fields that carry non-integer baryon or lepton number. For illustrative purposes, we will use the simplified model discussed most heavily (Model 1 in Ref.~\cite{Arnold:2012sd}), which we'll call the AFW1 model, to perform our calculation. Identical steps can be used to make predictions for other, more complicated BSM theories once a six-quark effective Hamiltonian $\mathcal{H}_{eff}^{n\bar{n}}$ has been determined for those theories.

The AFW1 model adds two new scalars to the standard model, $X_1$ and $X_2$, which transform as $X_1 \in (\bar{6},1,-1/3)$ and $X_2 \in (\bar{6},1,2/3)$ under $SU(3)_c \times SU(2)_L \times U(1)_Y$.  The $X_1$ and $X_2$ couplings to the SM right-handed fermions are given by $g_1^\prime$ and $g_2$, respectively, and an additional three-scalar coupling between two-$X_1$ and one-$X_2$  is given by $\lambda$.   
This model allows neutron-antineutron transitions at tree-level. The Hamiltonian operator $\mathcal{H}_{eff}^{n\bar{n}}$ is found by evaluating a tree-level Feynman diagram connecting six external quarks all carrying zero momentum. The resulting $\mathcal{H}_{eff}^{n\bar{n}}$ for the AFW1 model is presented in terms of fixed-flavor quark fields $u^\alpha_i$, $d^\alpha_i$ in Eq.~(12) of Ref.~\cite{Arnold:2012sd}, neglecting scalar couplings to left-handed quarks for simplicity. In the fixed-flavor and chiral operator bases, this Hamiltonian is given at tree-level by
\begin{equation}
\mathcal{H}_{eff}^{n\bar{n}} = - \frac{(g_1^{\prime 11})^2 g_2^{11} \lambda}{4 M_1^4 M_2^2}\mathcal{O}^2_{RRR}= \frac{(g_1^{\prime 11})^2 g_2^{11} \lambda}{16 M_1^4 M_2^2}\left[ Q_4 + \frac{3}{5}\tilde{Q}_1 \right],
\end{equation}
where $g_1^{\prime 11} = g_2^{11}$ are dimensionless couplings assumed to be $O(1)$ at a high scale $M$, and $\lambda$, $M_1$, and $M_2$ are massive couplings assumed to be $O(M)$. Perturbative corrections to this expression include $\ln(\mu^2/M^2)$ factors, and so to allow the validity of tree-level BSM matching just described $\mu=M$ is chosen. RG evolution is simplest in minimal subtraction schemes such as $\bar{\text{MS}}$ and, as a result, we formally prescribe that these corrections should be calculated in the $\bar{\text{MS}}$ scheme. With these renormalization choices and BSM naturalness assumptions, the AFW1 Hamiltonian can be expressed as
\begin{equation}
  \mathcal{H}_{eff}^{n\bar{n}} = \frac{1}{16 M^5}\left[ Q_4^{\bar{\text{MS}}}(M) + \frac{3}{5}\tilde{Q}_1^{\bar{\text{MS}}}(M) \right].
\end{equation}
The operator renormalization results of this work then allow the AFW1 Hamiltonian to be expressed as
\begin{equation}
  \mathcal{H}_{eff}^{n\bar{n}} = \frac{1}{16M^5}\left[ U_4(M,p_0)Q_4^{\text{RI}}(p_0) + \frac{3}{5}\tilde{U}_1(M,p_0)\tilde{Q}_1^{\text{RI}}(p_0) \right],
  \label{modelH}
\end{equation}
where $U_I(\mu,p_0)$ is the RG evolution and renormalization-scheme-matching factor appearing in Eq.~\eqref{master}. Explicit evaluation of $U_I(\mu,p_0)$ for arbitrary $\mu$, $p_0$ requires an accurate parameterization of $\alpha_s(\mu)$. For this we take the four-loop parametrization of $\alpha_s(\mu)$ in terms of $\Lambda_{N_f}^{\bar{\text{MS}}}$ and known $\beta$-function coefficients presented in Ref.~\cite{Bethke:2009jm}. The full RG evolution between $\mu$ and $p_0$ is included through a product of factors $U_I^{N_f}(\mu_1,\mu_2)$ where $N_f$ is varied across each quark mass threshold. Implicit $N_f$ dependence in the parametrization of $\alpha_s(\mu)$ in terms of the fit parameters $\Lambda_{N_f}^{\bar{\text{MS}}}$ must be included along with explicit $N_f$ dependence in $\beta_0$, $\beta_1$, and $\gamma_I^{(1)}$.

Preliminary lattice QCD neutron-antineutron simulations have been performed~\cite{Buchoff:2012bm} on anisotropic Wilson lattices with $390\  \text{MeV}$ pions. Updated values from these anisotropic lattices are given by\footnote{Lattice calculations with a chiral fermion discretization (domain-wall fermions) at near-physical pion masses ($140 \ \text{MeV}$) have been performed~\cite{Syritsyn:2015,Buchoff:INT2015} and a paper presenting these results is in progress~\cite{Buchoff:Progress}. There is currently no plan to publish the updated anisotropic Wilson lattice results due to the high level of computational complexity of the non-perturbative renormalization on these lattices.  }
\begin{equation}
  \mbraket{\bar{n}}{Q_4^{\text{RI}}(p_0)}{n} = (0.00 \pm 2.06)\times 10^{-5} \text{ GeV}^6, \hspace{20pt} \mbraket{\bar{n}}{\tilde{Q}_1^{\text{RI}}(p_0)}{n} = (-56.13 \pm 2.42)\times 10^{-5}\text{ GeV}^6,
  \label{latticeres}
\end{equation}
where the errors shown are purely statistical and fitting errors. It is important to note that these preliminary results include \textit{several significant sources of systematic uncertainty} that have not been quantified at this time.  These sources of systematics are the absence of RI-MOM non-perturbative renormalization,\footnote{Renormalization approximated by tadpole improved tree-level renormalization \cite{Lepage:1992xa,Patel:1992vu}.} unphysically large quark masses (pion masses of roughly 400 MeV), lattice spacing artifacts (which unphysically break chiral symmetry), and finite spatial extent artifacts.  All of these systematic uncertainties can be quantified and reduced with increased computing.\footnote{The required computational resources are expected to be similar to those used for lattice calculations of $B_K$ with physical pions and chiral fermion discretization \cite{Blum:2014tka}.}  For this reason, these results should only be viewed as \textit{an illustrative example} of how to combine the perturbative QCD operator renormalization results of this paper, BSM calculations of Wilson coefficients, and non-perturbatively renormalized lattice QCD matrix elements to arrive at a physical quantity that can be measured/bounded by experiment.

The experimental limit on $\delta m$ determined from Super K measurements of $\tau_{O^{16}}$ is \cite{Abe:2011ky,Phillips:2014fgb}
\begin{equation}
  |\delta m| < 2\times 10^{-33}\text{ GeV}.
  \label{expbound}
\end{equation}
Ref.~\cite{Arnold:2012sd} uses this limit and an estimate for the QCD matrix elements to relate this to a limit on the AFW1 model scale,
\begin{equation}
  M \gtrsim 500 \text{ TeV}.
\end{equation}
Using the operator renormalization factors of Table~\ref{tab:summary} and Eq.~\eqref{latticeres}, it is possible to express matrix elements of Eq.~\eqref{modelH} in terms of $M$ and known parameters. Substituting $M = 500\text{ TeV}$ into Eq.~\eqref{modelH},
\begin{equation}\begin{split}
  |\delta m| &= (\ \underbrace{6.74}_\text{LO}\quad\underbrace{-2.44}_\text{NLO}\quad\underbrace{+0.33}_\text{NNLO matching}\quad\underbrace{-0.94}_\text{NNLO running}\quad \underbrace{\pm 0.16}_\text{Lattice Statistical})\times 10^{-34} \text{ GeV} \\
  &= (3.68 \pm 0.16)\times 10^{-34} \text{ GeV},
\end{split}\end{equation}
gives rise to a $n\bar{n}$ vacuum transition time for the AFW1 model more than five times longer than the $\tau_{n\bar{n}}$ predicted by the QCD matrix element estimate used in Ref.~\cite{Arnold:2012sd}. Note that NLO one-loop running provides a multiplicative correction to the LO matrix element and is shown as an additive correction above only to illustrate the size of the correction for this example. Corrections beyond NLO can be organized as a perturbative power series in $\alpha_s(p_0)$. This perturbative series for $|\delta m|$ appears to be converging nicely, with $O(\alpha_s(p_0))$ NNLO corrections changing the NLO result by $7\%$. Assuming further corrections have the same rate of convergence, we expect unknown $O(\alpha_s(p_0)^2)$ N$^3$LO corrections to modify the NNLO result by $\sim 2\%$.

A constraint on $M$ can be derived by inverting the experimental bound Eq.~\eqref{expbound},
\begin{equation}\begin{split}
  M &> (\ \underbrace{402}_\text{LO}\quad \underbrace{-34}_\text{NLO}\quad \underbrace{+5}_\text{NNLO matching}\quad \underbrace{-16}_\text{NNLO running}\quad \underbrace{\pm 3}_\text{Lattice Statistical})\text{ TeV},\\
  M &> 357\pm 3 \text{ TeV}.
\end{split}\end{equation}
This constraint is nearly one-third weaker than the constraint estimated in Ref.~\cite{Arnold:2012sd}.

\section{Conclusion} \label{sec:concl}

To determine which BSM theories are able to produce the observed baryon asymmetry of our universe, it is essential that each make reliable predictions for $CP$ violating and baryon-number violating processes that can be probed experimentally. Theories with $\Delta B =2$ interactions can provide viable baryogenesis mechanisms while avoiding stringent experimental bounds on $\Delta B = 1$ proton decay rates. Some of these theories predict new physics in the 100-1000 TeV range that induce $n\bar{n}$ transitions that are just outside the reach of current experimental bounds. If these theories can be reliably constrained by experimental measurements of the $n\bar{n}$ vacuum transition time $\tau_{n\bar{n}}$, next-generation $n\bar{n}$ experiments can search for new physics appearing at scales comparable to, or higher than, scales probed in next-generation collider experiments.

Reliable predictions for $\delta m = 1/\tau_{n\bar{n}}$, the parameter governing the neutron-antineutron vacuum transition probability $P_{n\bar{n}} = \sin^2(|\delta m|t)$, can be made by perturbatively matching BSM theories to an effective field theory containing Standard Model operators. The six-quark operator matrix elements contributing to $\delta m$ can be calculated with lattice QCD at computationally accessible lattice matching scales of 2 GeV, and large logarithmic strong interaction corrections can be included using perturbative operator renormalization. At NLO, operator renormalization introduces known multiplicative corrections to six-quark operator matrix elements~\cite{Caswell:1982qs}. Further operator renormalization corrections are organized as a perturbative series in which the largest contributions arise from NNLO two-loop-running and one-loop-matching effects. These effects are calculated for the first time here and summarized in Table~\ref{tab:summary}. In addition, operator projectors needed for non-perturbative renormalization of $n\bar{n}$ operators and the chiral transformation properties of $n\bar{n}$ operators are presented.

Sec.~\ref{sec:pheno} discusses the calculation of $\delta m$ in a simplified model from Ref.~\cite{Arnold:2012sd} in order to illustrate how perturbative operator renormalization results are combined with lattice QCD six-quark operator matrix elements and experimental bounds on $\delta m$ to constrain the scale of new baryon-number violating physics.  For fixed BSM parameters, the $n\bar{n}$ vacuum transition time calculated with the perturbative operator renormalization results of this work and preliminary lattice QCD results is found to be more than a factor of five longer than was previously estimated. Several features of this simple example have generic implications for more complicated BSM theories and deserve explicit mention:
\begin{itemize}
	\item $\mathcal{H}_{eff}^{n\bar{n}}$ is described by a linear combination of multiple chiral basis operators that make very different contributions to $n\bar{n}$ matrix elements. Color and flavor structure matters when calculating the $n\bar{n}$ vacuum transition time predicted by a particular model.
	\item $Q_4$ has a large positive anomalous dimension and its contributions to $\delta m$ are suppressed compared to other operators.
	\item $Q_2$ is the only chiral basis operator with a negative anomalous dimension and its contributions are enhanced compared to those of other operators. It does not contribute in the simplified model considered.
	\item $\tilde{Q}_1$ arises in real models. $Q_1$ and $\tilde{Q}_1$ should be treated on equal footing, and $Q_1^{\bar{\text{MS}}}(\mu) \neq \tilde{Q}_1^{\bar{\text{MS}}}(\mu)$ must be remembered during BSM matching calculations. Identical considerations apply to $Q_3$ and $\tilde{Q}_3$.
  \item $Q_5$, $Q_6$, and $Q_7$ violate electroweak gauge invariance. They do not appear in the simplified model considered.
  \item Assuming $\mathcal{H}_{eff}^{n\bar{n}}$ can be expressed as linear combinations of $Q_1,\dots,Q_7,\;\tilde{Q}_1,\;\tilde{Q}_3$ and their parity conjugates without the use of (spin) Fierz relations, evanescent operators do not need to be explicitly included in tree-level BSM matching calculations. 
  \item At a fixed BSM scale of 500 TeV, NNLO effects correct the NLO+LO $\delta m$ prediction by $14\%$, within the generic range of $<26\%$. Assuming that further perturbative corrections have the same rate of convergence, unknown N$^3$LO corrections are estimated to change the NNLO+NLO+LO $\delta m$ prediction by $2\%$, within the generic estimate of $<7\%$.
  \item Perturbative corrections to BSM scale constraints are smaller than corrections to $\delta m$. In the simplified model considered, NNLO effects change the NLO+LO BSM scale constraint by $3\%$.
\end{itemize}

Without knowledge of NNLO two-loop-running and one-loop-matching factors, perturbative operator renormalization effects contribute large unquantified systematic uncertainties to BSM $n\bar{n}$ vacuum transition time predictions. Including NNLO effects and estimating the size of unknown N$^3$LO corrections turns perturbative operator renormalization into a few-percent-level uncertainty. This places perturbative QCD corrections to $\tau_{n\bar{n}}$ firmly under control. Electromagnetic one-loop-running corrections have also been calculated in Ref.~\cite{Caswell:1982qs}, though a complete electroweak one-loop-running calculation has not be performed. 

A complete lattice QCD determination of the $n\bar{n}$ matrix elements with controlled systematic uncertainties is necessary to remove the largest remaining Standard Model uncertainties present in BSM predictions of $\tau_{n\bar{n}}$~\cite{Buchoff:2012bm,Buchoff:Progress}. In particular, RI-MOM scheme non-perturbative renormalization factors should be calculated, continuum and infinite volume extrapolations should be performed, and $n\bar{n}$ matrix element calculations should be repeated with physical or near-physical pion masses. All of these systematic uncertainties can be removed using existing lattice QCD technology and computational resources available in the near future.

\begin{acknowledgments}
  We would like to thank Brian Tiburzi and Sergey Syritsyn for collaboration and input at the early stages of this work. We are especially thankful for Brian's notation and chiral operator constructions that made much of this work tractable. We are also very grateful to Sergey Syritsyn for detailed discussions of the consistency check that RI-MOM anomalous dimensions are identical for Fierz-conjugate operators and of vertex function momentum assignments. We would like to thank Chris Schroeder, Sergey Syritsyn, and Joe Wasem for collaboration on lattice calculations and for allowing the use of unpublished results from a preliminary calculation. We would like to thank Martin Savage for helpful advice throughout this work and Steve Sharpe for many useful discussions. We would also like to thank William Detmold and Zohreh Davoudi for helpful comments on a draft of this manuscript.  This work has been supported by the U. S. Department of Energy under Grant No. DE-FG02-00ER41132. Feynman diagrams were created with Jaxodraw~\cite{Binosi:2003yf}.
\end{acknowledgments}

\appendix
\section{Tensor Algebra}\label{app:algebra}

This appendix presents Fierz-type relations useful for resolving complicated spin, color, and flavor tensors in a desired tensor basis. All relations are derived using a well-known tensor reduction strategy: write the tensor $t^{ab}$ under consideration as a linear combination of chosen basis tensors $B_1^{ab}$, $B_2^{ab}$, \ldots with unknown coefficients $c_1$, $c_2$, \ldots, e.g.
\begin{equation}
  t^{ab} = c_1B_1^{ab} + c_2B_2^{ab}.
\end{equation}
It is often useful to chose basis tensors with definite index exchange symmetries. Contracting both sides of this equation with each basis tensor gives a system of equations
\begin{equation}
  \begin{pmatrix} B_1^{ab}t^{ab} \\ B_2^{ab}t^{ab} \end{pmatrix} = \begin{pmatrix} B_1^{ab}B_1^{ab} & B_1^{ab}B_2^{ab} \\ B_2^{ab}B_1^{ab} & B_2^{ab}B_2^{ab} \end{pmatrix} \begin{pmatrix} c_1 \\ c_2 \end{pmatrix},
  \label{ten_red}
\end{equation}
that can be readily solved for $c_1$, $c_2$. 

\subsection{Color Algebra}

There are five independent $\mathfrak{su}(3)_C$ tensors that can combine six quarks into color singlet operators,
\begin{equation}
  T^{SSS}_{\{ij\}\{kl\}\{mn\}},\hspace{20pt} T^{AAS}_{[ij][kl]\{mn\}},\hspace{20pt} T^{ASA}_{[ij]\{kl\}[mn]},\hspace{20pt} T^{SAA}_{\{ij\}[kl][mn]},\hspace{20pt} T^{AAA}_{[ij][kl][mn]}.
\end{equation}

These five basis tensors are constructed from
\begin{equation}
  \begin{split}
	T^{SSS}_{\{ij\}\{kl\}\{mn\}} &= \varepsilon_{ikm}\varepsilon_{jln} + \varepsilon_{jkm}\varepsilon_{iln} + \varepsilon_{ilm}\varepsilon_{jkn} + \varepsilon_{ikn}\varepsilon_{jlm},\\
	T^{AAS}_{[ij][kl]\{mn\}} &= \varepsilon_{ijm}\varepsilon_{kln} + \varepsilon_{ijn}\varepsilon_{klm},\\
	T^{AAA}_{[ij][kl][mn]} &= \varepsilon_{ijm}\varepsilon_{kln} - \varepsilon_{ijn}\varepsilon_{klm},
  \end{split}
\end{equation}
where $\varepsilon_{ijk}$ is the completely antisymmetric Levi-Civita tensor for $\mathfrak{su}(3)$. Each tensor is symmetrized $\{\;\}$ or antisymmetrized $[\;]$ in the three index pairs shown. This corresponds to combining the six $\mathbf{3}$ quarks as products of symmetrized $\mathbf{6}$ or antisymmetrized $\mathbf{\bar{3}}$ diquarks. These tensors also obey the diquark exchange symmetries
\begin{equation}
  \begin{split}
	T^{SSS}_{\{ij\}\{kl\}\{mn\}} &=  T^{SSS}_{\{kl\}\{ij\}\{mn\}} = T^{SSS}_{\{mn\}\{kl\}\{ij\}}\\
	T^{AAS}_{[ij][kl]\{mn\}} &=  T^{AAS}_{[kl][ij]\{mn\}}\\
	T^{AAA}_{[ij][kl][mn]} &= -T^{AAA}_{[kl][ij][mn]} = -T^{AAA}_{[ij][mn][kl]}.
  \end{split}
\end{equation}
The remaining basis tensors are defined by
\begin{equation}
  T^{ASA}_{[ij]\{kl\}[mn]} = T^{AAS}_{[ij][mn]\{kl\}},\hspace{20pt} T^{SAA}_{\{ij\}[kl][mn]} = T^{AAS}_{[mn][kl]\{ij\}}.
\end{equation}

When evaluating Feynman diagrams, one encounters contractions of the color tensors $T^{AAS}$ and $T^{SSS}$ present in $Q_I$ with the $\mathbf{su}(3)$ generators $t^A$. The resulting color tensors can always be expressed in terms of index permutations of the original color tensors. For most diagrams this is accomplished through the textbook identity
\begin{equation}
	t^A_{i^\prime i}t^A_{j^\prime j} = \frac{1}{2}\left( \delta_{i^\prime j}\delta_{j^\prime i} - \frac{1}{3}\delta_{i^\prime i}\delta_{j^\prime j}\right),
\end{equation}
where we assume the normalization $\tr(t^A t^B) = \frac{1}{2}\delta^{AB}$. Certain classes of diagrams involving three-gluon interactions require the additional identity
\begin{equation}
	f^{ABC} t_{i^\prime i}^A t_{j^\prime j}^B t_{k^\prime k}^C =  \frac{i}{4}\left(\delta_{i^\prime k}\delta_{j^\prime i}\delta_{k^\prime j} - \delta_{i^\prime j}\delta_{j^\prime k}\delta_{k^\prime i}\right),
\end{equation}
which we have derived by performing the tensor reduction of Eq. \eqref{ten_red} for a basis of Kronecker-delta products. The color structure produced by any diagram can therefore be determined from the relations of the generators above and color Fierz identities relating index-permuted tensors to the five basis tensors.

The symmetrized color tensor obeys the Fierz identity
\begin{equation}
  T^{SSS}_{\{kj\}\{il\}\{mn\}} =  -\frac{1}{2}T^{SSS}_{\{ij\}\{kl\}\{mn\}} - \frac{3}{2}T^{AAS}_{[ij][kl]\{mn\}}.
\end{equation}
The corresponding relations for interchange of any other index pair follow from the symmetries above. The mixed symmetry color tensor obeys the Fierz identities
\begin{equation}
  \begin{split}
  T^{AAS}_{[kj][il]\{mn\}} &= - \frac{1}{2}T^{SSS}_{\{ij\}\{kl\}\{mn\}} + \frac{1}{2}T^{AAS}_{[ij][kl]\{mn\}} \\\nonumber
  T^{AAS}_{[ij][ml]\{kn\}} &= -\frac{1}{2}T^{AAS}_{[ij][kl]\{mn\}} - \frac{1}{2}T^{ASA}_{[ij][mn]\{kl\}} + T^{AAA}_{[ij][kl][mn]}.
\end{split}
\end{equation}
All other index exchange relations follow by symmetry. The antisymmetrized color tensor obeys the Fierz identity
\begin{equation}
  T^{AAA}_{[kj][il][mn]} = \frac{1}{2}T^{ASA}_{[ij][mn]\{kl\}} - \frac{1}{2}T^{SAA}_{[mn][kl]\{ij\}},
\end{equation}
with all other relations again following by symmetry.

Color factors produced by any diagram can be expressed in this basis through repeated application of these Fierz identities or alternatively by direct tensor reduction of color factors in the forms of Table~\ref{color_factors}. We find the second approach more convenient at the two-loop level because it can be readily automated in a computer algebra program such as Mathematica.

\subsection{Dirac Algebra}\label{Dirac}

QCD loop diagrams introduce additional factors of $\gamma^\mu \gamma^\nu\cdots$ into each diquark. In a theory with massless quarks perturbative corrections will not modify the chirality of each diquark. We can therefore express any Dirac structure produced by a loop diagram as $CP_{\chi_1}\Gamma_1\otimes CP_{\chi_2}\Gamma_2\otimes CP_{\chi_3}\Gamma_3$, where the chirality labels are identical to those of the tree-level operator in question.  In $D=4$, a suitable basis of chirality preserving $\Gamma_1\otimes \Gamma_2\otimes \Gamma_3$ independent of quark momenta is given by
\begin{equation}
	1\otimes 1\otimes 1,\hspace{20pt} \sigma_{\mu\nu}\otimes\sigma_{\mu\nu}\otimes 1,\hspace{20pt} 1\otimes\sigma_{\mu\nu}\otimes\sigma_{\mu\nu}, \hspace{20pt} \sigma_{\mu\nu}\otimes 1\otimes\sigma_{\mu\nu},\hspace{20pt} i\sigma_{\mu\nu}\otimes\sigma_{\mu\rho}\otimes\sigma_{\nu\rho},
\end{equation}
where $\sigma_{\mu\nu} = \frac{i}{2}[\gamma_\mu, \gamma_\nu]$ and $1$ represents the $4\times 4$ identity matrix. An additional independent structure $\sigma_{\mu\nu}\sigma_{\rho\tau}\otimes \sigma_{\mu\rho}\otimes \sigma_{\nu\tau}$ is not produced in the diagrams considered here. When discussing these basis tensors we will often omit the Lorentz indices and write shorthand expressions like $\sigma\otimes \sigma\otimes 1$ and $\sigma\otimes \sigma\otimes \sigma$.

These basis structures provide a convenient orthogonal basis for tensor decompositions of two-loop Dirac structures. Operators built from these basis structures are not explicitly included in our physical operator basis. Using spin Fierz relations, each basis tensor can be related to a combination of index permutations of $1\otimes 1\otimes 1$ and therefore to the physical basis structures. Different techniques are required to find basis decompositions for Dirac structures produced in loop diagrams that are valid in $D$ dimensions. Useful discussions on the $D$-dimensional Dirac algebra needed for three- and four-quark operator renormalization can be found in Refs.~\cite{Buras:1998raa,Pivovarov:1991nk,Kraenkl:2011qb,Gracey:2012gx}, and in particular many results and techniques for $D$-dimensional tensor reduction can be found in Refs.~\cite{Dugan:1990df,Vasiliev:1995qj}.

As discussed at length in Sec.~\ref{sec:matchingev}, spin Fierz relations are broken in dimensional regularization since the 16 Dirac matrices $\{1,\gamma_5,\gamma_\mu,\sigma_{\mu\nu}\}$ (with $\mu<\nu$) are not a complete basis for Dirac matrices in general $D$. Spin Fierz relations should instead be considered prescriptions for defining evanescent operators built from the difference between the LHS and RHS of these identities, see Refs.~\cite{Buras:1998raa,Buras:1989xd}. One has the freedom to add $O(\varepsilon)$ terms in defining this prescription, for example
\begin{equation}
  [CP_\chi \sigma_{\mu\nu}]^{\alpha\beta} [CP_{\chi^\prime} \sigma_{\mu\nu}]^{\gamma\delta} \rightarrow \delta_{\chi\chi^\prime}\left( (8+a_1\varepsilon)[CP_\chi]^{\alpha\delta}[CP_\chi]^{\gamma\beta} - (4+a_2\varepsilon)[CP_\chi]^{\alpha\beta}[CP_\chi]^{\gamma\delta} \right).
  \label{2sigmaev}
\end{equation}
The $O(\varepsilon^0)$ coefficients can be calculated by performing a tensor reduction in $D=4$. Our basis of evanescent operators is explicitly defined in Appendix~\ref{app:evops}. When applying prescriptions such as Eq.~\eqref{2sigmaev} to define evanescent operators, finite matching factors and Wilson coefficients will depend on the chosen $a_1,a_2$. Basis dependence cancels so that physical quantities such as $\mathcal{H}_{eff}^{n\bar{n}}$ are independent of $a_1,a_2$. Relations between renormalized matrix elements calculated with different one-loop evanescent bases follow from general considerations of renormalization scheme dependence, see Refs.~\cite{Buras:1992tc,Herrlich:1994kh}. Alternative schemes for defining evanescent operators can be found in Refs.~\cite{Dugan:1990df,Pivovarov:1991nk,Kraenkl:2011qb}.

Additional Dirac structures appear in two-loop diagrams that are independent in general $D$. These must be treated with analogous spin Fierz evanescent operator prescriptions, such as
\begin{equation}
  CP_{\chi}\sigma_{\rho\tau}\sigma_{\mu\nu}\otimes CP_{\chi^\prime}\sigma_{\mu\nu}\sigma_{\rho\tau} \rightarrow \delta_{\chi\chi^\prime}\left( (48 + b_1\varepsilon) CP_{\chi}\otimes CP_{\chi} + (8 + b_2\varepsilon )CP_{\chi}\sigma_{\mu\nu}\otimes CP_{\chi} \sigma_{\mu\nu}\right),
	\label{4sigmaev}
\end{equation}
where $b_1$ and $b_2$ are arbitrary parameters used to specify a basis for two-loop evanescent counterterms. Freedom to specify $b_1,b_2$ and other two-loop spin Fierz prescriptions suggests there is an additional ambiguity in $\gamma^{(1)}$ besides the choice of $a_1,a_2$ that determines the one-loop evanescent counterterms. This suggestion is false.\footnote{It is for this reason that we do not consider a tensor reduction technique such as the ``Greek projections'' used in Ref.~\cite{Buras:1998raa} that commutes with algebraic relations valid in $D$-dimensions. The Greek projections provide algebraicly consistent continuations of spin-Fierz relations between Dirac structures of the form $\Gamma\otimes \Gamma^\prime$ to $D$-dimensions and for example specify $b_1=-80$, $b_2=-12$ in Eq.~\eqref{4sigmaev}. Straightforward generalizations of the Greek projections can relate structures of the form $\Gamma_1\otimes\Gamma_2\otimes\Gamma_3$. However, there is no straightforward extension of the Greek projections for Eq.~\eqref{2sigmaev} unless $\sigma\otimes\sigma$ operators are included in the physical basis as in Ref.~\cite{Buras:2000if}. Since the $n\bar{n}$ basis of interest for many BSM models includes the scalar diquark operators considered here, Greek projections do not give a useful way to define our one-loop evanescent counterterm basis. After choosing this one-loop evanescent basis, $\gamma^{(1)}$ is fully determined and we have no further need to establish concrete evanescent basis conventions.} Since one-loop-matching factors are independent of the $b$'s, there is no way for $\gamma^{(1)}$ to depend on the $b$'s while keeping $\mathcal{H}^{n\bar{n}}_{eff}$ independent of this arbitrary basis choice. Independence of $\gamma_I^{(1)}$ on the $b$'s is proven for four-quark operators in Ref.~\cite{Herrlich:1994kh}. The proof only relies on cancellation of non-local divergences and the factor of $1/2$ multiplying evanescent counterterm diagrams in Eq.~\eqref{evCTprescrip} and applies to our six-quark operators without modification. We have explicitly verified that cancellation of $b$'s dependence occurs diagram-by-diagram between two-loop diagrams and one-loop evanescent counterterm diagrams in our calculation.

In addition to Eq.~\eqref{4sigmaev}, two-loop diagram evaluation requires the $D=4$ spin Fierz identities
\begin{equation}
  \begin{split}
  CP_{\chi_1} \sigma_{\rho\tau}\sigma_{\mu\nu}\otimes CP_{\chi_2} \sigma_{\rho\tau}\otimes CP_{\chi_3} \sigma_{\mu\nu} \stackrel{D=4}{=}& \Delta_\chi \left(4 CP_\chi \otimes CP_\chi\sigma_{\mu\nu}\otimes CP_\chi\sigma_{\mu\nu}\right.\\
  &\hspace{30pt} \left.- 4iCP_\chi\sigma_{\mu\nu}\otimes CP_\chi\sigma_{\mu\rho}\otimes CP_\chi\sigma_{\nu\rho}\right),
  \label{4sigmaev2}
\end{split}
\end{equation}
where
\begin{equation}
  \Delta_\chi \equiv \delta_{\chi_1\chi_2}\delta_{\chi_2\chi_3}
\end{equation}
vanishes unless all three diquarks have identical chirality. Relating the above Dirac basis tensors to permutations of $1\otimes 1\otimes 1$ also requires
\begin{equation}
  \begin{split}
  i[CP_{\chi_1}\sigma_{\mu\nu}]^{\alpha\beta}[CP_{\chi_2} \sigma_{\mu\rho}]^{\gamma\delta}[CP_{\chi_3} \sigma_{\nu\rho}]^{\eta\zeta} \stackrel{D=4}{=}& \Delta_\chi \left( 16 [CP_\chi]^{\alpha\zeta} [CP_\chi]^{\gamma\beta} [CP_\chi]^{\eta\delta} \right.\\
  &\hspace{20pt} - 8[CP_\chi]^{\alpha\delta} [CP_\chi]^{\gamma\beta} [CP_\chi]^{\eta\zeta} - 8 [CP_\chi]^{\alpha\beta} [CP_\chi]^{\gamma\zeta} [CP_\chi]^{\eta\delta} \\
  &\hspace{20pt} \left. - 8[CP_\chi]^{\alpha\zeta}[CP_{\chi}]^{\gamma\delta} [CP_\chi]^{\eta\beta}+ 8 [CP_\chi]^{\alpha\beta} [CP_\chi]^{\gamma\delta} [CP_\chi]^{\eta\zeta} \right).
\end{split}
\label{sigma3}
\end{equation}
Other useful formulae are derived by combining Eq.~\eqref{2sigmaev}~-~\eqref{sigma3}. A particularly useful identity is
\begin{equation}
  \delta_{\chi_1\chi_2}1^{\alpha\delta}\sigma^{\gamma\beta}\sigma^{\eta\zeta} \stackrel{D=4}{=} \Delta_\chi \left[ \frac{1}{2}1\otimes\sigma\otimes\sigma + \frac{1}{2}\sigma\otimes 1\otimes\sigma - \frac{1}{2}\sigma\otimes\sigma\otimes\sigma \right]^{\alpha\beta\gamma\delta\eta\zeta}.
  \label{nasty}
\end{equation}

Fierz relations involving $\fs{p}$ are also useful when computing evanescent counterterm diagrams
\begin{subequations}
  \label{pfierz}
  \begin{align}
	\delta_{\chi\chi^\prime}\frac{1}{p^2}(CP_\chi \gamma_\mu \fs{p})\otimes (CP_\chi \fs{p}\gamma_\mu) \stackrel{D=4}{=}& \delta_{\chi\chi^\prime}(CP_\chi)\otimes (CP_\chi) + \frac{1}{4}(CP_\chi\sigma_{\mu\nu})\otimes (CP_\chi \sigma_{\mu\nu}),\\
	\frac{1}{p^2}(CP_\chi \sigma_{\mu\nu} \gamma_\lambda \fs{p})\otimes (CP_{\chi^\prime} \sigma_{\mu\nu}\fs{p}\gamma_\lambda) \stackrel{D=4}{=}& 12\delta_{\chi\chi^\prime}(CP_\chi)\otimes(CP_{\chi^\prime}) - (CP_\chi \sigma_{\mu\nu})\otimes(CP_{\chi^\prime} \sigma_{\mu\nu}),\\
	\frac{1}{p^2}(CP_\chi \sigma_{\mu\nu} \gamma_\lambda \fs{p})\otimes (CP_{\chi^\prime} \fs{p}\gamma_\lambda\sigma_{\mu\nu}) \stackrel{D=4}{=}& 12\delta_{\chi\chi^\prime}(CP_\chi)\otimes(CP_{\chi^\prime}) + 3(CP_\chi \sigma_{\mu\nu})\otimes(CP_{\chi^\prime} \sigma_{\mu\nu}).
  \end{align}
\end{subequations}
Additional Fierz relations are useful for diagram classes 34-45, in particular
\begin{subequations}
	\label{pfierz2}
	\begin{align}
		P_{\chi}^{\alpha\zeta}[P_{\chi^\prime} \gamma_\mu\fs{p}]^{\gamma\delta}[P_{\chi} \fs{p}\gamma_\mu]^{\eta\beta} \stackrel{D=4}{=}& \frac{1}{2}[P_{\chi} \gamma_\mu\fs{p}]^{\alpha\beta}[P_{\chi^\prime}\fs{p}\gamma_\mu]^{\gamma\delta}P_\chi^{\eta\zeta} + \frac{1}{8}[P_\chi \sigma_{\mu\nu}\gamma_\rho \fs{p}]^{\alpha\beta}[P_{\chi^\prime} \fs{p}\gamma_\rho]^{\gamma\delta}[P_\chi \sigma_{\mu\nu}]^{\eta\zeta},\\
		[\gamma_\mu\fs{p} P_{\chi}]^{\alpha\zeta}[P_{\chi^\prime} \fs{p}\gamma_\mu]^{\gamma\delta}P_{\chi}^{\eta\beta} \stackrel{D=4}{=}& \frac{1}{2}[P_{\chi} \gamma_\mu\fs{p}]^{\alpha\beta}[P_{\chi^\prime}\fs{p}\gamma_\mu]^{\gamma\delta}P_\chi^{\eta\zeta} + \frac{1}{8}[P_\chi \gamma_\rho \fs{p}\sigma_{\mu\nu}]^{\alpha\beta}[P_{\chi^\prime} \fs{p}\gamma_\rho]^{\gamma\delta}[P_\chi \sigma_{\mu\nu}]^{\eta\zeta}.
		\end{align}
	\end{subequations}
Additional identities are found by permutation of the tensor product structures $\Gamma_1\otimes\Gamma_2\otimes\Gamma_3$ appearing on both sides of the above equations. For example, applying the permutation $\Gamma_1\otimes\Gamma_2\otimes\Gamma_3 \rightarrow \Gamma_2\otimes\Gamma_1\otimes\Gamma_3$ to the LHS of either equation leads to a new identity with the RHS modified by $1\otimes\sigma\otimes\sigma \leftrightarrow \sigma\otimes 1\otimes\sigma$, $1\otimes 1\otimes 1$ and $\sigma\otimes\sigma\otimes 1$ left unchanged, and $\sigma\otimes\sigma\otimes\sigma \rightarrow - \sigma\otimes\sigma\otimes \sigma$. Identities involving general permutations of $\Gamma_1\otimes\Gamma_2\otimes\Gamma_3$ are constructed analogously, and in particular $\sigma\otimes\sigma\otimes \sigma$ will change sign under any permutation of $\Gamma_1\otimes\Gamma_2\otimes\Gamma_3$ with odd signature. All other Dirac structures produced by two-loop diagrams can be related to those above and our basis structures by algebra valid in general $D$.

\subsection{Flavor Algebra}\label{sec:flavor}

A convenient basis for $\mathfrak{su}(2)_\chi$ tensors is given by
\begin{equation}
	\tau^2,\; \tau^2\tau^A,\; \tau^2\tau^A\tau^B,\dots
\end{equation}
where the $\tau^A$ are normalized as Pauli matrices $\tr(\tau^A\tau^B) = 2$.

After applying the $\sigma\otimes \sigma$ spin Fierz identity of Eq.~\eqref{2sigmaev}, the resulting spin-singlet diquarks no longer have their flavor indices contracted with one of the basis structures above. Flavor (as well as color) Fierz relations are useful in relating the resulting structures to the original operator basis. One-loop diagrams involving flavor singlet diquarks require
\begin{equation}
  \tau^2_{ad}\tau^2_{cb} = \frac{1}{2}\tau^2_{ab}\tau^2_{cd} + \frac{1}{2}(\tau^2\tau^A)_{ab}(\tau^2\tau^A)_{cd}.
\end{equation}
Similarly, flavor vector-singlet structures require
\begin{equation}
	\tau^2_{ad}(\tau^2\tau^A)_{cb} = \frac{1}{2}\tau^2_{ab}(\tau^2\tau^A)_{cd} + \frac{1}{2}(\tau^2\tau^B)_{ab}(\tau^2\tau^A \tau^B)_{cd}.
\end{equation}
Finally, flavor vector-vector structures require
\begin{equation}
  \begin{split}
	(\tau^2 \tau^A)_{ad}(\tau^2\tau^B)_{cb} =& \frac{1}{2}\left\lbrace(\tau^2 \tau^A)_{ab}(\tau^2\tau^B)_{cd} + (\tau^2 \tau^B)_{ab}(\tau^2\tau^A)_{cd} + i\varepsilon^{ABC}\left[ (\tau^2\tau^C)_{ab}\tau^2_{cd} - \tau^2_{ab}(\tau^2\tau^C)_{cd} \right]\right.\\
	&\hspace{10pt}+ \left. \delta^{AB}\left[\tau^2_{ab}\tau^2_{cd} - (\tau^2 \tau^C)_{ab}(\tau^2\tau^C)_{cd}\right]\right\rbrace.
  \end{split}
  \label{flavorvecvec}
\end{equation}
Eq.~\eqref{flavorvecvec} implies in particular that symmetric traceless tensors are Fierz self-conjugate.



\section{Two-Loop Integrals}\label{app:integrals}

When evaluating simple Feynman diagrams, one can often perform Dirac ``numerator algebra'' that reduces the diagram to a simple product of a Dirac structure times a scalar integral. When evaluating diagrams with gluon propagators connecting quarks in separate spin-singlet diquarks, this is not possible. One is forced to work with tensor integrals that contain free Lorentz indices contracted with structures such as $\sigma_{\mu\nu}\otimes\sigma_{\rho\tau}$. In this case, tensor reduction techniques similar to those described in Appendix \ref{app:algebra} can be used to express tensor integrals in terms of linear combinations of scalar integrals. In our calculation of the diagrams of Fig.~\ref{diagrams}, the complete set of two-loop tensor integrals appearing in these diagrams was organized according to the propagator powers and loop-momentum vectors appearing. Each tensor integral was then expressed as a linear combination of basic tensors and two-loop scalar integrals by tensor reduction techniques. The two-loop scalar integrals were recursively evaluated as described below and the results tabled for use in tensor integral evaluation. Computer algebra was essential for this process and performed using Mathematica scripts written by the authors.

There exists a vast literature on evaluation of multi-loop tensor and scalar integrals. References to reviews and original literature are given below, and it should be emphasized that none of the techniques reviewed in this appendix are novel. Our aim is simply to consolidate known techniques needed for two-loop anomalous dimension calculations without detailing the additional complications and generalizations needed for more complex higher-order calculations. 

\subsection{Two-Loop Scalar Integrals}

We are only concerned here with calculating the $1/\bar{\varepsilon}$ single- and double-pole pieces of two-loop diagrams. This allows for substantial simplifications. In particular, external momenta can be freely chosen diagram by diagram. To see this, note that in a renormalizable theory, these pole pieces can be at most polynomial in external momenta. After factoring out possible overall dimensionful factors common to all diagrams, this means the pole pieces are independent of external momenta. This holds for individual diagrams as long as they contain no subdivergences, and therefore for general two-loop diagrams as long as one-loop counterterm diagrams cancelling all subdivergences are included~\cite{Collins:105730}. We may therefore freely choose a different momentum routing convenient for each two-loop diagram under consideration as long as the same routing is used in all corresponding one-loop counterterm diagrams.

The only caveat to this statement is that the choice of external momentum routing must not introduce IR divergences. For instance, if in a massless theory one sets all external momenta to zero then all integrals vanish identically in dimensional regularization. This means IR divergences have been introduced that are regulated as $1/\bar{\varepsilon}$ poles and cancel all of the original UV divergences~\cite{Vladimirov:1979zm}. We are interested in the counterterms needed to cancel UV divergences only, and so we must use care to choose momentum routings free of IR divergences. See Ref.~\cite{Steinhauser:2002rq} for a detailed review of this ``infrared rearrangement'' trick; for our purposes it is enough to note that IR divergences can be found and avoided through standard power counting arguments used to determine a diagram's degree of UV divergence~\cite{Collins:105730}.

For all diagrams in Fig.~\ref{diagrams}, a momentum routing can be chosen so that the only scalar integrals appearing are of the form
\begin{equation}
  T(n_1,n_2,n_3,n_4,n_5) = \mu^{4\varepsilon}\int\frac{d^Dkd^Dq }{(2\pi)^{2D}} \frac{1}{(p+k)^{2n_1}(p+q)^{2n_2}k^{2n_3}q^{2n_4}(k-q)^{2n_5}},
  \label{2loopscalar}
\end{equation}
where $p$ is an arbitrary external momenta that serves as an IR regulator and we are suppressing omnipresent $i\epsilon$ terms in factors such as $(k^2+i\epsilon)^{n_3}$. If one of the propagator factors does not appear (one $n_i$ equals zero), then the two loop integral can be expressed as a product of one loop integrals. The second loop includes non-integer propagator powers, but can still be evaluated through the textbook formula
\begin{equation}
  \begin{split}
  I(\alpha,\beta) &= \mu^{2\varepsilon}\int \frac{d^Dk}{(2\pi)^D} \frac{1}{(p+k)^{2\alpha}k^{2\beta} } \\
  &= \frac{(p^2)^{2-\alpha-\beta-\varepsilon}}{(4\pi)^{2-\varepsilon}} \left[\frac{\Gamma(\alpha+\beta-2+\varepsilon)\Gamma(2-\alpha-\varepsilon)\Gamma(2-\beta-\varepsilon)}{\Gamma(4-\alpha-\beta-2\varepsilon)\Gamma(\alpha)\Gamma(\beta)}\right].
\end{split}
\end{equation}
Scalar two-loop integrals with (at least) one zero argument are given by
\begin{equation}
  \begin{split}
  T(n_1,n_2,n_3,n_4,0) &= I(n_3,n_1)I(n_4,n_2),\\
  T(0,n_2,n_3,n_4,n_5) &= (p^2)^{n_3+n_5-2+\varepsilon} I(n_3,n_5)I(n_3+n_4+n_5-2+\varepsilon,n_2),\\
  T(n_1,n_2,0,n_4,n_5) &= (p^2)^{n_1+n_5-2+\varepsilon} I(n_1,n_5)I(n_4,n_1+n_2+n_5-2+\varepsilon).
\end{split}
\label{intbase}
\end{equation}
The cases of $n_2=0$ and $n_4=0$ can be found by the $(n_1,n_3)\leftrightarrow(n_2,n_4)$ symmetry of $T(n_1,n_2,n_3,n_4,n_5)$.

This leaves the case of non-vanishing $n_1,\ldots,n_5$. This case can be evaluated recursively through the ``integration by parts'' technique of Refs.~\cite{Chetyrkin:1981qh,Tkachov:1984xk}, see Ref.~\cite{Grozin:2003ak} for a review. The starting point for this technique is the observation that there are no surface terms when integrating a total derivative in dimensional regularization~\cite{Collins:105730}, that is
\begin{equation}
  0 = \mu^{4\varepsilon} \int \frac{d^Dkd^Dq}{(2\pi)^{2D}} \left(\frac{\partial}{\partial q^\mu} a^\mu(k,q,p)\right)
\label{intbyparts}
\end{equation}
where $a^\mu(k,q,p)$ is an arbitrary vector that may depend on loop and external momenta. Useful identities are generated by taking $a_\mu$ to be a loop momentum vector times the integrand of Eq.~\eqref{2loopscalar}. Consider in particular
\begin{equation}
  \begin{split}
  &\frac{\partial}{\partial q^\mu}\left[\frac{(k-q)^\mu}{(p+k)^{2n_1}(p+q)^{2n_2}k^{2n_3}q^{2n_4}(k-q)^{2n_5}}  \right]\\
  &\hspace{20pt}= \left[ -D -\frac{2n_2(k-q)\cdot (p+q)}{(p+q)^2} - \frac{2n_4(k-q)\cdot q}{q^2} + 2n_5 \right]\frac{1}{(p+k)^{2n_1}(p+q)^{2n_2}k^{2n_3}q^{2n_4}(k-q)^{2n_5}}.
\end{split}
\label{intbypartsderiv}
\end{equation}
Next, re-write all scalar products appearing in Eq.~\eqref{intbypartsderiv} in terms of linear combinations of $p^2$ and denominator factors, for instance
\begin{equation}
  \begin{split}
  2(k-q)\cdot (p+q) &= (k+p)^2 - (k-q)^2 - (p+q)^2,\\
  2(k-q)\cdot q &= k^2 - (k-q)^2 - q^2.
\end{split}
\end{equation}
This allows us to express Eq.~\eqref{intbyparts} as
\begin{equation}
  0 = \left[ 2n_5 + n_2 + n_4 - D + n_2\mathbf{2}^+(\mathbf{5}^- - \mathbf{1}^-) + n_4\mathbf{4}^+(\mathbf{5}^- - \mathbf{3}^-) \right]T(n_1,n_2,n_3,n_4,n_5),
\end{equation}
where we define
\begin{equation}
  \mathbf{1}^\pm T(n_1,n_2,n_3,n_4,n_5) = T(n_1\pm 1,n_2,n_3,n_4,n_5),
\end{equation}
etc. This identity is sufficient to derive a recursive solution for $T(n_1,n_2,n_3,n_4,n_5)$ with all $n_i$ non-zero,
\begin{equation}
  T(n_1,n_2,n_3,n_4,n_5) = \frac{1}{D - n_2 - n_4 - 2n_5}\left[ n_2\mathbf{2}^+(\mathbf{5}^- - \mathbf{1}^-) + n_4\mathbf{4}^+(\mathbf{5}^- - \mathbf{3}^-) \right]T(n_1,n_2,n_3,n_4,n_5).
  \label{intrecursion}
\end{equation}
This recursion terminates when each integral on the RHS has at least one $n_i$ zero and the base case Eq.~\eqref{intbase} can be applied. Many other integration by parts identities and more powerful recursive algorithms can be constructed but are not needed for the calculation at hand. For further discussions of more general one-loop scalar integrals see Refs.~\cite{Hooft:1978xw,Davydychev:1992xr}. For further discussions of two-loop scalar integral evaluation see Refs.~\cite{Usyukina:1992jd,Davydychev:1992mt} and the review Ref.~\cite{Grozin:2003ak}.

\subsection{Two-Loop Tensor Integrals}

Two-loop tensor integrals can be expressed in terms of scalar integrals through tensor reduction techniques. Consider for example the rank two integral
\begin{equation}
  T^2_{\mu\nu}(n_1,n_2,n_3,n_4,n_5) = \mu^{4\varepsilon}\int \frac{d^Dkd^D q }{(2\pi)^{2D}}\frac{k_\mu k_\nu} {(p+k)^{2n_1}(p+q)^{2n_2}k^{2n_3}q^{2n_4}(k-q)^{2n_5}}.
\end{equation}
By Lorentz invariance, the integral can be expressed as a linear combination
\begin{equation}
  T^2_{\mu\nu} = T^2_\delta g_{\mu\nu} + T^2_\alpha \frac{1}{p^2} p_\mu p_\nu.
\end{equation}
Contracting both sides with these same tensors gives the system of equations
\begin{equation}
  \begin{pmatrix} T^2_\delta \\ T^2_\alpha \end{pmatrix} = \begin{pmatrix} 4-2\varepsilon & 1 \\ 1 & 1 \end{pmatrix}^{-1} \begin{pmatrix} g^{\mu\nu}T^2_{\mu\nu} \\ \frac{1}{p^2}p^\mu p^\nu T^2_{\mu\nu}\end{pmatrix}.
\end{equation}
The contractions of the RHS can be reduced to linear combinations of scalar integrals by re-writing tensor products in terms of differences of propagator factors as before,
\begin{equation}
  \begin{split}
    g^{\mu\nu}T^2_{\mu\nu}(n_1,n_2,n_3,n_4,n_5) &= \mathbf{3}^-T(n_1,n_2,n_3,n_4,n_5)\\
    \frac{1}{p^2}p^\mu p^\nu T^2_{\mu\nu}(n_1,n_2,n_3,n_4,n_5) &= \frac{1}{2p^2}p^\mu\left[ \mathbf{1}^- - \mathbf{3}^- - p^2\right]T^1_\mu(n_1,n_2,n_3,n_4,n_5).
  \end{split}
\end{equation}
This final formula does not apply to the cases of $n_1=0$ and $n_3=0$. These must be treated separately, and a general method can be constructed by first performing a tensor reduction of a one-loop subintegral. This problem is systematically considered in Ref.~\cite{Weiglein:1993hd}. The following recipe is sufficient for the integrals considered in this work: first evaluate the one-loop integral for the loop momentum that only appears in two propagators by a one-loop tensor reduction. A change of variables may be useful to ensure there is only one ``external momentum'' scale (which may be a linear combination of $p_\mu$ and the other loop momentum) that needs to be included in the one-loop tensor reduction. The second integral will then be another one-loop tensor integral involving a single scale that can be readily evaluated. For further discussion of tensor integral reduction techniques, see Ref.\cite{Passarino:1978jh,Davydychev:1991va} and for a review see Ref.~\cite{Denner:1991kt}.

\section{Evanescent Operators}\label{app:evops}

The $\bar{\text{MS}}$ and RI-MOM renormalization schemes are fully defined by the renormalization conditions of Sec.~\ref{sec:renormschemes} and specification of the one-loop evanescent counterterms appearing in Sec.~\ref{sec:matchingev}. Two-loop $\bar{\text{MS}}$ anomalous dimensions, one-loop RI-MOM matching factors, and Wilson coefficients from one-loop BSM matching all separately depend on the basis chosen for evanescent operator counterterms. In particular, loop-level BSM matching calculations must use the same evanescent basis used in this work. Our basis includes the following evanescent operators needed as one-loop counterterms to $Q_I$,
\begin{subequations}\label{evbasis}
\begin{align}
	E_1^a &= (\psi CP_R \sigma_{\mu\nu} i\tau_2\psi)(\psi CP_R \sigma_{\mu\nu} i\tau_2\psi)(\psi CP_R i\tau_2\tau_+ \psi)T^{SSS} - 12Q_1,\\\nonumber
	E_1^b &= (\psi CP_R  i\tau_2\psi)(\psi CP_R \sigma_{\mu\nu} i\tau_2\psi)(\psi CP_R\sigma_{\mu\nu} i\tau_2\tau_+ \psi)T^{ASA}\\\nonumber
	&\hspace{20pt}+ (\psi CP_R\sigma_{\mu\nu}  i\tau_2\psi)(\psi CP_R  i\tau_2\psi)(\psi CP_R\sigma_{\mu\nu} i\tau_2\tau_+ \psi)T^{SAA} - 8Q_1,\\\nonumber\\
	E_2^a &= (\psi CP_L \sigma_{\mu\nu} i\tau_2\psi)(\psi CP_R \sigma_{\mu\nu} i\tau_2\psi)(\psi CP_R i\tau_2\tau_+ \psi)T^{SSS},\\\nonumber
	E_2^b &= (\psi CP_L  i\tau_2\psi)(\psi CP_R \sigma_{\mu\nu} i\tau_2\psi)(\psi CP_R\sigma_{\mu\nu} i\tau_2\tau_+ \psi)T^{ASA}\\\nonumber
	&\hspace{20pt}+ (\psi CP_L\sigma_{\mu\nu}  i\tau_2\psi)(\psi CP_R  i\tau_2\psi)(\psi CP_R\sigma_{\mu\nu} i\tau_2\tau_+ \psi)T^{SAA} - 4Q_2,\\\nonumber\\
	E_3^a &= (\psi CP_L \sigma_{\mu\nu} i\tau_2\psi)(\psi CP_L \sigma_{\mu\nu} i\tau_2\psi)(\psi CP_R i\tau_2\tau_+ \psi)T^{SSS} - 12Q_1,\\\nonumber
	E_3^b &= (\psi CP_L  i\tau_2\psi)(\psi CP_L \sigma_{\mu\nu} i\tau_2\psi)(\psi CP_R\sigma_{\mu\nu} i\tau_2\tau_+ \psi)T^{ASA}\\\nonumber
	&\hspace{20pt} + (\psi CP_L\sigma_{\mu\nu}  i\tau_2\psi)(\psi CP_L  i\tau_2\psi)(\psi CP_R\sigma_{\mu\nu} i\tau_2\tau_+ \psi)T^{SAA},\\\nonumber\\
	E_4 &= (\psi CP_R \sigma_{\mu\nu} i\tau_2\tau_{3}\psi) (\psi CP_R \sigma_{\mu\nu} i\tau_2\tau_3\psi)(\psi CP_R i\tau_2\tau_{+}\psi)T^{AAS}  \\\nonumber
	&\hspace{20pt}  - \frac{1}{5}(\psi CP_R \sigma_{\mu\nu} i\tau_2\tau_{A}\psi)(\psi CP_R \sigma_{\mu\nu} i\tau_2\tau_A\psi)(\psi CP_R i\tau_2\tau_{+}\psi)T^{AAS}  \\\nonumber
	&\hspace{20pt}+ (\psi CP_R  i\tau_2\tau_{3}\psi)(\psi CP_R \sigma_{\mu\nu} i\tau_2\tau_3\psi)(\psi CP_R\sigma_{\mu\nu} i\tau_2\tau_{+}\psi)T^{SAA} \\\nonumber
	&\hspace{20pt} - \frac{1}{5} (\psi CP_R  i\tau_2\tau_{A}\psi)(\psi CP_R \sigma_{\mu\nu} i\tau_2\tau_A\psi)(\psi CP_R\sigma_{\mu\nu} i\tau_2\tau_{+}\psi)T^{SAA} \\\nonumber
	&\hspace{20pt}+ (\psi CP_R\sigma_{\mu\nu}  i\tau_2\tau_{3}\psi)(\psi CP_R  i\tau_2\tau_3\psi) (\psi CP_R\sigma_{\mu\nu} i\tau_2\tau_{+}\psi)T^{ASA}\\\nonumber
	&\hspace{20pt}- \frac{1}{5}(\psi CP_R\sigma_{\mu\nu}  i\tau_2\tau_{3}\psi)(\psi CP_R  i\tau_2\tau_3\psi)(\psi CP_R\sigma_{\mu\nu} i\tau_2\tau_{+}\psi)T^{ASA} - 12Q_4,\\\nonumber\\
	E_5 &= (\psi CP_R \sigma_{\mu\nu} i\tau_2\tau_{-}\psi)(\psi CP_L \sigma_{\mu\nu} i\tau_2\tau_+ \psi)(\psi CP_L i\tau_2\tau_{+}\psi)T^{AAS}  \\\nonumber
	&\hspace{20pt}+ (\psi CP_R  i\tau_2\tau_{-}\psi)(\psi CP_L \sigma_{\mu\nu} i\tau_2\tau_+\psi)(\psi CP_L\sigma_{\mu\nu} i\tau_2\tau_{+}\psi)T^{SAA}\\\nonumber
	&\hspace{20pt}+ (\psi CP_R\sigma_{\mu\nu}  i\tau_2\tau_{-}\psi)(\psi CP_L  i\tau_2\tau_+\psi)(\psi CP_L\sigma_{\mu\nu} i\tau_2\tau_{+}\psi)T^{ASA} - 4Q_5,\\\nonumber\\
	\tilde{E}_1 &= \frac{1}{3}(\psi CP_R \sigma_{\mu\nu} i\tau_2\tau_{A}\psi)(\psi CP_R \sigma_{\mu\nu} i\tau_2\tau_A\psi)(\psi CP_R i\tau_2\tau_{+}\psi)T^{AAS}  \\\nonumber
	&\hspace{20pt}+ \frac{1}{3}(\psi CP_R  i\tau_2\tau_{A}\psi)(\psi CP_R \sigma_{\mu\nu} i\tau_2\tau_A\psi)(\psi CP_R\sigma_{\mu\nu} i\tau_2\tau_{+}\psi)T^{SAA}\\\nonumber
	&\hspace{20pt}+ \frac{1}{3}(\psi CP_R\sigma_{\mu\nu}  i\tau_2\tau_{A}\psi)(\psi CP_R  i\tau_2\tau_A\psi)(\psi CP_R\sigma_{\mu\nu} i\tau_2\tau_{+}\psi)T^{ASA} + \frac{4}{3}\tilde{Q}_1,\\\nonumber\\
	\tilde{E}_3 &= \frac{1}{3}(\psi CP_L \sigma_{\mu\nu} i\tau_2\tau_{A}\psi)(\psi CP_L \sigma_{\mu\nu} i\tau_2\tau_A\psi)(\psi CP_R i\tau_2\tau_{+}\psi)T^{AAS}  \\\nonumber
	&\hspace{20pt}+ \frac{1}{3}(\psi CP_L  i\tau_2\tau_{A}\psi)(\psi CP_L \sigma_{\mu\nu} i\tau_2\tau_A\psi)(\psi CP_R\sigma_{\mu\nu} i\tau_2\tau_{+}\psi)T^{SAA}\\\nonumber
	&\hspace{20pt}+ \frac{1}{3}(\psi CP_L\sigma_{\mu\nu}  i\tau_2\tau_{A}\psi)(\psi CP_L  i\tau_2\tau_A\psi)(\psi CP_R\sigma_{\mu\nu} i\tau_2\tau_{+}\psi)T^{ASA} + 4\tilde{Q}_3.\\\nonumber
\end{align}
\end{subequations}
Above we have grouped evanescent operators that make similar contributions to one-loop counterterm diagrams, see Tables \ref{results_A}, \ref{results_S}. 

The coefficients of $Q_I$ appearing above are determined by demanding that the RHS vanish in $D=4$. We have calculated them using two independent methods for verification: first by pen and paper application of the spin-color-flavor Fierz relations derived in Appendix~\ref{app:algebra} and second by automated Mathematica application of the operator projectors of Eq.~\ref{projectordef} to explicit vertex functions constructed for each structure. It is straightforward to verify that all other structures produced in amplitudes for $d$ = 1-3 vanish by quark exchange antisymmetry.

When constructing one-loop counterterm diagrams for two-loop diagrams $d$ = 4-6, 16-24, and 32-45 containing $d$ = 1-3 subdiagrams, it is useful to employ a different evanescent operator basis $E_I^\prime$. The $E_I^\prime$ basis is defined by demanding that the prescription 
\begin{equation}
  [CP_\chi \sigma_{\mu\nu}]^{\alpha\beta} [CP_{\chi^\prime} \sigma_{\mu\nu}]^{\gamma\delta} \stackrel{D=4}{=} \delta_{\chi\chi^\prime}\left( 8[CP_\chi]^{\alpha\delta}[CP_\chi]^{\gamma\beta} - 4[CP_\chi]^{\alpha\beta}[CP_\chi]^{\gamma\delta} \right),
\end{equation}
always provides valid operator identities in general $D$ when $E_I^\prime$ operators are included. Applying this prescription to the amplitudes for $d$ = 1-3 provides an explicit construction of the $E_I^\prime$ operators
\begin{subequations}\label{evbasisprime}
\begin{align}
	E_1^{a\prime} &= (\psi CP_R \sigma_{\mu\nu} i\tau_2\psi)(\psi CP_R \sigma_{\mu\nu} i\tau_2\psi)(\psi CP_R i\tau_2\tau_+ \psi)T^{SSS}\\\nonumber
	&\hspace{10pt} -  8(\psi^\alpha [CP_R]^{\alpha\delta} i\tau_2\psi^\beta)(\psi^\gamma [CP_R]^{\gamma\beta} i\tau_2\psi^\delta)(\psi^\eta [CP_R]^{\eta\zeta} i\tau_2\tau_+ \psi^\zeta)T^{SSS}\\\nonumber
	&= E_1^a + 6(Q_1 - \tilde{Q}_1)\\
	E_3^{a\prime} &= (\psi CP_L \sigma_{\mu\nu} i\tau_2\psi)(\psi CP_L \sigma_{\mu\nu} i\tau_2\psi)(\psi CP_R i\tau_2\tau_+ \psi)T^{SSS}\\\nonumber
	&\hspace{10pt} -  8(\psi^\alpha [CP_L]^{\alpha\delta} i\tau_2\psi^\beta)(\psi^\gamma [CP_L]^{\gamma\beta} i\tau_2\psi^\delta)(\psi^\eta [CP_R]^{\eta\zeta} i\tau_2\tau_+ \psi^\zeta)T^{SSS}\\\nonumber
	&= E_3^a + 6(Q_3 - \tilde{Q}_3) \\
	\tilde{E}_1^\prime &= \frac{1}{3}(\psi CP_R \sigma_{\mu\nu} i\tau_2\tau_{A} \psi)(\psi CP_R \sigma_{\mu\nu} i\tau_2\tau_A \psi)(\psi CP_R i\tau_2\tau_{+} \psi)T^{AAS}\\\nonumber
	&\hspace{10pt}+ \frac{1}{3}(\psi CP_R i\tau_2\tau_{A} \psi)(\psi CP_R \sigma_{\mu\nu} i\tau_2\tau_A \psi)(\psi CP_R\sigma_{\mu\nu}  i\tau_2\tau_{+} \psi)T^{SAA}\\\nonumber
	&\hspace{10pt}+ \frac{1}{3}(\psi CP_R\sigma_{\mu\nu}  i\tau_2\tau_{A} \psi)(\psi CP_R i\tau_2\tau_A \psi)(\psi CP_R\sigma_{\mu\nu}  i\tau_2\tau_{+} \psi)T^{ASA}\\\nonumber
	&\hspace{10pt} -  \frac{8}{3}(\psi^\alpha [CP_R]^{\alpha\delta} i\tau_2\tau_{A} \psi^\beta)(\psi^\gamma [CP_R]^{\gamma\beta} i\tau_2\tau_A \psi^\delta)(\psi^\eta [CP_R]^{\eta\zeta} i\tau_2\tau_{+} \psi^\zeta)T^{AAS}\\\nonumber
	&\hspace{10pt} -  \frac{8}{3}(\psi^\alpha [CP_R]^{\alpha\beta} i\tau_2\tau_{A} \psi^\beta)(\psi^\gamma [CP_R]^{\gamma\zeta} i\tau_2\tau_A \psi^\delta)(\psi^\eta [CP_R]^{\eta\delta} i\tau_2\tau_{+} \psi^\zeta)T^{SAA}\\\nonumber
	&\hspace{10pt} -  \frac{8}{3}(\psi^\alpha [CP_R]^{\alpha\zeta} i\tau_2\tau_{A} \psi^\beta)(\psi^\gamma [CP_R]^{\gamma\delta} i\tau_2\tau_A \psi^\delta)(\psi^\eta [CP_R]^{\eta\beta} i\tau_2\tau_{+} \psi^\zeta)T^{ASA}\\\nonumber
	&= \tilde{E}_1 + \frac{10}{3}(Q_1 - \tilde{Q}_1)\\
	\tilde{E}_3^\prime &= \frac{1}{3}(\psi CP_L \sigma_{\mu\nu} i\tau_2\tau_A \psi)(\psi CP_L \sigma_{\mu\nu} i\tau_2\tau_A \psi)(\psi CP_R i\tau_2\tau_+ \psi)T^{AAS}\\\nonumber
	&\hspace{10pt} -  \frac{8}{3}(\psi^\alpha [CP_L]^{\alpha\delta} i\tau_2\tau_A \psi^\beta)(\psi^\gamma [CP_L]^{\gamma\beta} i\tau_2\tau_A \psi^\delta)(\psi^\eta [CP_R]^{\eta\zeta} i\tau_2\tau_+ \psi^\zeta)T^{AAS}\\\nonumber
	&= \tilde{E}_3 + 2(Q_3 - \tilde{Q}_3)
\end{align}
\end{subequations}
with all other $E_I^\prime$ equal to the corresponding $E_I$. The $E_I^\prime$ basis is convenient for two-loop diagram evaluation, but is cumbersome for RG evolution because it includes one-loop mixing between $Q_1$ and $\tilde{Q}_1$ and between $Q_3$ and $\tilde{Q}_3$.

After evaluating two-loop diagrams in the $E_I^\prime$ basis to determine the loop coefficients $[L_{tot}^\prime]_{II}^{(2),1}$ defined in Eq.~\eqref{Ltotprime}, a change of basis to the $E_I$ basis can be performed to recover the coefficients $[L_{tot}]_{II}^{(2),1}$ directly appearing in the anomalous dimension formula Eq.~\eqref{gamma1fin}. Reading off the coefficients $L_{IE_I}^{(1)}$ from the entries for $d$ = 1-3 in Tables~\ref{results_A}-\ref{results_S}, the necessary change of basis formulas to recover $[L_{tot}]_{II}^{(2),1}$ from $[L_{tot}^\prime]_{II}^{(2),1}$ are found to be
\begin{equation}
	\begin{split}
		[L_{tot}]_{11}^{(2),1} &= [L_{tot}^\prime]_{11}^{(2),1} - \frac{3}{2}(r_1^{(0)} - \tilde{r}_1^{(0)})\\
		[L_{tot}]_{33}^{(2),1} &= [L_{tot}^\prime]_{33}^{(2),1} - \frac{3}{2}(r_3^{(0)} - \tilde{r}_3^{(0)})\\
		[\tilde{L}_{tot}]_{11}^{(2),1} &= [\tilde{L}_{tot}^\prime]_{11}^{(2),1} - \frac{5}{2}(r_1^{(0)} - \tilde{r}_1^{(0)})\\
		[\tilde{L}_{tot}]_{33}^{(2),1} &= [\tilde{L}_{tot}^\prime]_{33}^{(2),1} - \frac{3}{2}(r_3^{(0)} - \tilde{r}_3^{(0)}).
	\end{split}\label{EprimeE}
\end{equation}
One-loop matching results give $r_1^{(0)} - \tilde{r}_1^{(0)} = 7 = r_3^{(0)} - \tilde{r}_3^{(0)}$.


\section{Diagram Results}\label{app:tables}

After choosing a convenient momentum routing, each two-loop diagram in Fig.~\ref{diagrams} and associated counterterm diagrams are evaluated in terms of a tensor integral contracted with a Dirac tensor, as discussed in Appendix~\ref{app:integrals}. The amplitude for each diagram is represented by a complicated combination of color factors and Dirac structures that can be simplified with the tensor reduction techniques of Appendix~\ref{app:algebra}. We find it convenient to first perform a Dirac tensor reduction using the prescription of Eq.~\eqref{2sigmafierzprescrip} with $a_1=a_2=0$ (that is, working in the $E_I^\prime$ basis) that allows us to express the diagrams in a given class as individual color factors times a common combination of Dirac basis structures. These Dirac structures are shown in Table~\ref{diag_results}. We then perform color tensor reductions of the color factors of Table~\ref{color_factors}. This allows each diagram amplitude to be expressed as a combined spin-color tensor. While there are 25 distinct spin-color tensors that can be built from the basis tensors of Appendix~\ref{app:algebra}, most vanish by quark exchange antisymmetry when contracted with external quark fields and flavor tensors to form operator corrections.

When contracted with $(\psi i\tau_2 \psi)(\psi i\tau_2 \psi)(\psi i\tau_2 \tau_A \psi)$, the only spin-color tensors that give non-vanishing contributions are
\begin{equation}
  (1\otimes 1\otimes 1)T^{AAS},\hspace{20pt} (\sigma\otimes\sigma\otimes 1)T^{SSS},\hspace{20pt} (1\otimes\sigma\otimes\sigma)T^{ASA},\hspace{20pt} (\sigma\otimes 1\otimes\sigma)T^{AAS}.
\end{equation}
The overall contribution from each diagram class to $\mathcal{M}^A_d$ is represented by a linear combination of these four spin-color tensors in Table~\ref{results_A}. Analogously, when contracted with $(\psi i\tau_2\tau_A \psi)(\psi i\tau_2 \tau_B \psi)(\psi i\tau_2 \tau_C \psi)$, the only spin-color tensors that give non-vanishing contributions are
\begin{equation}
  (1\otimes 1\otimes 1)T^{SSS},\hspace{20pt} (\sigma\otimes\sigma\otimes 1)T^{AAS},\hspace{20pt} (1\otimes\sigma\otimes\sigma)T^{SAA},\hspace{20pt} (\sigma\otimes 1\otimes \sigma)T^{ASA},\hspace{20pt} (\sigma\otimes\sigma\otimes\sigma)T^{AAA}.
\end{equation}
The overall contribution from each diagram class to $\mathcal{M}^S_d$ is represented by a linear combination of these five spin-color tensors in Table~\ref{results_S}.

After adding the amplitudes $\mathcal{M}^A_d$ and $\mathcal{M}^S_d$ from each diagram class, we consider each $Q_I$ independently form operator corrections by Eq.~\eqref{rosetta}. The resulting operator corrections can be expressed as a simple multiple of $Q_I$ in $D=4$ through application of either the operator projectors of Eq.~\eqref{projectors} or the relations of Appendix~\ref{app:evops} and
\begin{equation}
  \begin{split}
	  & \frac{i}{3}(\psi CP_R i\tau_2\tau_{A}\sigma_{\mu\nu} \psi)(\psi CP_R\sigma_{\mu\rho} i\tau_2\tau_A \psi)(\psi CP_R\sigma_{\nu\rho} i\tau_2\tau_{+} \psi)T^{AAA} \\
  & \stackrel{D=4}{=} -\frac{8}{3}(\psi CP_R i\tau_2\tau_{A} \psi)(\psi CP_R i\tau_2\tau_A \psi)(\psi CP_R i\tau_2\tau_{+} \psi)T^{SSS},
\end{split}
\end{equation}
and
\begin{equation}
  \begin{split}
	& i(\psi CP_R i\tau_2\tau_{\{3}\sigma_{\mu\nu} \psi)(\psi CP_R\sigma_{\mu\rho} i\tau_2\tau_3 \psi)(\psi CP_R\sigma_{\nu\rho} i\tau_2\tau_{+\}} \psi)T^{AAA} \\
  & \hspace{40pt} - \frac{1}{5}i(\psi CP_R i\tau_2\tau_{\{A}\sigma_{\mu\nu} \psi)(\psi CP_R\sigma_{\mu\rho} i\tau_2\tau_A \psi)(\psi CP_R\sigma_{\nu\rho} i\tau_2\tau_{+\}} \psi)T^{AAA} \\
  & \stackrel{D=4}{=} 4\left[ \vphantom{\frac{1}{5}} (\psi CP_R i\tau_2\tau_{\{3}\sigma_{\mu\nu} \psi)(\psi CP_R i\tau_2\tau_3 \psi)(\psi CP_R i\tau_2\tau_{+\}} \psi)T^{SSS}\right. \\
  & \hspace{40pt} \left. - \frac{1}{5}(\psi CP_R i\tau_2\tau_{\{A} \psi)(\psi CP_R i\tau_2\tau_A \psi)(\psi CP_R i\tau_2\tau_{+\}} \psi)T^{SSS}\right] .
\end{split}
\end{equation}
The result is an operator correction proportional to $Q_I$ plus irrelevant evanescent contributions. The $1/\bar{\varepsilon}$ pole coefficient of this amplitude represents $L_{II}^{(2),2}/\bar{\varepsilon}^2 + [L_{tot}^\prime]_{II}^{(2),1}/\bar{\varepsilon}$. After including the change of evanescent basis factors in Eq.~\eqref{EprimeE}, $\gamma_I^{(1)}$ is immediately given by Eq.~\eqref{gamma1fin}.

\renewcommand{\arraystretch}{1.15}

\begin{table}
\begin{tabular}{|c|c||c|c||c|c||c|c||c|c||c|c|}
    \hline $d$ & $N_d$ & \multicolumn{2}{c||}{$1\otimes 1 \otimes 1$} & \multicolumn{2}{c||}{$\sigma\otimes\sigma\otimes 1$} & \multicolumn{2}{c||}{$1\otimes\sigma\otimes\sigma$} & \multicolumn{2}{c||}{$\sigma\otimes 1\otimes \sigma$} & \multicolumn{2}{c|}{$\sigma\otimes\sigma\otimes \sigma$}\\\hline
    & & $1/\bar{\varepsilon}^2$ & $1/\bar{\varepsilon}$ & $1/\bar{\varepsilon}^2$ & $1/\bar{\varepsilon}$ & $1/\bar{\varepsilon}^2$ & $1/\bar{\varepsilon}$ & $1/\bar{\varepsilon}^2$ & $1/\bar{\varepsilon}$ & $1/\bar{\varepsilon}^2$ & $1/\bar{\varepsilon}$ \\\hline\hline 
   1 & 3 & - & $-4$ & - & $0$ & - & $0$ & - & $0$ & - & $0$ \\\hline
   2 & 6 & - & $-1$ & - & $1/4$ & - & $0$ & - & $0$ & - & $0$ \\\hline
   3 & 6 & - & $-1$ & - & $-1/4$ & - & $0$ & - & $0$ & - & $0$ \\\hline\hline
   4 & 3 & $-8$ & $8$ & $0$ & $0$ & $0$ & $0$ & $0$ & $0$ & $0$ & $0$ \\\hline
   5 & 6 & $(-1- 3\delta_\chi)/2$ & $(5 + 3\delta_\chi)/4$ & $1/2$ & $-1/2$ & $0$ & $0$ & $0$ & $0$ & $0$ & $0$\\\hline
   6 & 6 & $(-1 - 3\delta_\chi)/2$ & $(5 + 10\delta_\chi)/4$ & $-1/2$ & $15/16$ &  $0$ & $0$ & $0$ & $0$ & $0$ & $0$\\\hline
   7 & 3 & $0$ & $-2$ & $0$ & $0$ & $0$ & $0$ & $0$ & $0$ & $0$ & $0$  \\\hline
   8 & 6 & $0$ & $-2 + 3\delta_\chi$ & $0$ & $1/4$ & $0$ & $0$ & $0$ & $0$ & $0$ & $0$\\\hline
   9 & 6 & $0$ & $-2 + 3\delta_\chi$ & $0$ & $-1/4$ & $0$ & $0$ & $0$ & $0$ & $0$ & $0$\\\hline
   10 & 6 & $2$ & $2$ & $0$ & $0$ & $0$ & $0$ & $0$ & $0$ & $0$ & $0$\\\hline
   11 & 12 & $1/2$ & $0$ & $-1/8$ & $-5/16$ & $0$ & $0$ & $0$ & $0$ & $0$ & $0$ \\\hline
   12 & 12 & $1/2$ & $0$ & $1/8$ & $5/16$ & $0$ & $0$ & $0$ & $0$ & $0$ & $0$\\\hline
   13 & 6 & $-2$ & $1$ & $0$ & $0$ & $0$ & $0$ & $0$ & $0$ & $0$ & $0$\\\hline
   14 & 12 & $-1/2$ & $0$ & $1/8$ & $1/16$ & $0$ & $0$ & $0$ & $0$ & $0$ & $0$\\\hline
   15 & 12 & $-1/2$ & $0$ & $-1/8$ & $-1/16$ & $0$ & $0$ & $0$ & $0$ & $0$ & $0$\\\hline
   16 & 12 & $-2$ & $1+7\delta_\chi/4$ & $0$ & $-1/16$ & $0$ & $0$ & $0$ & $0$ & $0$ & $0$\\\hline
   17 & 12 & $-2$ & $-1$ & $1/2$ & $1/2$ & $0$ & $0$ & $0$ & $0$ & $0$ & $0$\\\hline
   18 & 12 & $-2$ & $1-7\delta_\chi/4$ & $0$ & $1/16$ & $0$ & $0$ & $0$ & $0$ & $0$ & $0$\\\hline
   19 & 12 & $-2$ & $-1$ & $-1/2$ & $-1/2$ & $0$ & $0$ & $0$ & $0$ & $0$ & $0$\\\hline
   20 & 12 & $-1/2 + 3\delta_\chi/2$ & $(3 - 3\delta_\chi)/4$ & $-1/4$ & $1/8$ & $0$ & $0$ & $0$ & $0$ & $0$ & $0$\\\hline
   21 & 12 & $-1/2+3\delta_\chi/2$ & $(3 - 10\delta_\chi)/4$ & $1/4$ & $-9/16$ & $0$ & $0$ & $0$ & $0$ & $0$ & $0$\\\hline
   22 & 3 & $-16$ & $0$ & $0$ & $0$ & $0$ & $0$ & $0$ & $0$ & $0$ & $0$\\\hline
   23 & 3 & $-1-3\delta_\chi$ & $0$ & $1$ & $0$ & $0$ & $0$ & $0$ & $0$ & $0$ & $0$ \\\hline
   24 & 3 & $-1-3\delta_\chi$ & $7\delta_\chi/2$ & $-1$ & $7/8$ & $0$ & $0$ & $0$ & $0$ & $0$ & $0$\\\hline
   25 & 6 & $-6$ & $5$ & $0$ & $0$ & $0$ & $0$ & $0$ & $0$ & $0$ & $0$\\\hline
   26 & 12 & $-3/2$ & $1/2$ & $3/8$ & $1/16$ & $0$ & $0$ & $0$ & $0$ & $0$ & $0$ \\\hline
   27 & 12 & $-3/2$ & $1/2$ & $-3/8$ & $-1/16$ & $0$ & $0$ & $0$ & $0$ & $0$ & $0$\\\hline
   28 & 12 & $0$ & $0$ & $0$ & $3/4$ & $0$ & $0$ & $0$ & $0$ & $0$ & $0$\\\hline
   29 & 3 & $15/2 - N_f$ & $ -13 + 4N_f/3 $ & $0$ & $0$ & $0$ & $0$ & $0$ & $0$ & $0$ & $0$\\\hline
   3$0$ & 6 & $0$ & $0$ & $-5/8 + N_f/12$ & $17/48 - N_f/72$ & $0$ & $0$ & $0$ & $0$ & $0$ & $0$\\\hline
   31 & 6 & $0$ & $0$ & $5/8 - N_f/12$ & $-17/48 + N_f/72$ & $0$ & $0$ & $0$ & $0$ & $0$ & $0$\\\hline
   32 & 6 & $-4$ & $0$ & $1$ & $0$ & $0$ & $0$ & $0$ & $0$ & $0$ & $0$\\\hline
   33 & 6 & $-4$ & $0$ & $-1$ & $0$ & $0$ & $0$ & $0$ & $0$ & $0$ & $0$\\\hline
   34 & 6 & $-1/2$ & $0$ & $1/8$ & $1/16$ & $-\Delta_\chi/8$ & $-1/16 + \Delta_\chi/16$ & $1/8$ & $-1/16$ & $-1/8$ & $0$ \\\hline
   35 & 6 & $-1/2$ & $0$ & $1/8$ & $1/16$  & $\Delta_\chi/8$ & $1/16 - \Delta_\chi/16$ & $-1/8$ & $1/16$ & $1/8$ & $0$ \\\hline
   36 & 6 & $-1/2$ & $0$ & $1/8$ & $1/16$ & $-\Delta_\chi/8$ & $-1/16 + \Delta_\chi/16$ & $1/8$ & $-1/16$ & $1/8$ & $0$ \\\hline
   37 & 6 & $-1/2$ & $0$ & $1/8$ & $1/16$ & $\Delta_\chi/8$ & $1/16 - \Delta_\chi/16$ & $-1/8$ & $1/16$ & $-1/8$ & $0$ \\\hline
   38 & 6 & $-1/2$ & $0$ & $-1/8$ & $-1/16$ & $-\Delta_\chi/8$ & $-1/16+\Delta_\chi/16$ & $-1/8$ & $1/16$ & $-1/8$ & $0$ \\\hline
   39 & 6 & $-1/2$ & $0$ & $-1/8$ & $-1/16$ & $\Delta_\chi/8$ & $1/16 - \Delta_\chi/16$ & $1/8$ & $-1/16$ & $1/8$ & $0$ \\\hline
   40 & 6 & $-1/2$ & $0$ & $-1/8$ & $-1/16$ & $-\Delta_\chi/8$ & $-1/16 + \Delta_\chi/16$ & $-1/8$ & $1/16$ & $1/8$ & $0$ \\\hline
   41 & 6 & $-1/2$ & $0$ & $-1/8$ & $-1/16$ & $\Delta_\chi/8$ & $1/16 - \Delta_\chi/16$ & $1/8$ & $-1/16$ & $-1/8$ & $0$ \\\hline
   42 & 6 & $-1$ & $0$ & $1/4$ & $0$ & $-\Delta_\chi/4$ & $0$ & $1/4$ & $0$ & $-1/4$ & $0$ \\\hline
   43 & 6 & $-1$ & $0$ & $1/4$ & $0$ & $\Delta_\chi/4$ & $0$ & $-1/4$ & $0$ & $1/4$ & $0$ \\\hline
   44 & 6 & $-1$ & $0$ & $-1/4$ & $0$ & $\Delta_\chi/4$ & $0$ & $1/4$ & $0$ & $1/4$ & $0$ \\\hline
   45 & 6 & $-1$ & $0$ & $-1/4$ & $0$ & $-\Delta_\chi/4$ & $0$ & $-1/4$ & $0$ & $-1/4$ & $0$ \\\hline
   46 & 8 & $0$ & $0$ & $0$ & $0$ & $0$ & $0$ & $0$ & $0$ & $0$ & $3/8$ \\\hline
\end{tabular}
\caption{NDR $1/\bar{\varepsilon}$ pole structure of the diagram amplitudes in Feynman gauge without color factors. $d$ labels the diagrams classes of Fig.~\ref{diagrams}, $N_d$ is the number of diagrams within class $d$, $\delta_{\chi}\equiv \delta_{\chi_{_1}\chi_{_2}}$ and $\Delta_\chi \equiv \delta_{\chi_{_1}\chi_{_2}}\delta_{\chi_{_2}\chi_{_3}}$. Evanescent counterterms in the $E_I^\prime$ basis defined by applying Eq.~\eqref{evCTprescrip} to diverent subdiagrams are included in these results.}
\label{diag_results}
\end{table}

\begin{table}
\begin{tabular}{|c||c|}
\hline $d$ & Color Factor \\\hline\hline
1 & $\frac{1}{2}\left[T_{(ji)(kl)\{mn\}}-\frac{1}{3}T_{(ij)(kl)\{mn\}}\right]$ \\\hline
2 &  $\frac{1}{2}\left[T_{(il)(kj)\{mn\}}-\frac{1}{3}T_{(ij)(kl)\{mn\}}\right]$ \\\hline
3 &  $\frac{1}{2}\left[T_{(ik)(jl)\{mn\}}-\frac{1}{3}T_{(ij)(kl)\{mn\}}\right]$\\\hline\hline
4 &  $\frac{1}{4}\left[-\frac{2}{3}T_{(ji)(kl)\{mn\}}+\frac{10}{9}T_{(ij)(kl)\{mn\}}\right]$ \\\hline
5 &  $\frac{1}{4}\left[-\frac{2}{3}T_{(il)(kj)\{mn\}}+\frac{10}{9}T_{(ij)(kl)\{mn\}}\right]$ \\\hline
6 &  $\frac{1}{4}\left[-\frac{2}{3}T_{(ik)(jl)\{mn\}}+\frac{10}{9}T_{(ij)(kl)\{mn\}}\right]$ \\\hline
7 &  $\frac{1}{4}\left[\frac{7}{3}T_{(ji)(kl)\{mn\}}+\frac{1}{9}T_{(ij)(kl)\{mn\}}\right]$ \\\hline
8 &  $\frac{1}{4}\left[\frac{7}{3}T_{(il)(kj)\{mn\}}+\frac{1}{9}T_{(ij)(kl)\{mn\}}\right]$ \\\hline
9 &  $\frac{1}{4}\left[\frac{7}{3}T_{(ik)(jl)\{mn\}}+\frac{1}{9}T_{(ij)(kl)\{mn\}}\right]$ \\\hline
10 &  $-\frac{1}{12}\left[T_{(ji)(kl)\{mn\}}-\frac{1}{3}T_{(ij)(kl)\{mn\}}\right]$ \\\hline
11 &  $-\frac{1}{12}\left[T_{(il)(kj)\{mn\}}-\frac{1}{3}T_{(ij)(kl)\{mn\}}\right]$ \\\hline
12 &  $-\frac{1}{12}\left[T_{(ik)(jl)\{mn\}}-\frac{1}{3}T_{(ij)(kl)\{mn\}}\right]$ \\\hline
13 &  $\frac{2}{3}\left[T_{(ji)(kl)\{mn\}}-\frac{1}{3}T_{(ij)(kl)\{mn\}}\right]$ \\\hline
14 &  $\frac{2}{3}\left[T_{(il)(kj)\{mn\}}-\frac{1}{3}T_{(ij)(kl)\{mn\}}\right]$ \\\hline
15 &  $\frac{2}{3}\left[T_{(ik)(jl)\{mn\}}-\frac{1}{3}T_{(ij)(kl)\{mn\}}\right]$ \\\hline
16 &  $\frac{1}{4}\left[T_{(jl)(ki)\{mn\}}-\frac{1}{3}T_{(ji)(kl)\{mn\}}-\frac{1}{3}T_{(il)(kj)\{mn\}}+\frac{1}{9}T_{(ij)(kl)\{mn\}}\right]$ \\\hline
17 &   $\frac{1}{4}\left[T_{(li)(kj)\{mn\}}-\frac{1}{3}T_{(ji)(kl)\{mn\}}-\frac{1}{3}T_{(il)(kj)\{mn\}}+\frac{1}{9}T_{(ij)(kl)\{mn\}}\right]$  \\\hline
18 & $\frac{1}{4}\left[T_{(jk)(il)\{mn\}}-\frac{1}{3}T_{(ji)(kl)\{mn\}}-\frac{1}{3}T_{(ik)(jl)\{mn\}}+\frac{1}{9}T_{(ij)(kl)\{mn\}}\right]$ \\\hline
19 &  $\frac{1}{4}\left[T_{(ki)(jl)\{mn\}}-\frac{1}{3}T_{(ji)(kl)\{mn\}}-\frac{1}{3}T_{(ik)(jl)\{mn\}}+\frac{1}{9}T_{(ij)(kl)\{mn\}}\right]$ \\\hline
20 &   $\frac{1}{4}\left[T_{(ik)(lj)\{mn\}}-\frac{1}{3}T_{(ik)(jl)\{mn\}}-\frac{1}{3}T_{(il)(kj)\{mn\}}+\frac{1}{9}T_{(ij)(kl)\{mn\}}\right]$ \\\hline
21 &  $\frac{1}{4}\left[T_{(il)(jk)\{mn\}}-\frac{1}{3}T_{(ik)(jl)\{mn\}}-\frac{1}{3}T_{(il)(kj)\{mn\}}+\frac{1}{9}T_{(ij)(kl)\{mn\}}\right]$ \\\hline
22 &  $\frac{1}{4}\left[T_{(ji)(lk)\{mn\}}-\frac{1}{3}T_{(ij)(lk)\{mn\}}-\frac{1}{3}T_{(ji)(kl)\{mn\}}+\frac{1}{9}T_{(ij)(kl)\{mn\}}\right]$ \\\hline
23 &  $\frac{1}{4}\left[T_{(kl)(ij)\{mn\}}-\frac{1}{3}T_{(il)(kj)\{mn\}}-\frac{1}{3}T_{(kj)(il)\{mn\}}+\frac{1}{9}T_{(ij)(kl)\{mn\}}\right]$ \\\hline
24 &  $\frac{1}{4}\left[T_{(lk)(ji)\{mn\}}-\frac{1}{3}T_{(lj)(ki)\{mn\}}-\frac{1}{3}T_{(ik)(jl)\{mn\}}+\frac{1}{9}T_{(ij)(kl)\{mn\}}\right]$  \\\hline
25 &  $-\frac{3}{4}\left[T_{(ji)(kl)\{mn\}}-\frac{1}{3}T_{(ij)(kl)\{mn\}}\right]$ \\\hline
26 &  $-\frac{3}{4}\left[T_{(il)(kj)\{mn\}}-\frac{1}{3}T_{(ij)(kl)\{mn\}}\right]$ \\\hline
27 &  $-\frac{3}{4}\left[T_{(ik)(jl)\{mn\}}-\frac{1}{3}T_{(ij)(kl)\{mn\}}\right]$\\\hline
28 &  $\frac{1}{4}\left[T_{(jl)(ki)\{mn\}} - T_{(li)(kj)\{mn\}}\right]$ \\\hline
29 & $\frac{1}{2}\left[T_{(ji)(kl)\{mn\}}-\frac{1}{3}T_{(ij)(kl)\{mn\}}\right]$ \\\hline
30 &  $\frac{1}{2}\left[T_{(il)(kj)\{mn\}}-\frac{1}{3}T_{(ij)(kl)\{mn\}}\right]$\\\hline
31 &  $\frac{1}{2}\left[T_{(ik)(jl)\{mn\}}-\frac{1}{3}T_{(ij)(kl)\{mn\}}\right]$\\\hline
32 &  $\frac{1}{4}\left[T_{(il)(kj)\{nm\}}-\frac{1}{3}T_{(il)(kj)\{mn\}}-\frac{1}{3}T_{(ij)(kl)\{nm\}}+\frac{1}{9}T_{(ij)(kl)\{mn\}}\right]$ \\\hline
33 &  $\frac{1}{4}\left[T_{(lj)(ki)\{nm\}}-\frac{1}{3}T_{(lj)(ki)\{mn\}}-\frac{1}{3}T_{(ij)(kl)\{nm\}}+\frac{1}{9}T_{(ij)(kl)\{mn\}}\right]$ \\\hline
34 &  $\frac{1}{4}\left[T_{(in)(kj)\{ml\}}-\frac{1}{3}T_{(il)(kj)\{mn\}}-\frac{1}{3}T_{(in)(kl)\{mj\}}+\frac{1}{9}T_{(ij)(kl)\{mn\}}\right]$ \\\hline
35 &  $\frac{1}{4}\left[T_{(im)(kj)\{ln\}}-\frac{1}{3}T_{(il)(kj)\{mn\}}-\frac{1}{3}T_{(im)(kl)\{jn\}}+\frac{1}{9}T_{(ij)(kl)\{mn\}}\right]$  \\\hline
36 &  $\frac{1}{4}\left[T_{(mj)(il)\{kn\}}-\frac{1}{3}T_{(kj)(il)\{mn\}}-\frac{1}{3}T_{(mj)(kl)\{in\}}+\frac{1}{9}T_{(ij)(kl)\{mn\}}\right]$ \\\hline
37 &  $\frac{1}{4}\left[T_{(nj)(il)\{mk\}}-\frac{1}{3}T_{(kj)(il)\{mn\}}-\frac{1}{3}T_{(nj)(kl)\{mi\}}+\frac{1}{9}T_{(ij)(kl)\{mn\}}\right]$ \\\hline
38 &  $\frac{1}{4}\left[T_{(im)(jl)\{kn\}}-\frac{1}{3}T_{(ik)(jl)\{mn\}}-\frac{1}{3}T_{(im)(kl)\{jn\}}+\frac{1}{9}T_{(ij)(kl)\{mn\}}\right]$ \\\hline
39 &  $\frac{1}{4}\left[T_{(in)(jl)\{mk\}}-\frac{1}{3}T_{(ik)(jl)\{mn\}}-\frac{1}{3}T_{(in)(kl)\{mj\}}+\frac{1}{9}T_{(ij)(kl)\{mn\}}\right]$ \\\hline
40 &  $\frac{1}{4}\left[T_{(nj)(ki)\{ml\}}-\frac{1}{3}T_{(lj)(ki)\{mn\}}-\frac{1}{3}T_{(nj)(kl)\{mi\}}+\frac{1}{9}T_{(ij)(kl)\{mn\}}\right]$  \\\hline
41 &  $\frac{1}{4}\left[T_{(mj)(ki)\{ln\}}-\frac{1}{3}T_{(lj)(ki)\{mn\}}-\frac{1}{3}T_{(mj)(kl)\{in\}}+\frac{1}{9}T_{(ij)(kl)\{mn\}}\right]$ \\\hline
42 &  $\frac{1}{4}\left[T_{(ml)(kj)\{in\}}-\frac{1}{3}T_{(il)(kj)\{mn\}}-\frac{1}{3}T_{(mj)(kl)\{in\}}+\frac{1}{9}T_{(ij)(kl)\{mn\}}\right]$ \\\hline
43 &  $\frac{1}{4}\left[T_{(nl)(kj)\{mi\}}-\frac{1}{3}T_{(il)(kj)\{mn\}}-\frac{1}{3}T_{(nj)(kl)\{mi\}}+\frac{1}{9}T_{(ij)(kl)\{mn\}}\right]$\\\hline
44 &  $\frac{1}{4}\left[T_{(mk)(jl)\{in\}}-\frac{1}{3}T_{(ik)(jl)\{mn\}}-\frac{1}{3}T_{(mj)(kl)\{in\}}+\frac{1}{9}T_{(ij)(kl)\{mn\}}\right]$ \\\hline
45 &  $\frac{1}{4}\left[T_{(nk)(jl)\{mi\}}-\frac{1}{3}T_{(ik)(jl)\{mn\}}-\frac{1}{3}T_{(nj)(kl)\{mi\}}+\frac{1}{9}T_{(ij)(kl)\{mn\}}\right]$\\\hline
46 &  $\frac{1}{4}\left[T_{(in)(kj)\{ml\}}-T_{(il)(kn)\{mj\}}\right]$ \\\hline
\end{tabular}
\caption{Single diagram color factors corresponding to the explicit diagrams in Fig.~\ref{diagrams}.  This table alleviates potential sign and coefficient ambiguity in Table~\ref{diag_results} due to choice of color terms factored out. $T_{(ij)(kl)\{mn\}}$ represents $T^{AAS}_{[ij][kl]\{mn\}}$ or $T^{SSS}_{\{ij\}\{kl\}\{mn\}}$ depending on which operator is inserted in the diagram.}
\label{color_factors}
\end{table}

\begin{table} 
  \begin{tabular}{|c||c|c||c|c||c|c|}
	\hline $ d $ & \multicolumn{2}{c||}{$(1\otimes 1 \otimes 1)T^{AAS}$} & \multicolumn{2}{c||}{$(\sigma\otimes\sigma\otimes 1)T^{SSS} $} & \multicolumn{2}{c|}{$(1\otimes\sigma\otimes\sigma)T^{ASA} + (\sigma\otimes 1\otimes \sigma)T^{SAA}$} \\\hline
	& $1/\bar{\varepsilon}^2$ & $1/\bar{\varepsilon}$ & $\hspace{25pt}1/\bar{\varepsilon}^2\hspace{25pt}$ & $1/\bar{\varepsilon}$ & $1/\bar{\varepsilon}^2$ & $1/\bar{\varepsilon}$ \\\hline\hline 
   1  & - & $4$ & - & 0 & - & 0 \\\hline
   2  & - & $3/2$ & - & $-1/8$ & - & $-1/8$ \\\hline
   3  & - & $3/2$ & - & $-1/8$ & - & $-1/8$ \\\hline\hline
   4  & $-8$ & $8$ & 0 & 0 & 0 & 0  \\\hline
   5  & $ (-11 - 7\delta_A^1 - 13\delta_A^2)/12$ & $(55+7\delta_A^1+13\delta_A^2)/24$ & $1/12$ & $-1/12$ & $1/12$ & $-1/12$\\\hline
   6  & $(-11-7\delta_A^1-13\delta_A^2)/12$ & $(165+70\delta_A^1+130\delta_A^2)/72$ & $1/12$ & $-5/32$ & $1/12$ & $-5/32$\\\hline
   7  & 0 & $1$ & 0 & 0 & 0 & 0 \\\hline
   8  & 0 & $(10+23\delta_A^1-19\delta_A^2)/12$ & 0 & $-7/48$ & 0 & $-7/48$ \\\hline
   9  & 0 & $(10+23\delta_A^1-19\delta_A^2)/12$ & 0 & $-7/48$ & 0 & $-7/48$ \\\hline
   10  & $2/3$ & $2/3$ & 0 & 0 & 0 & 0 \\\hline
   11  & $1/4$ & 0 & $-1/48$ & $-5/96$ & $-1/48$ & $-5/96$ \\\hline
   12  & $1/4$ & 0 & $-1/48$ & $-5/96$ & $-1/48$ & $-5/96$ \\\hline 
   13  & $16/3$ & $-8/3$ & 0 & 0 & 0 & 0 \\\hline
   14  & $2$ & 0 & $-1/6$ & $-1/12$ & $-1/6$ & $-1/12$ \\\hline
   15  & $2$ & 0 & $-1/6$ & $-1/12$ & $-1/6$ & $-1/12$ \\\hline
   16  & $-2/3$ & $(24-28\delta_A^1 + 35\delta_A^2)/72$ & 0 & $1/48$ & 0 & $-1/96$ \\\hline
   17  & $-2/3$ & $-1/3$ & $1/3$ & $1/3$ & $1/12$ & $1/12$ \\\hline
   18  & $-2/3$ & $(24+28\delta_A^1 - 35\delta_A^2)/72$ & 0 & $1/48$ & 0 & $-1/96$ \\\hline
   19  & $-2/3$ & $-1/3$ & $1/3$ & $1/3$ & $1/12$ & $1/12$ \\\hline
   20  & $(-1-13\delta_A^1 + 8\delta_A^2)/12$ & $(3+13\delta_A^1-8\delta_A^2)/24$ & $1/8$ & $-1/16$ & 0 & 0 \\\hline
   21  & $(-1-13\delta_A^1 + 8\delta_A^2)/12$ & $(9+130\delta_A^1 - 80\delta_A^2)/72$ & $1/8$ & $-9/32$ & 0 & 0 \\\hline
   22  & 0 & 0 & 0 & 0 & 0 & 0 \\\hline
   23  & $(-5-7\delta_A^1 - 4\delta_A^2)$ & $0$ & $1/12$ & $0$ & $1/3$ & $0$ \\\hline
   24  & $(-5-7\delta_A^1 - 4\delta_A^2)$ & $(49\delta_A^1 + 28\delta_A^2)/72$ & $1/12$ & $-7/96$ & $1/3$ & $-7/24$ \\\hline
   25  & $-18$ & $15$ & 0 & 0 & 0 & 0 \\\hline
   26  & $-27/4$ & $9/4$ & $9/16$ & $3/32$ & $9/16$ & $3/32$ \\\hline
   27  & $-27/4$ & $9/4$ & $9/16$ & $3/32$ & $9/16$ & $3/32$ \\\hline
   28  & 0 & 0 & 0 & $-3/4$ & 0 & 0 \\\hline
   29  & $-15/2 + N_f$ & $ 13 - 4N_f/3 $ & 0 & 0 & 0 & 0 \\\hline
   30  & 0 & 0 & $5/16 - N_f/24$ & $-17/96 + N_f/144$ & $5/16-N_f/24$ & $-17/96+N_f/144$ \\\hline
   31  & 0 & 0 & $5/16 - N_f/24$ & $-17/96 + N_f/144$ & $5/16-N_f/24$ & $-17/96 + N_f/144$\\\hline
   32  & $-14/3$ & 0 & $-1/6$ & 0 & $1/3$ & 0 \\\hline
   33  & $-14/3$ & 0 & $-1/6$ & 0 & $1/3$ & 0 \\\hline
   34  & $1/48$ & 0 & $(10-3\Delta_\chi)/192$ & $(-1+\Delta_\chi)/128$ & $(10-3\Delta_\chi)/192$ & $(-1+\Delta_\chi)/128$ \\\hline
   35  & $1/48$ & 0 & $(10-3\Delta_\chi)/192$ & $(-1+\Delta_\chi)/128$ & $(10-3\Delta_\chi)/192$ & $(-1+\Delta_\chi)/128$ \\\hline
   36  & $1/48$ & 0 & $(10-3\Delta_\chi)/192$ & $(-1+\Delta_\chi)/128$ & $(10-3\Delta_\chi)/192$ & $(-1+\Delta_\chi)/128$ \\\hline
   37  & $1/48$ & 0 & $(10-3\Delta_\chi)/192$ & $(-1+\Delta_\chi)/128$ & $(10-3\Delta_\chi)/192$ & $(-1+\Delta_\chi)/128$ \\\hline
   38  & $1/48$ & 0 & $(10-3\Delta_\chi)/192$ & $(-1+\Delta_\chi)/128$ & $(10-3\Delta_\chi)/192$ & $(-1+\Delta_\chi)/128$ \\\hline
   39  & $1/48$ & 0 & $(10-3\Delta_\chi)/192$ & $(-1+\Delta_\chi)/128$ & $(10-3\Delta_\chi)/192$ & $(-1+\Delta_\chi)/128$ \\\hline
   40  & $1/48$ & 0 & $(10-3\Delta_\chi)/192$ & $(-1+\Delta_\chi)/128$ & $(10-3\Delta_\chi)/192$ & $(-1+\Delta_\chi)/128$ \\\hline
   41  & $1/48$ & 0 & $(10-3\Delta_\chi)/192$ & $(-1+\Delta_\chi)/128$ & $(10-3\Delta_\chi)/192$ & $(-1+\Delta_\chi)/128$ \\\hline
   42  & $-11/24$ & 0 & $(10-3\Delta_\chi)/96$ & 0 & $(-2-9\Delta_\chi)/96$ & 0 \\\hline
   43  & $-11/24$ & 0 & $(10-3\Delta_\chi)/96$ & 0 & $(-2-9\Delta_\chi)/96$ & 0 \\\hline
   44  & $-11/24$ & 0 & $(10-3\Delta_\chi)/96$ & 0 & $(-2-9\Delta_\chi)/96$ & 0 \\\hline
   45  & $-11/24$ & 0 & $(10-3\Delta_\chi)/96$ & 0 & $(-2-9\Delta_\chi)/96$ & 0 \\\hline
   46  & 0 & 0 & 0 & 0 & 0 & 0 \\\hline
\end{tabular}
\caption{Pole structure for $T^{AAS}$ operators $Q_1,\;Q_2,\;Q_3$ in the $E_I^\prime$ evanescent basis. As before $d$ labels the diagram class. Unlike Table~\ref{diag_results}, the remaining columns include the total contribution from all diagrams in the class. Only spin-color tensors with index symmetries appropriate for these operators are shown. $\delta_A^1 \equiv \delta_{\chi_1\chi_2}$, $\delta_A^2 \equiv \delta_{\chi_2\chi_3}+\delta_{\chi_1\chi_3}$.}
\label{results_A}
\end{table}

\begin{table} 
\begin{tabular}{|c||c|c||c|c||c|c|}
  \hline $\multirow{2}{*}{$d$}$ & \multicolumn{2}{c||}{\multirow{2}{*}{$(1\otimes 1 \otimes 1)T^{SSS}$}} & \multicolumn{2}{c||}{$(\sigma\otimes\sigma\otimes 1)T^{AAS} $} & \multicolumn{2}{c|}{\multirow{2}{*}{$(\sigma\otimes\sigma\otimes\sigma)T^{AAA}$}}\\
  & \multicolumn{2}{c||}{} & \multicolumn{2}{c||}{$+ (1\otimes\sigma\otimes\sigma)T^{SAA} + (\sigma\otimes 1\otimes \sigma)T^{ASA}$} & \multicolumn{2}{c|}{} \\\hline
  & $1/\bar{\varepsilon}^2$ & $1/\bar{\varepsilon}$ & $\hspace{25pt}1/\bar{\varepsilon}^2\hspace{25pt}$ & $1/\bar{\varepsilon}$ & $1/\bar{\varepsilon}^2$ & $1/\bar{\varepsilon}$ \\\hline\hline 
   1  & - & $-4$ & - & 0 & - & 0  \\\hline
   2  & - & $5/2$ & - & $-3/8$ & - & 0 \\\hline
   3  & - & $5/2$ & - & $-3/8$ & - & 0 \\\hline\hline
   4  & $-8/3$ & $8/3$ & 0 & 0 & 0 & 0 \\\hline
   5  & $(-13-13\delta_S)/12$ & $(65+13\delta_S)/24$ & $1/4$ & $-1/4$ & 0 & 0 \\\hline
   6  & $(-13-13\delta_S)/12$ & $(195+130\delta_S)/72$ & $1/4$ & $-15/32$ &  0 & 0 \\\hline
   7  & 0 & $-11/3$ & 0 & 0 & 0 & 0 \\\hline
   8  & 0 & $(38-19\delta_S)/12$ & 0 & $-7/16$ & 0 & 0 \\\hline
   9  & 0 & $(38-19\delta_S)/12$ & 0 & $-7/16$ & 0 & 0 \\\hline
   10  & $-2/3$ & $-2/3$ & 0 & 0 & 0 & 0 \\\hline
   11  & $5/12$ & 0 & $-1/16$ & $-5/32$ & 0 & 0 \\\hline
   12  & $5/12$ & 0 & $-1/16$ & $-5/32$ & 0 & 0 \\\hline
   13  & $-16/3$ & $8/3$ & 0 & 0 & 0 & 0 \\\hline
   14  & $10/3$ & 0 & $-1/2$ & $-1/4$ & 0 & 0 \\\hline 
   15  & $10/3$ & 0 & $-1/2$ & $-1/4$ & 0 & 0  \\\hline
   16  & $10/3$ & $(-60-35\delta_S)/36$ & 0 & $-1/8$ & 0 & 0 \\\hline
   17  & $10/3$ & $5/3$ & $-1/2$ & $-1/2$ & 0 & 0 \\\hline
   18  & $10/3$ & $(-60+35\delta_S)/36$ & 0 & $-1/8$ & 0 & 0 \\\hline
   19  & $10/3$ & $5/3$ & $-1/2$ & $-1/2$ & 0 & 0 \\\hline
   20  & $(1-\delta_S)/12$ & $(-3+\delta_S)/24$ & $-3/8$ & $3/16$ & 0 & 0 \\\hline
   21  & $(1-\delta_S)/12$ & $(-9+10\delta_S)/72$ & $-3/8$ & $27/32$ & 0 & 0 \\\hline
   22  & $-16/3$ & 0 & 0 & 0 & 0 & 0 \\\hline
   23  & $(-13-13\delta_S)/12$ & $0$ & $1/4$ & $0$ & 0 & 0 \\\hline
   24  & $(-13-13\delta_S)/12$ & $91\delta_S/72$ & $1/4$ & $-7/32$ & 0 & 0 \\\hline
   25  & $18$ & $-15$ & 0 & 0 & 0 & 0 \\\hline
   26  & $-45/4$ & $15/4$ & $27/16$ & $9/32$ & 0 & 0 \\\hline
   27  & $-45/4$ & $15/4$ & $27/16$ & $9/32$ & 0 & 0 \\\hline
   28  & 0 & 0 & 0 & $9/4$ & 0 & 0 \\\hline
   29  & $15/2 - N_f$ & $ -13 + 4N_f/3 $ & 0 & 0 & 0 & 0 \\\hline
   30  & 0 & 0 & $15/16 - N_f/8$ & $-17/32 + N_f/48$ & 0 & 0 \\\hline
   31  & 0 & 0 & $15/16 - N_f/8$ & $-17/32 + N_f/48$ & 0 & 0 \\\hline
   32  & $10/3$ & 0 & $-1/2$ & 0 & 0 & 0 \\\hline
   33  & $10/3$ & 0 & $-1/2$ & 0 & 0 & 0 \\\hline
   34  & $-25/48$ & 0 & $(10-3\Delta_\chi)/64$ & $(-3+3\Delta_\chi)/128$ & $-9/32$ & 0 \\\hline
   35  & $-25/48$ & 0 & $(10-3\Delta_\chi)/64$ & $(-3+3\Delta_\chi)/128$ & $-9/32$ & 0 \\\hline
   36  & $-25/48$ & 0 & $(10-3\Delta_\chi)/64$ & $(-3+3\Delta_\chi)/128$ & $-9/32$ & 0 \\\hline
   37  & $-25/48$ & 0 & $(10-3\Delta_\chi)/64$ & $(-3+3\Delta_\chi)/128$ & $-9/32$ & 0 \\\hline
   38  & $-25/48$ & 0 & $(10-3\Delta_\chi)/64$ & $(-3+3\Delta_\chi)/128$ & $-9/32$ & 0 \\\hline
   39  & $-25/48$ & 0 & $(10-3\Delta_\chi)/64$ & $(-3+3\Delta_\chi)/128$ & $-9/32$ & 0 \\\hline
   40  & $-25/48$ & 0 & $(10-3\Delta_\chi)/64$ & $(-3+3\Delta_\chi)/128$ & $-9/32$ & 0 \\\hline
   41  & $-25/48$ & 0 & $(10-3\Delta_\chi)/64$ & $(-3+3\Delta_\chi)/128$ & $-9/32$ & 0 \\\hline
   42  & $-25/24$ & 0 & $(10-3\Delta_\chi)/32$ & 0 & $-9/16$ & 0 \\\hline
   43  & $-25/24$ & 0 & $(10-3\Delta_\chi)/32$ & 0 & $-9/16$ & 0 \\\hline
   44  & $-25/24$ & 0 & $(10-3\Delta_\chi)/32$ & 0 & $-9/16$ & 0 \\\hline
   45  & $-25/24$ & 0 & $(10-3\Delta_\chi)/32$ & 0 & $-9/16$ & 0 \\\hline
   46  & 0 & 0 & 0 & 0 & 0 & $9/4$ \\\hline
\end{tabular}
\caption{Pole structure for $T^{SSS}$ operators $Q_4,\;Q_5,\;\tilde{Q}_1,\;\tilde{Q}_3$ analogous to Table~\ref{results_A}. $\delta_S \equiv \delta_{\chi_1\chi_2} + \delta_{\chi_2\chi_3}+\delta_{\chi_1\chi_3}$.}
\label{results_S}
\end{table}

\newpage 
\bibliography{NNbar} 


\end{document}